# High-throughput discovery of metal oxides with high thermoelectric performance via interpretable feature engineering on small data

**Shengluo Ma[1], Yongchao Rao[1], Xiang Huang[1], Shenghong Ju[1, 2, *]**

[1] China-UK Low Carbon College, Shanghai Jiao Tong University, Shanghai, 201306, China

[2] Materials Genome Initiative Center, School of Material Science and Engineering, Shanghai Jiao Tong University, Shanghai 200240, China

## ABSTRACT

In this work, we have proposed a data-driven screening framework combining the interpretable machine learning with high-throughput calculations to identify a series of metal oxides that exhibit both high-temperature tolerance and high power factors. Aiming at the problem of weak generalization ability of small data with power factors at high temperatures, we employ symbolic regression for feature creation which enhances the robustness of the model while preserving the physical meaning of features. 33 candidate metal oxides are finally targeted for high-temperature thermoelectric applications from a pool of 48,694 compounds in the Materials Project database. The Boltzmann transport theory is utilized to perform electrical transport properties calculations at 1,000 K. The relaxation time is approximated by employing constant electron-phonon coupling based on the deformation potential theory. Considering band degeneracy, the electron group velocity is obtained using the momentum matrix element method, yielding 28 materials with power factors greater than 50 $\mu Wcm^{-1}K^{-2}$. The high-throughput framework we proposed is instrumental in the selection of metal oxides for high-temperature thermoelectric applications. Furthermore, our data-driven analysis and transport calculation suggest that metal oxides rich in elements such as cerium (Ce), tin (Sn), and lead (Pb) tend to exhibit high power factors at high temperatures.

[*] Corresponding author: shenghong.ju@sjtu.edu.cn.

# 1. INTRODUCTION

The capture and recovery of waste heat is an end-of-pipe solution for enhancing energy efficiency. Selecting thermoelectric (TE) materials to convert heat into electricity is the most direct approach to implementing this solution[1-3]. Given that waste heat is often abundant in industrial processes, automobile exhaust, and other medium to high temperature environments exceeding 400K, more stringent requirements are placed on the design of TE materials[4, 5]. For instance, conventional TE materials containing elements such as Tellurium (Te) and Antimony (Sb) may face the risk of oxidation when exposed to high-temperature air[6]. Typically, half-Heusler alloys[7-10], skutterudites[7, 10, 11], and Zintl[7, 9, 11] compounds are suitable for mid-temperature applications ranging from 400K to 900K. Furthermore, rare earth chalcogenides[12], borides[12] and metal oxides[13-17] (including oxide perovskites[18, 19]) can be applied at a high temperature around 1000K[5]. From a commercial perspective, the elemental abundance and production cost of TE materials are also important[4]. Metal oxides are a preferred choice for commercial applications in high-temperature TEs due to their resistance to oxidation, cost-effectiveness, and ease of large-scale synthesis[20].

The TE properties are assessed by the figure of merit, $ZT = S^2T\sigma/(\kappa_e+\kappa_L)$. Here, $S$ represents the Seebeck coefficient, $\sigma$ denotes the electrical conductivity, $\kappa_e$ and $\kappa_L$ refer to the electronic and lattice thermal conductivity respectively, and $T$ signifies the absolute temperature. Notably, $S^2\sigma$, also known as the power factor (PF), serves as a comprehensive indicator of the electrical transport properties. Nevertheless, the maximal optimization of $ZT$ remains a challenge due to the interplay among transport parameters. Therefore, metal oxides offer the following advantages as candidates: Firstly, they have a wide range of electronic, chemical and mechanical properties that are favorable for screening[13, 17]. Secondly, by employing nanostructure engineering[21-23], metal oxides can achieve high carrier concentration, high PF and low thermal conductivity properties through the control of electrons and phonons. These effective doping mechanisms further enhance the TE properties of metal oxides[4, 13, 14, 24].

High-throughput (HTP) screening has become an emerging tool for finding promising TE materials. Zhu et al.[25] predicted the high TE performance of triangular and quadrilateral $XYZ_2$ compounds by screening over 9000 materials from the Materials Project (MP)[26] database using PF calculations. Similarly, Ricci and Chen et al.[27, 28] evaluated the electrical transport properties of nearly 48,000 compounds from the MP database using the constant relaxation time approximation (CRTA). Wang et al.[29] investigated thousands of compounds in the AFLOW repository and analyzed the characteristics of materials with high PFs. Carrete et al.[30] screened 75 nanograined compounds from AFLOW and estimated their *ZT* via the constant mean free path approximation (CMFPA). However, both the CRTA and CMFPA treatments neglect the material dependence of charge carrier scatterings, thus raising doubts about their accuracy. Electron-phonon scattering serves as a crucial mechanism for measuring TE properties, particularly at high temperatures. Accurate predictions of the electronic relaxation time can be obtained through ab initio calculations of electron-phonon coupling[31], but this method is extremely time-consuming and not conducive to HTP calculations. Recently, electrical transport calculations based on deformation potential (DP) theory[32] have been used in HTP work, where the electron-phonon coupling strength is evaluated by $(\boldsymbol{E}_{\mathrm{DP}})^2/\boldsymbol{Y}$, with $\boldsymbol{E}_{\mathrm{DP}}$ denoting the DP constant and $\boldsymbol{Y}$ representing Young's modulus. Jia et al.[33] identified 50 promising TE materials among 243 chalcogenides using this constant electron-phonon coupling approximation (CEPCA) through DP theory. Xi et al.[34] listed a series of novel high-performance TE chalcogenides through CEPCA applied in HTP workflow based on the Materials Informatics Platform (MIP). Jin et al.[35] screened more than 600 potential TE compounds through CEPCA in the MIP. CEPCA can be perceived as a method compatible with HTP, taking into consideration material dependence without imposing excessive computational demand. Currently, HTP screening of metal oxide for high-temperature TE application scenarios (especially around 1000 K) has not been reported. The challenge resides in assessing material stability under high-temperature conditions. Furthermore, the lack of scale-consistent data regarding high-temperature TE properties hinders the execution of virtual screening, even though machine learning (ML) can be applied to small data[36, 37].

Herein, we designed an HTP framework that combines interpretable ML for virtual screening and DFT-based electrical transport calculations via CEPCA. Through this framework, we identified a series of metal oxides with potentially high PFs in high-temperature waste heat recovery. Melting points[38] are crucial for applications in high-temperature scenarios. We evaluated the high-temperature tolerance properties by accessing the melting point prediction model[38], which combined the Graph Neural Network (GNN) and residual neural network (ResNet) architectures. Due to the limited availability of *ZT* data at high temperatures (approximately 1000K), we chose PF as a screening target[15]. To begin with, we collected a small dataset of PFs for 67 metal-like crystal materials containing metal oxides at different temperatures from 500K to 1000K. To address the issue of overfitting in ML models trained on small datasets, we proposed a feature creation technique based on symbolic regression[39-41] (SR) in addition to the original descriptor down-selection[40]. This technique generated training features that exhibit a larger correlation with the PFs, thereby enhancing the robustness of the trained models. We validated the effectiveness of the technique from two perspectives: model training performance and feature contribution of Shapley Additive exPlanations[42] (SHAP). The results demonstrated that SR has the advantages of strong targeting and high interpretability in feature creation. We employed three PF prediction models and the melting point prediction model to conduct virtual screening on 48,694 metal oxides in the MP database. Metal oxides containing elements Cerium (Ce) or Tin (Sn) tend to have high PFs in high-temperature scenarios according to the data-driven analysis. The PFs of 33 target materials within the ML model intersection were calculated by Boltzmann transport theory using CEPCA. In contrast to the previous HTP work[35] that approximated the Young's modulus with bulk modulus, here we calculated the Young's modulus directly. Then we use the deep core states of the atoms[43, 44] as reference energy levels to fit the resulting DPs under different strains. From the results, 28 metal oxides exhibit PFs exceeding 50 $\mu Wcm^{-1}K^{-2}$ at 1000K. Finally, we analyze the influence of different metal elements in various metal oxides on high-temperature TE. Overall, our proposed HTP framework facilitates the screening of metal oxides in high-temperature TE applications.

## 2. COMPUTATIUONAL METHODS

### *2.1 Electrical Transport Properties Calculation based on DP theory*

All calculations were conducted using the Vienna ab initio Simulation Package (VASP) with the projector augmented wave (PAW) method, which is based on density functional theory (DFT)[45, 46]. The Perdew-Burke-Ernzerhof (PBE) generalized gradient approximation (GGA)[47] was employed as the exchange-correlation functional in the underlying calculations related to the structure and elastic constants. All the self-consistent calculations were performed using a plane-wave energy cutoff of 600 eV and an energy convergence criterion of $10^{-6}$ eV. All the atomic positions were fully relaxed until the calculated Hellmann−Feynman force on each atom was less than $10^{-2}$ eV/Å. The electronic structures and transport properties were calculated using the Strongly Constrained and Appropriately Normed (SCAN) semilocal density functional[48], and the Hubbard-U correction with Dudarev's approach was applied. This approach has demonstrated its effectiveness in other TE materials and the setup of Hubbard U values referenced the work done by Yao et al[49].

Electrical transport properties were calculated based on the Boltzmann transport theory[50]. The electrical conductivity $\sigma$, Seebeck coefficient $S$, and electronic thermal conductivity $\kappa_e$ were expressed as:

$$\sigma_{\alpha\beta}(\mu,T)=\frac{1}{V}\sum_{n\mathbf{k}}v_{n\mathbf{k}\alpha}v_{n\mathbf{k}\beta}\tau_{n\mathbf{k}}\left[-\frac{\partial f_{\mu}(\varepsilon_{n\mathbf{k}},T)}{\partial\varepsilon_{n\mathbf{k}}}\right] \tag{1}$$

$$S_{\alpha\beta}(\mu,T)=\frac{\sigma_{\alpha\beta}(\mu,T)^{-1}}{eTV}\sum_{n\mathbf{k}}v_{n\mathbf{k}\alpha}v_{n\mathbf{k}\beta}\tau_{n\mathbf{k}}(\mu-\varepsilon_{n\mathbf{k}})\left[-\frac{\partial f_{\mu}(\varepsilon_{n\mathbf{k}},T)}{\partial\varepsilon_{n\mathbf{k}}}\right] \tag{2}$$

$$\begin{aligned}\kappa_{e\alpha\beta}(\mu,T)=&\frac{1}{e^{2}TV}\sum_{n\mathbf{k}}v_{n\mathbf{k}\alpha}v_{n\mathbf{k}\beta}\tau_{n\mathbf{k}}(\mu-\varepsilon_{n\mathbf{k}})^{2}\left[-\frac{\partial f_{\mu}(\varepsilon_{n\mathbf{k}},T)}{\partial\varepsilon_{n\mathbf{k}}}\right]\\&-\frac{1}{eV}\sum_{n\mathbf{k}}v_{n\mathbf{k}\alpha}v_{n\mathbf{k}\gamma}\tau_{n\mathbf{k}}(\mu-\varepsilon_{n\mathbf{k}})\left[-\frac{\partial f_{\mu}(\varepsilon_{n\mathbf{k}},T)}{\partial\varepsilon_{n\mathbf{k}}}\right]S_{\gamma\beta}(\mu,T)\end{aligned} \tag{3}$$

Here, $\varepsilon_{n\mathbf{k}}$ is the band energy corresponding to band index $n$ and reciprocal coordinate $\mathbf{k}$, $\tau_{n\mathbf{k}}$ and $v_{n\mathbf{k}}$ are, respectively, the electronic relaxation time and group velocity. $\mu$, $T$, $V$, $f_{\mu}$, and $e$ are the Fermi level, the absolute temperature, the volume of unit cell, the Fermi-Dirac distribution, and the electron charge, respectively.

The relaxation time was treated by the constant electron-phonon coupling approximation (CEPCA) based on the DP method[32]. Specifically, the scattering from long-wave LA phonons can be described and evaluated by calculating the two parameters of DP and Young's modulus. Elastic constant matrices for all materials were calculated via the stress-strain relationship and Young's modulus was obtained by the Voigt-Reuss-Hill (VRH) theory[51]. In this work, the deformation potentials for the VBM and CBM were considered separately, which were fitted in a strain range of $\pm 0.3$, with an interval of 0.1. The reference level for the DP calculations was the deep core states of the atoms, and the R-squared accuracy of all linear fits is greater than 0.9. The DP calculation details can be found in Supplementary Note 7. The relaxation time $\tau_{n\mathbf{k}}$ is calculated using the following formula:

$$\tau_{n\mathbf{k}}^{-1}=\frac{2\pi k_{B}TE_{DP}^{2}}{V\hbar Y}\sum_{m\mathbf{k}'}\delta(\varepsilon_{n\mathbf{k}}-\varepsilon_{m\mathbf{k}'}) \tag{4}$$

where $E_{DP}$ is the deformation potential of the band edge, $Y$ is the Young's modulus, and $\delta(\varepsilon_{n\mathbf{k}}-\varepsilon_{m\mathbf{k}'})$ adopts the form of Gaussian function.

Considering the band degeneracy, the electron group velocity was treated by the momentum matrix method[52-54]. The electron group velocity[55] corresponding to energy $\mu$ is calculated as shown below:

$$v_e(\mu,T)=\frac{\sqrt{\sum_{n\mathbf{k}}\left(v_{n\mathbf{k}x}^2+v_{n\mathbf{k}y}^2+v_{n\mathbf{k}z}^2\right)\left[-\frac{\partial f_\mu(\varepsilon_{n\mathbf{k}},T)}{\partial\varepsilon_{n\mathbf{k}}}\right]}}{\sum_{n\mathbf{k}}\left[-\frac{\partial f_\mu(\varepsilon_{n\mathbf{k}},T)}{\partial\varepsilon_{n\mathbf{k}}}\right]} \quad (5)$$

It should be noted that $v_e$ represents the average of the anisotropic electron group velocitiy $v_{\mathbf{k}x}$, $v_{\mathbf{k}y}$, and $v_{\mathbf{k}z}$. The relaxation time calculations via CEPCA and the electron group velocity solution are conducted using TransOpt 2.0[53] software. The details of electrical transport properties calculation can be found in Supplementary Table S8.

*2.2 Data Collection and Feature Preparation for Power Factors*

PF data for 67 metal-like (including metal oxides) materials with non-zero band gaps were collected. The total 402 PFs data were obtained from 500K to 1000K by CRTA with a constant scattering time of $10^{-14}$ s, following a unified DFT computational standard[56]. PF data for ML can be downloaded from the https://github.com/SJTU-MI/HTPS4HTTEMOs. For the initial descriptors, a series of material features in the MP database[26] were accessed through the pymatgen API[57]. Then we processed the features and obtained 20 descriptors that encompass the crystal structure, computational information, and fundamental properties. The 290 compositional descriptors were calculated through Xenonpy[58] software. The temperature corresponding to PF was also considered as an input descriptor. A total of 311 descriptors were collected, and detailed information can be found in Supplementary Note 2.

*2.3 Feature Engineering via Symbolic Regression*

The SR formula-building strategy[39, 40] was taken into our feature creation process and implemented in gplearn[59] software. Considering that the genetic optimization of SR is not suitable for large dimensions, 18 descriptors after the descriptor down-selection were utilized as input variables (see Supplementary Table S4). SR was performed on all PF data using Pearson, Spearman, and Distance coefficients as fitness functions, to obtain three new sets of descriptors that exhibited higher correlations. Further, the grid search strategy with the hyperparameters and metrics as listed in Table 1 was applied to determine the mathematical formulas. We listed 12 formulas at the Pareto front that were identified by the bivariate density distribution approach. More information about SR can be found in Supplementary Note 4.

**Table 1:** Setup of hyperparameters in gplearn software for SR

| Parameter | Value | Combination |
|---|---|---|
| Generations | 300 | 1 |
| Population size in every generation | 5000 | 1 |
| Probability of crossover (pc) | [0.30, 0.85], (step = 0.05) | 477 |
| Probability of subtree mutation (ps) | [(1-pc)/3, (1-pc)/2] (step = 0.01) | |
| Probability of hoist mutation (ph) | [(1-pc)/3, (1-pc)/2] (step = 0.01) | |
| Probability of point mutation (pp) | 1-pc-ps-ph | |
| Function set | $\{+, -, \times, \div, \sqrt{x}, \ln x, \lvert x \rvert, -x, 1/x\}$ | 1 |
| Parsimony coefficient | auto | 1 |
| Metric | Pearson, Spearman, Distance Cor. | 3 |
| Stopping criteria | 0.90 | 1 |
| Random_state | 0, 1, 2 | 3 |
| Init_depth | [2, 6], [4, 8], [6, 10] | 3 |

### *2.4 Construction and Analysis of ML Models*

The ML models of Random Forest (RF), Extreme Gradient Boosting (XGBoost), and Multi-Layer Perceptron (MLP) were implemented by using Scikit-learn[60]. The training set and test set were randomly divided in a 9:1 ratio. Hyperparameter optimization was conducted using the Bayesian Optimization package[61], which is a global optimization tool, to achieve a high prediction accuracy ($R^2$). The Gaussian process and acquisition function were initially trained with 20 random pairs of parameters, and the ideal parameters for each ML model were determined after 200 optimization iterations.

To explain the association of the descriptors created by SR with PF, the SHAP[42] toolkit was used to assess the feature importance based on the RF model. The SHAP analysis is a method based on game-theoretic Shapley values to interpret the contribution of features for ML predictions.

### *2.5 Screening from Melting Point Prediction Models*

Melting points were utilized to exclude metal oxides that are not stable under high-temperature operating conditions. Our high-throughput requirements were met by a melting point model[38] trained using a database of around 10,000 compounds developed by Hong et al. The melting point model was accessed using an API to predict the melting temperature of over two thousand candidate materials. Metal oxides with melting points greater than 1200K were retained. To enable HTP prediction, we wrote a script that retrieves and processes material information from the MP database, generating a JSON file as output. This JSON file can be used to input melting point models and obtain results via API commands. The script is available from the https://github.com/SJTU-MI/HTPS4HTTEMOs.

## 3. RESULTS AND DISCUSSION

### *3.1 Screening of High-temperature TE Metal Oxides*

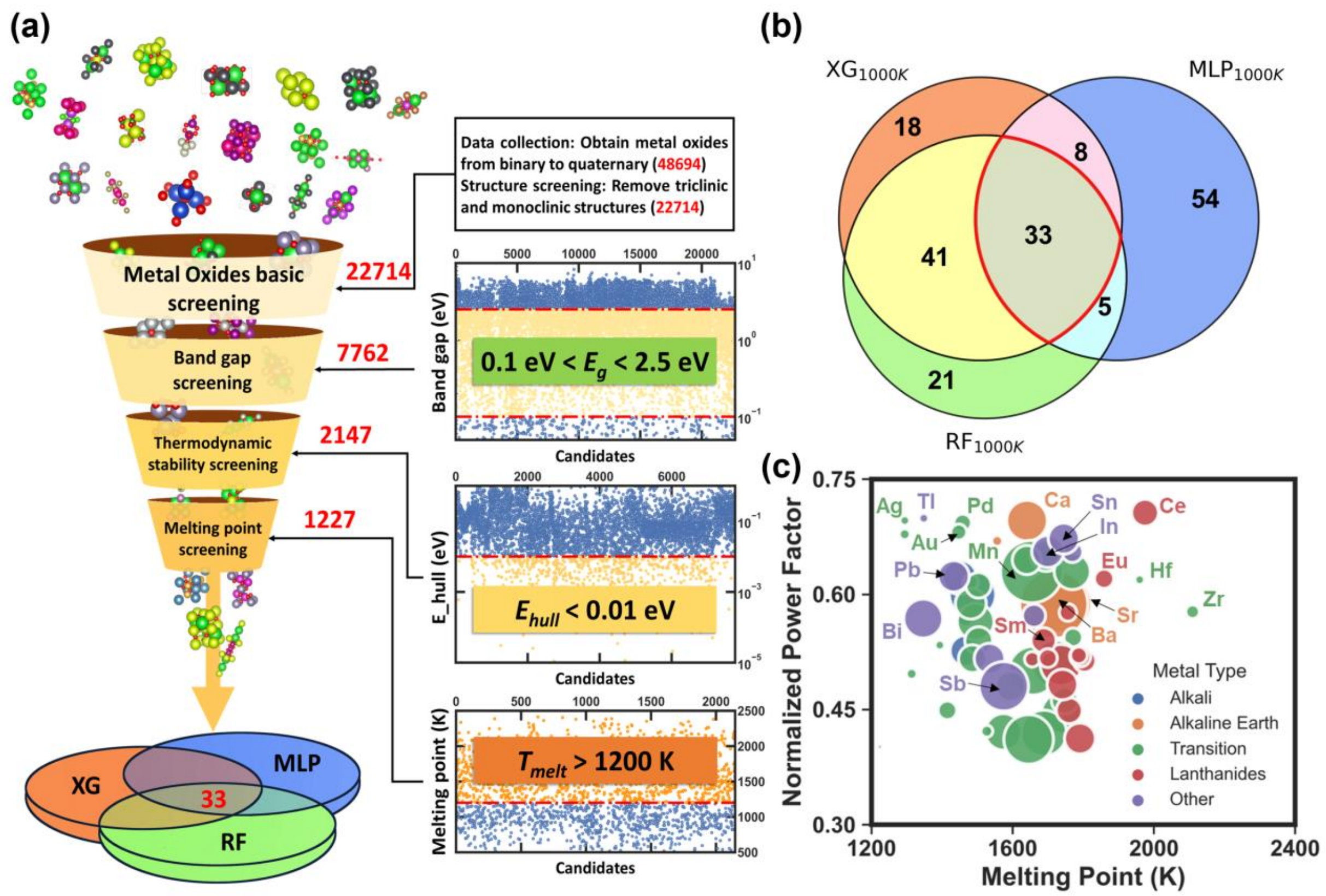

**Fig. 1.** Data-driven virtual screening and analysis for high-temperature TE metal oxides. (a) HTP virtual screening framework for metal oxides. (b) Multi-model ensemble screening results for top 100 materials with PF ranking at 1000K. (c) Statistical Analysis for normalized PF vs. melting point of 1227 high-temperature TE candidate metal oxides shown via scatter plot.

Fig. 1a illustrates the HTP screening process conducted in this work. The details are as follows: (1) We retrieved a total of 48,694 binary to quaternary metal oxides from the MP database as screening candidates, excluding materials containing highly radioactive metal elements. (2) Subsequently, we removed crystal structures with weak symmetry such as triclinic and monoclinic[62], leaving 22,714. (3) Considering the band gap correction for final HTP calculations, we retained 7762 candidates with band gaps (as obtained from GGA-based DFT in the MP database) ranging from 0.1 to 2.5 eV. (4) Then we limited our study to thermodynamically stable compounds with energy above the convex hull less than 0.01 eV per atom, thus, the number of compounds was reduced to 2147. (5) To meet the requirement of high-temperature (around 1000K) applications for TE materials, we utilized the melting points prediction model[38] to screen 1227 metal oxides with melting temperatures above 1200K, considering prediction errors and fluctuations in operating temperature. (6) Finally, we predicted the PFs of 1227 materials using the three machine learning models previously trained on a small dataset (to be detailed later) and obtained the intersection of the top 100 candidate sets ranked by each model as the final screening result.

Multi-model ensemble screening is a comprehensive strategy to enhance the accuracy of selection[63]. Fig. 1b presents the top 100 candidate rankings predicted by three different machine learning models at a temperature of 1000K, with a total of 33 materials lying at the intersection of all three models' predictions. Given that this intersection is derived from the predictions of three independent ML models, the tightness of the intersection indirectly indicates the strong generalization capability of our screening models, even when trained on small data. The intersection between the XGBoost and RF models is larger than their respective intersections with the MLP model. This is because both XGBoost and RF models have evolved from tree-based theories[64-66], and their hierarchical tree-like learning strategies lead to more similar prediction outcomes. Similarly, we also conducted ensemble screening at a temperature of 900K, as shown in Supplementary Fig. S1. The intersection and union of the models were nearly identical to those at 1000K, with only very few materials differing. For instance, the quantity of materials in the central intersection at 900K matches that at 1000K. However, there is a difference in one material between the two sets, as shown in Supplementary Table S1. Such subtle differences arise because all three models incorporate temperature contributions, which allow for considering the TE responses of various materials at different temperatures.

Additionally, given that the presence of different metal elements can significantly influence the properties of metal oxides, we conducted a statistical analysis on the collection of 1227 compounds incorporating the melting points model and three PFs model from a data-driven perspective. Fig. 1c displays the overall distribution of different metal elements in the screening targets for metal oxides, where the center of the scatter points represents the average melting point and normalized PF for all compounds containing the corresponding element, and the size of the scatter points indicates the number of materials in the selection results. Supplementary Fig. S2 comprehensively shows the prediction of different materials by each element under the selection models. Compounds in the collection containing elements such as Zr, Ce, Hf, Eu, etc., tend to have higher melting points. From the perspective of elemental categories, these elements possess high-energy d and f orbital electron subshells, which leads to stronger bonding with oxygen ions and thus results in high melting point characteristics[67]. In terms of PF, oxides containing elements such as Ce, Ag, Tl, Ca rank higher. However, due to the complex coupling mechanisms in TE performance, further transport calculations are needed for analysis. Additionally, the scatter points for elements like Ba, Mn, Sr are quite large, indicating their high prevalence in the oxide collection, and thus increasing their probability as one of the elemental combinations in candidate materials. Overall, metal oxides containing Ce and Sn are preferred choices for high-temperature TE materials. Firstly, these compounds are more likely to exhibit high PF in high-temperature applications. Secondly, they have an advantage in terms of the number of candidates within the screening set.

### *3.2 Feature Engineering and ML Model Training for Power Factors*

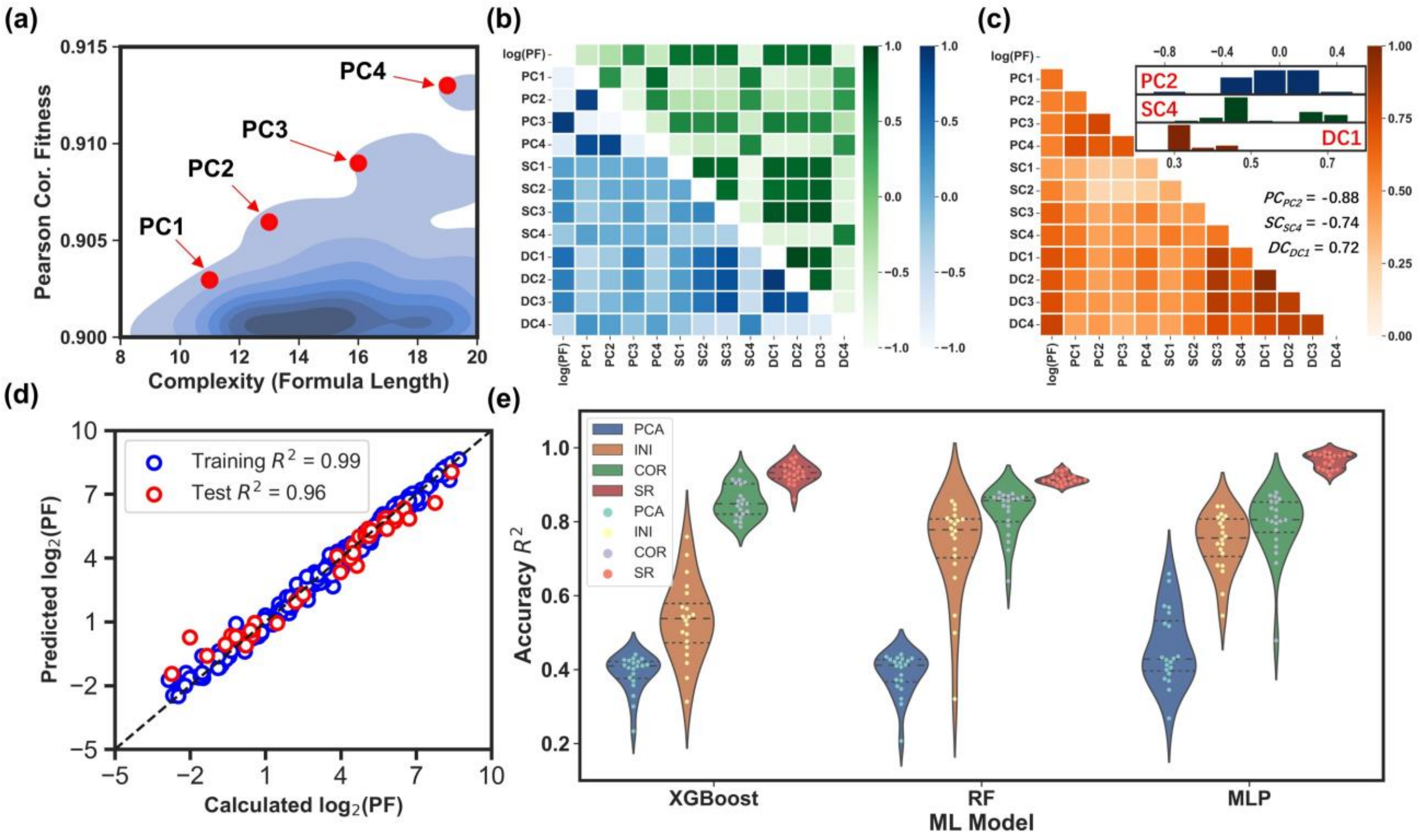

**Fig. 2.** Feature creation via SR and ML Model Training for PFs. (a) Pareto front from Pearson Cor. fitness vs. complexity of SR formulas shown via density plot. (b) and (c) are heat maps of the new descriptors for Pearson (blue part), Spearman (green part), and Distance (orange part) correlations. (d) Accuracy of RF model based on SR descriptors, where training $R^2$ is 0.98 and test $R^2$ is 0.96. (e) Accuracy of ML models at different feature engineering processes, including initial (INI), mathematical correlation (COR) coefficients filtering, and feature creation via SR stages. And, an additional PCA approach was applied to compare.

Metal oxides have a broader definition compared to previous HTP work of TE materials in fixed structures or systems. Different metal elements form compounds with oxygen and exhibit various crystal structures, resulting in a wide range of metal oxides. Here, we focus on screening binary to quaternary metal oxides. This is because the increase in the number of metal elements is accompanied by the emergence of the high entropy effect[68], which is not conducive to digging out some potential laws in our work. We limited the metal elements to between the second to sixth groups of the periodic table, as shown in Supplementary Fig. S2. We collected 402 PFs data for 67 crystal materials containing the aforementioned metal elements in the temperature range of 500K to 1000K.

Feature engineering for metal oxide PF data involves four main steps: collecting initial features, removing features with low variance, filtering features through three correlation coefficients[40], and creating better features via symbolic regression. The initial descriptors consist of 20 descriptors obtained from the MP database[26] and followed by post-processing, 290 compositional descriptors calculated through Xenonpy[58, 69] software, and the temperature descriptor for PF calculation, totaling 311. Further details can be found in Supplementary Note 2. We removed descriptors with a variance less than 0.01, as these descriptors are unsuitable for fitting machine learning models[40]. In the end, 297 descriptors were retained. In the field of materials informatics, there has been great interest in mathematical correlations between descriptors and material properties. To assess the linear, monotonic, and nonlinear relationships, we developed an evaluation system based on Pearson, Spearman, and Distance correlations. As illustrated in Supplementary Fig. S3, we have delineated the probability distribution of three correlation coefficients for different descriptors to the PF. The Pearson correlations (PC) predominantly fall within the ±0.2 range, while the Spearman correlations (SC) are generally confined within ±0.3. Additionally, the Distance correlations (DC) are universally below 0.35. This evidences a tepid correlation between descriptors and the PF, primarily due to the excessive redundancy in the metal oxide small data. This redundancy is attributable to the fact that the same material, despite varying temperatures, continues to be characterized by a consistent set of recorded descriptors such as elemental composition and crystal structure. However, the considerable redundancy within the dataset substantially impairs the robustness of predictive models in small data[36]. To address this, we implemented a weight assignment mechanism[40], as documented in our prior work, for the subsequent phase of descriptor selection. In this study, equal weight was given to the three correlation assessments, with selection thresholds set at 0.166, 0.276, and 0.302, respectively, which are also marked out in Supplementary Fig. S3. A total of 35 descriptors met one of the correlation selection criteria, while 18 descriptors satisfied at least two of these conditions. The detailed descriptors, optimized across various stages of low-variance and correlation-based selection, are documented in Supplementary Note 3. The employment of these two methods ensures the preservation of the intrinsic meaning of our original descriptors throughout the dimensionality reduction process, facilitating the exploration of relationships inherent in the raw data, as opposed to methodologies like Principal Component Analysis[70] (PCA). Moreover, the correlation-based selection has highlighted descriptors that are directly and indirectly related to PF, such as electronic orbital matrices and electronegativity. However, these descriptors are derived from elemental composition rather than directly calculated from physical processes, like the sparse matrix in electronic orbitals does not promote a strong correlation with electrical transport properties.

After completing the feature selection, the robustness of the ML models could not be effectively improved, as evidenced by the comparison of descriptor sets in Fig. 2e. The primary reason is the persistent lack of features strongly correlated with the PF. To tackle this issue, we employed a reverse mind by utilizing symbolic regression (SR)[39] to feature creation rather than conventional feature interpretation. SR has traditionally been used to formulate explanations for different descriptors in ML models. However, our focus is on creating formulaic descriptors that exhibit stronger correlations with the PF through SR. In contrast to explanations, the difference lies in the configuration of the objective function: the former starts with fitting the target attribute, whereas the latter sets the loss function to reflect the correlation fitness. We conducted a grid search for hyperparameter optimization in SR on all PF data and the 18 retained features, with the detailed hyperparameter settings provided in Table 1. Feature creation was carried out based on three correlations, and the Pareto front[40, 71, 72] was identified using the bivariate density distribution method (as shown in Fig. 2a), resulting in a total of 12 formulaic descriptors. As can be seen from Fig. 2b-c, the 12 descriptors generated through SR show significant improvement across all three correlations. Here, considering the breadth of data coverage and to prevent feature collinearity, we selected three descriptors, PC2, SC4, and DC1, for inclusion in the subsequent training of the ML models. Details related to SR can be found in Supplementary Note 4.

Fig. 2d presents the results of the RF[64, 65] model trained with the SR set. This set comprises 18 descriptors after correlation-based filtering and 3 new descriptors created by SR. The accuracy $R^2$ of the training and test sets are 0.98 and 0.96. To verify the scalability of the new descriptors, we deployed XGBoost[66] and MLP[73] models for training (see Supplementary Note 5). The XGBoost model demonstrated training $R^2$ of 0.99 and test $R^2$ of 0.96, while the MLP model showed $R^2$ of 0.99 and 0.97, respectively. The performance of ML models with different descriptor sets, evaluated by $R^2$, is shown in Fig. 2e (with RMSE presented in Supplementary Fig. S9). Overall, all three models trained with the SR set basically exhibited higher accuracy than other descriptor combinations, and the boxplot distribution indicates that the SR descriptors significantly enhanced the models' robustness, reducing overfitting. This validates the effectiveness of feature creation via SR and underscores the potential of SR descriptors in various ML model applications. Furthermore, due to the formulaic creation, the inherent meaning represented by the data has not been altered. This is beneficial for our subsequent analysis of the data from a physical perspective.

*3.3 Physical Insights from the Interpretable ML Model*

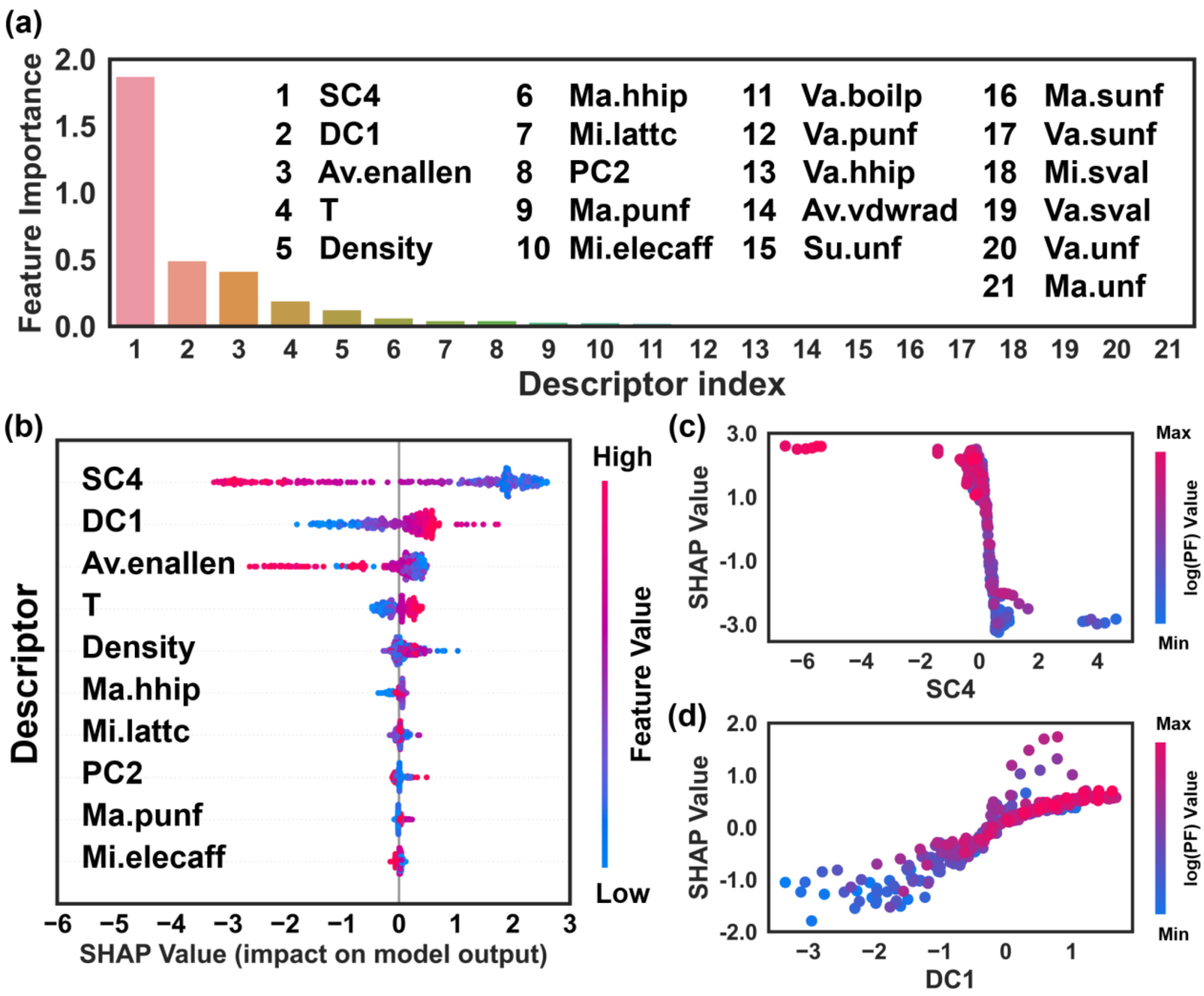

**Fig. 3.** Analysis of feature importance using SHAP on RF model trained by SR descriptors set. (a) Mean absolute SHAP values for 21 descriptors. (b) Represent the SHAP values of each descriptor related to the PF data in a beeswarm plot. (c) and (d) SHAP values of DC1 and PC2 in a dependence plot.

Fig. 3 summarizes the feature contributions to the RF model trained on the SR sets using SHAP. The SHAP approach attempts to resolve the unexplainable black-box challenge of ML models by calculating the marginal contribution of features to the model output. Fig. 3a shows the importance and ranking of features based on the mean absolute SHAP values. Among them, SC4 and DC1 are identified as the top two features, representing the features created based on monotonicity and nonlinearity, respectively. In addition, the feature importance ranking based on impurity carried out from the tree model (see Supplementary Fig. S11) is similar to the above results. This further demonstrates the effectiveness of SR in feature creation compared to previous machine learning modeling efforts. Fig. 3b lists the distribution of SHAP values for the top ten ranked features, where the color of data points in the beeswarm plot indicates the magnitude of the feature values. Overall, the SHAP values for higher-ranked features have a broader distribution, but the relationship between SHAP values and feature values is not simply linear, particularly for features with lower rankings. This is due to the complex nonlinear interactions among features affecting the PF or the excessive localization of discrete features, as shown in Supplementary Fig. S12.

Herein, we discuss the feature dependence of descriptors SC4 and DC1 through the SHAP analysis. The reason is that they rank top two and still retain the original meaning of the data. The distribution range of SHAP values in SC4 is significantly larger compared to other features (Fig. 3b), suggesting that SC4 has the greatest predictive impact on the PF. From Fig. 3c, it can be observed that the standardized values of SC4 are negatively correlated with the SHAP values, indicating that as SC4 increases, the predicted value of the PF tends to decrease. Conversely, DC1 exhibits a generally positive correlation trend in its SHAP values distribution (Fig. 3d). The SHAP impacts of these two descriptors are consistent with the correlation analysis (see Supplementary Table S6).

Compared to other descriptors, the SHAP value distributions of SC4 and DC1, created by SR, are more uniform and less concentrated, with fewer nonlinear interactions on model output. Reflected in the sub-features of the SR formulas (see Supplementary Table S6), this provides physical causality references for our selection of elements or materials. For SC4, elements with a greater number of unfilled electrons tend to have a higher PF[74], as an increase in the number of unfilled electrons typically indicates more carriers, which usually leads to higher electrical conductivity. The d orbitals in transition metals and f orbitals in lanthanide series are more likely to have unfilled electrons, making them potential TE metal oxides[17]. Regarding DC1, compounds with higher electron affinity and larger lattice constants are more favorable for enhancing the prediction of PF. Electron affinity, in simple terms, refers to the ability of an atom to attract and accept electrons, which is related to electrical transport, although this influence is complex[75]. Moreover, the lattice constant, as a parameter describing the distance between atoms, is also intricately linked to the calculation of electrical transport[76]. We observed the sub-features of lattice constant and temperature that coexist in both SC4 and DC1. These sub-features consistently demonstrate the same trend under different formulas, highlighting the reliability and consistency in SR. Overall, SR can establish a stronger response relationship with the PF by the formulaic combination of features that have physical associations but are discrete in data, such as electronic orbital descriptors.

*3.4 HTP Electrical Transport Calculations Based on Deformation Potential*

Considering that the screening results at 1000K and 900K were nearly identical, we combined them to identify a total of 34 candidate materials with high-temperature resistance. Due to the presence of transition metals and lanthanide elements in these materials, the strong correlation interactions on the d and f orbitals cannot be overlooked. Hence, we implemented the DFT energy correction and got the adjusted band gap[77, 78]. Additionally, considering the optimal band gap range for TE materials at high temperatures[79], materials with excessively large band gaps (greater than 3.5 eV) were excluded. We conducted HTP DFT calculations on the electrical transport properties at 1000K for the 33 suitable materials (comprising 28 non-magnetic systems and 5 magnetic systems) based on the Boltzmann transport theory[50, 80].

Taking computational convenience into account, previous researchers like Jin et al.[35] used the bulk modulus as a substitute for Young's modulus to evaluate the strength of electron-phonon coupling. However, Young's modulus, as per formula $\boldsymbol{Y} = 9\boldsymbol{BG}/(3\boldsymbol{B}+\boldsymbol{G})$, is influenced not only by the bulk modulus $\boldsymbol{B}$ but also by the shear modulus $\boldsymbol{G}$, which in turn affects the calculation of the electron-phonon coupling constant. We calculated the elastic constant matrices of all materials based on the stress-strain relationship and obtained Young's modulus. Building upon this, we incorporated a strain range gradient of ±0.3 to further calculate the band structures of the materials and obtain the values for the Valence Band Maximum (VBM) and Conduction Band Minimum (CBM). During the calculation, the reference energy levels for P-type and N-type were taken from the 1s energy levels of the atoms that contribute significantly to the corresponding VBM and CBM, which are the core energy levels[43, 44]. All deformation potentials have an R-squared value greater than 0.9 for linear fitting, and the related fitting details can be found in Supplementary Fig. S13. Furthermore, in Supplementary Fig. S14, we compared the deformation potentials obtained using the bare eigenvalues of the electronic structures for some compounds from the MatHub-3d database[35]. The deformation potentials from both methods are distributed near the equal-value diagonals, demonstrating the viability of our approach. Fig. 4a shows the distribution of deformation potentials at the VBM and CBM for the candidate materials. As shown, the range of deformation potentials at the VBM is between -6 and 4.3 eV, which is more concentrated and slightly narrower than that at the CBM (-7.4 to 7.8 eV). There are no deformation potentials at the CBM with an absolute value less than 1. Overall, the distribution of deformation potentials is centered around the value of 0 but is not symmetrically distributed, with both types of deformation potentials showing a non-smooth gradient descent into negative values, particularly at the VBM. Detailed information on Young's modulus and deformation potentials of the materials can be referred to in Supplementary Table S8.

The constant relaxation time approximation overlooks the material dependency of relaxation time, however, electron-phonon scattering is a crucial mechanism in the evaluation of TE material performance, especially at high temperatures. We used the DP theory[32] to calculate the electron-phonon coupling constant and effectively estimate the relaxation time under the constant electron-phonon coupling approximation. As shown in Supplementary Fig. S15, we compared the electrical transport calculations under the constant relaxation time approximation and the constant electron-phonon coupling approximation using $Cu_2O$ as an example[35, 81]. Accurate estimation of relaxation time is beneficial for high-throughput calculations in assessing TE material performance. Similarly, given that band degeneracy[82] exists in most of the computational candidate systems (the details of band structure can be referred to Supplementary Note 9), the use of the momentum matrix element method[52] to calculate electron group velocities can effectively avoid inaccuracies in solving band gradients at band degeneracy points. Fig. 4b presents the maximum PF in relation to hole (electron) concentration at 1000K under P-type and N-type doping for each material. In total, 28 materials exhibited a PF exceeding 50 $\mu Wcm^{-1}K^{-2}$, with 21 materials meeting the criteria under P-type doping and 17 under N-type, and 10 materials exceeded the 50 $\mu Wcm^{-1}K^{-2}$ threshold for both P-type and N-type. The results verify the reliability of our high-throughput screening of metal oxides from a computational standpoint. Fig. 4b also lists the top 5 materials with the highest PF under P-type and N-type doping (PFs greater than 400 $\mu Wcm^{-1}K^{-2}$). It can be seen that in P-type, each material contains the element Ce, and in N-type, there are two materials containing the element Sn, consistent with our previous analysis from a data-driven perspective. Most materials contain elements Ba and Sr, which is inseparable from the high occupancy rate mentioned earlier. From the analysis of band structures and density of states (see Supplementary Note 9), Ba and Sr are not the main contributing atoms at the VBM and CBM. As alkaline earth metals, they exhibit a very stable +2 oxidation state compared to transition metals and lanthanides. However, regarding the influence of elements, we still need to analyze and interpret from a computational perspective rather than making qualitative judgments.

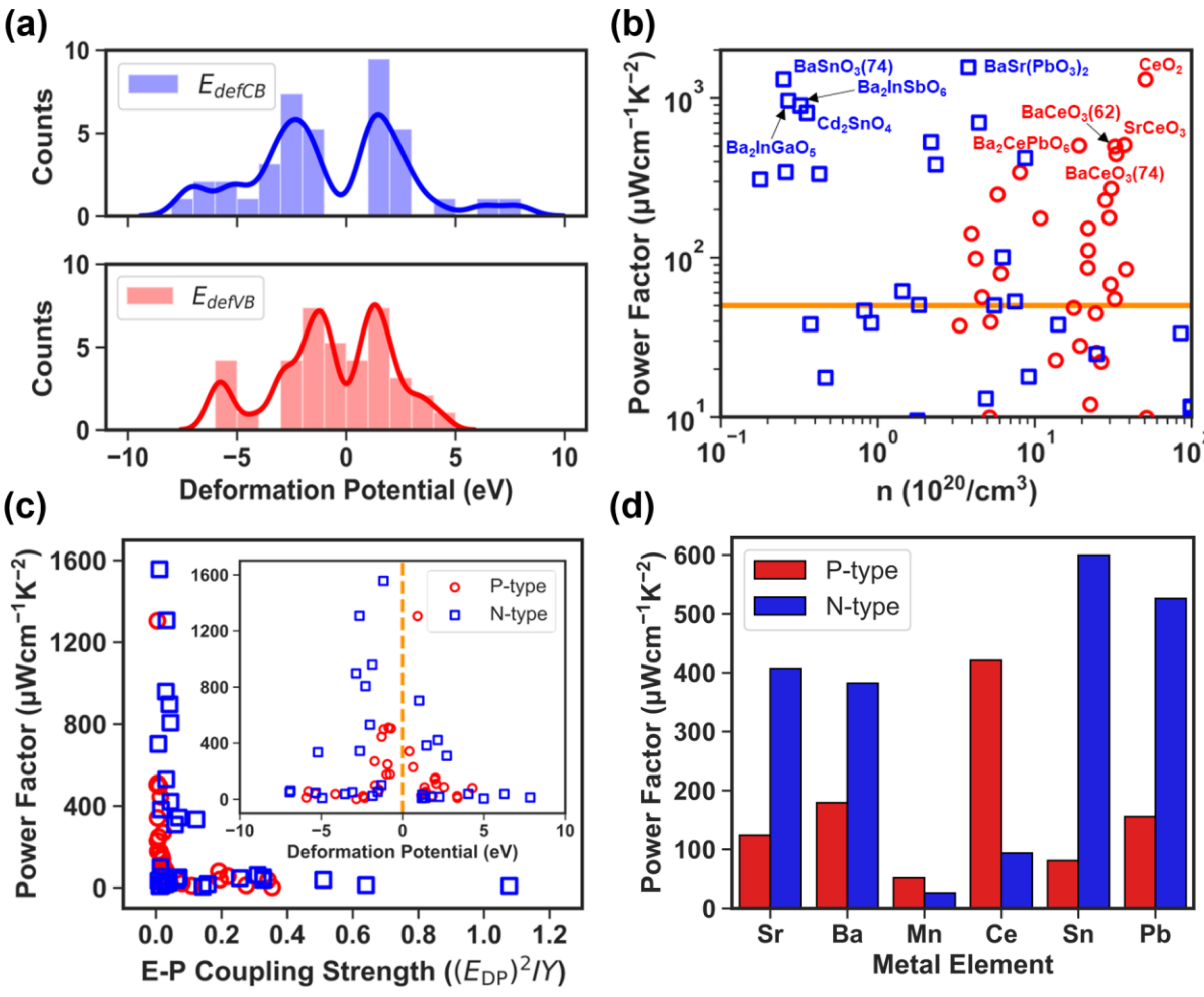

**Fig. 4.** Electrical transport calculation and analysis based on deformation potential (DP) theory. (a) DP distribution of 33 candidate materials at VBM and CBM. (b) Calculated maximum PF values in P-type and N-type. (c) PF and electron-phonon coupling strength evaluated by $(\boldsymbol{E}_{\mathrm{DP}})^2/\boldsymbol{Y}$ in P-type and N-type, the abscissa of the sub-figure is DP. (d) Statistical performance of PF for different metal elements in P-type and N-type.

The sub-figure in Fig. 4c shows the relationship between the PFs and deformation potentials (DPs) of 33 materials at 1000K for both P-type and N-type. It can be observed that the closer the DP is to zero, the higher the likelihood of a high PF. This is due to the small absolute value of DP leading to a smaller $(\boldsymbol{E}_{\mathrm{DP}})^2/\boldsymbol{Y}$, which often can be simply equated to the strength of electron-phonon coupling, subsequently resulting in a larger relaxation time and affecting the PF. Therefore, the DP and Young's modulus, both of which depend on the material, influence the PF. As can be seen in Fig. 4c, there is a clear negative correlation between electron-phonon coupling strength and PF. Materials with a high PF always have a smaller $(\boldsymbol{E}_{\mathrm{DP}})^2/\boldsymbol{Y}$. Thus, a small deformation potential and a large Young's modulus can serve as preliminary criteria for the high-throughput screening of TE materials in the future. The estimation of the strength of electron-phonon coupling is just one factor that affects the relaxation time. The band structure of the material itself also influences the relaxation time and electron group velocity, contributing to electrical transport, while the interactions between different atoms constitute a complex process.

We performed a mean statistical analysis based on the composition of metal elements for the electrical transport calculation results of the 33 candidate materials. The statistical subjects were metal elements that appeared more than three times, and the statistical content focused on the average transport properties of all materials containing those elements. Considering the reliability of the statistical analysis, we averaged the electrical transport performance within a ±0.1 eV energy range around the energy level corresponding to the maximum PFs for both P-type and N-type materials. As shown in Fig. 4d, we present the statistical mean of the P-type and N-type PFs corresponding to different elements, with the elements being ordered according to their position in the periodic table. More statistical details of transport calculation can be found in Supplementary Note 10. The alkaline earth metal elements Sr and Ba appeared 23 times in the candidate group, but they are not the key elements that determine the TE performance (as mentioned earlier). Mn and Ce, which are transition metals and lanthanides, respectively, have a greater impact in P-type than in N-type, especially Ce. Element Ce has a larger PF, which is inseparable from its greater electron group velocity in P-type compared to Mn. On the contrary, Sn and Pb, as post-transition metals, exhibit a phenomenon where the N-type PF is greater than the P-type. This is related to their typically large relaxation time and group velocity in N-type, resulting in a larger sigma. Therefore, metal oxides applied to high-temperature TE can tend to choose Ce, Sn and Pb as constituent elements. For example, the candidate material $Ba_2CePbO_6$ has potential P-type and N-type TE transport properties, with the maximum PF being 504.1 $\mu Wcm^{-1}K^{-2}$ for P-type and 703.1 $\mu Wcm^{-1}K^{-2}$ for N-type.

## 4. CONCLUSION

In this work, we have developed a material HTP screening framework that combines the interpretable ML models with HTP DFT calculations to select metal oxides for high-temperature TE materials. During the process, the construction of the PF ML model was severely overfitted due to the influence of a small dataset. We employed SR for feature creation, which enhanced the robustness of the model and was significantly superior to traditional descriptor combinations. The results demonstrate that SR has strong objectives and high interpretability advantages in feature creation. Ultimately, we integrated the PF prediction model with the melting point prediction model to conduct a virtual screening of 48,694 metal oxides in the Materials Project database. Leveraging the electrical transport calculations based on the deformation potential theory, the PFs at 1000K were evaluated for 33 high-temperature-resistant metal oxides, as selected by the ML model. Furthermore, our combined data-driven analysis and transport calculations indicate that metal oxides rich in elements such as cerium (Ce), tin (Sn), and lead (Pb) tend to exhibit high PFs at high temperatures. In conclusion, the HTP framework we have proposed facilitates the selection and discovery of metal oxides suitable for high-temperature TE applications. The compounds identified could potentially be high-temperature TE candidate materials, meriting further investigation.

## DATA AVAILABILITY

The authors declare that the data supporting the findings of this study are available in the GitHub repository: https://github.com/SJTU-MI/HTPS4HTTEMOs or from the corresponding authors on reasonable request.

## CODE AVAILABILITY

The codes for feature engineering, ML model training, SHAP analysis, melting point model access and MP database virtual screening are available in the GitHub repository: https://github.com/SJTU-MI/HTPS4HTTEMOs. Detailed descriptions can be found in the Methods Section and Supplementary Information.

## ACKNOWLEDGEMENTS

This work was supported by the National Natural Science Foundation of China (No. 52006134), the Shanghai Key Fundamental Research Grant (No. 21JC1403300). The computations in this paper were run on the π 2.0 cluster supported by the Center for High-Performance Computing at Shanghai Jiao Tong University.

## AUTHOR CONTRIBUTIONS

S.J. conceived the idea and supervised the project. S.M. designed the HTP framework, executed the feature engineering, trained and evaluated the machine learning models, as well as performed the DFT calculations. Y.R. completed the collection of PF data. X.H. participated in the discussion. All authors modified and approved the manuscript.

## COMPETING INTERESTS

Authors declare that they have no competing interests.

## ADDITIONAL INFORMATION

The supporting information is available free of charge.

## Supplementary Material for

# High-throughput discovery of metal oxides with high thermoelectric performance via interpretable feature engineering on small data

**Shengluo Ma[1], Yongchao Rao[1], Xiang Huang[1], Shenghong Ju[1, 2, *]**

[1] China-UK Low Carbon College, Shanghai Jiao Tong University, Shanghai, 201306, China

[2] Materials Genome Initiative Center, School of Material Science and Engineering, Shanghai Jiao Tong University, Shanghai 200240, China

[*] Corresponding author: shenghong.ju@sjtu.edu.cn.

**Supplementary Note 1. Screening of high-temperature thermoelectric (TE) metal oxides**

We have positioned the compounds with differences under varying temperature sets in Table S1 at the final spots of the table. Figure S2 presents the detailed predictions of the machine learning model in the form of a periodic table, listing the metal elements that are selectable in our work. The horizontal axis in the periodic table represents the predicted melting point, while the vertical axis represents the normalized power factor (PF) across different machine learning (ML) models.

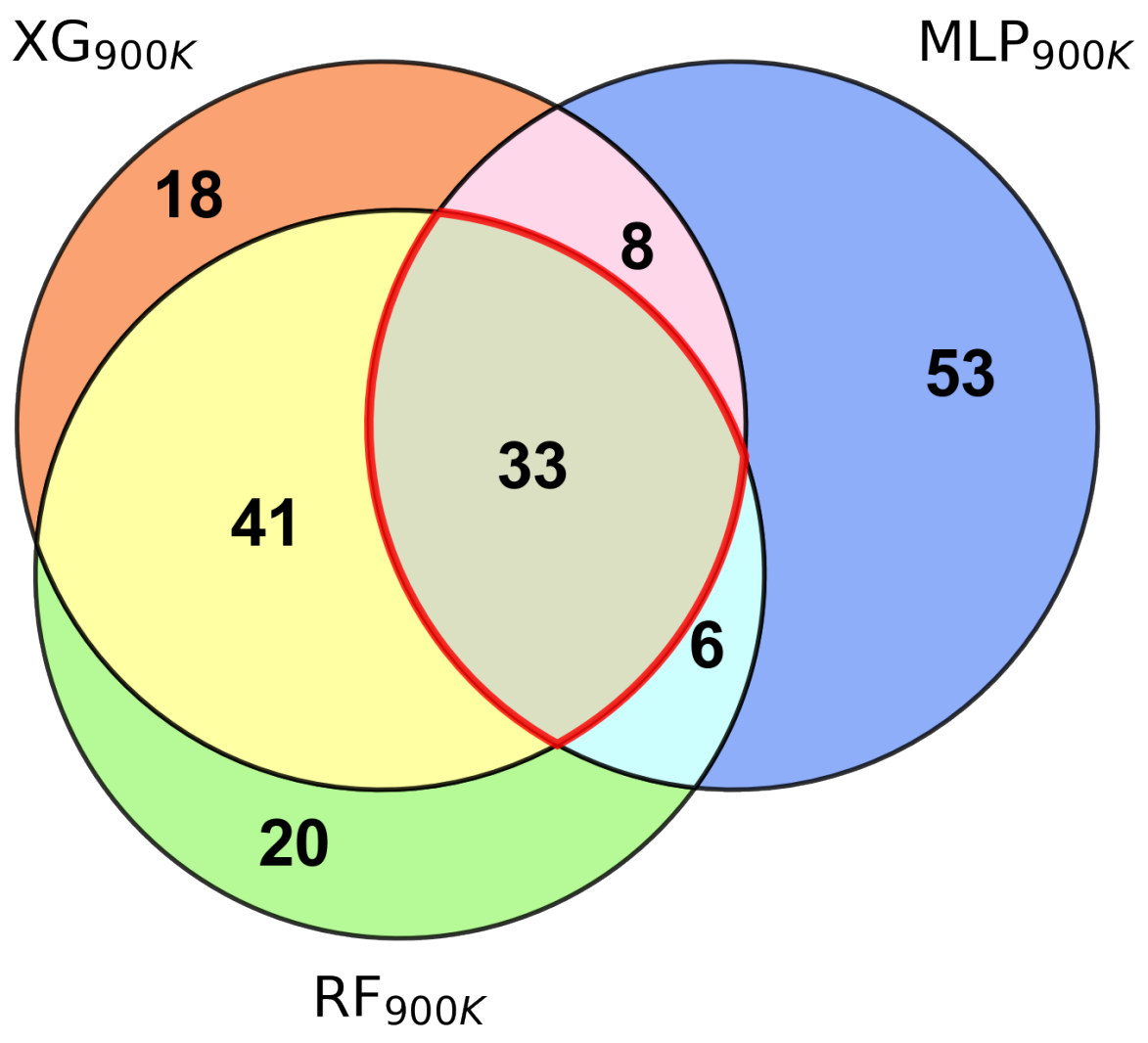

**Figure S1:** Multi-model screening for top 100 materials with PF ranking at 900K.

**Table S1:** The compounds MPID list of screening at 1000K and 900K

| Compounds MPID list at 1000K | | | Compounds MPID list at 900K | | |
|---|---|---|---|---|---|
| mp-1178513 | mp-1206536 | mp-1214389 | mp-1178513 | mp-1206536 | mp-1214389 |
| mp-1227843 | mp-1228392 | mp-15743 | mp-1227843 | mp-1228392 | mp-15743 |
| mp-16033 | mp-18971 | mp-19009 | mp-16033 | mp-18971 | mp-19009 |
| mp-20098 | mp-20194 | mp-20489 | mp-20098 | mp-20194 | mp-20489 |
| mp-20882 | mp-2097 | mp-22203 | mp-20882 | mp-2097 | mp-22203 |
| mp-22230 | mp-22428 | mp-23091 | mp-22230 | mp-22428 | mp-23091 |
| mp-2898 | mp-3163 | mp-3187 | mp-2898 | mp-3163 | mp-3187 |
| mp-3316 | mp-361 | mp-4359 | mp-3316 | mp-361 | mp-4359 |
| mp-4900 | mp-545603 | mp-546152 | mp-4900 | mp-545603 | mp-546152 |
| mp-554354 | mp-561553 | mp-5966 | mp-554354 | mp-561553 | mp-5966 |
| mp-9297 | mp-9298 | mp-1106089 | mp-9297 | mp-9298 | mp-20546 |

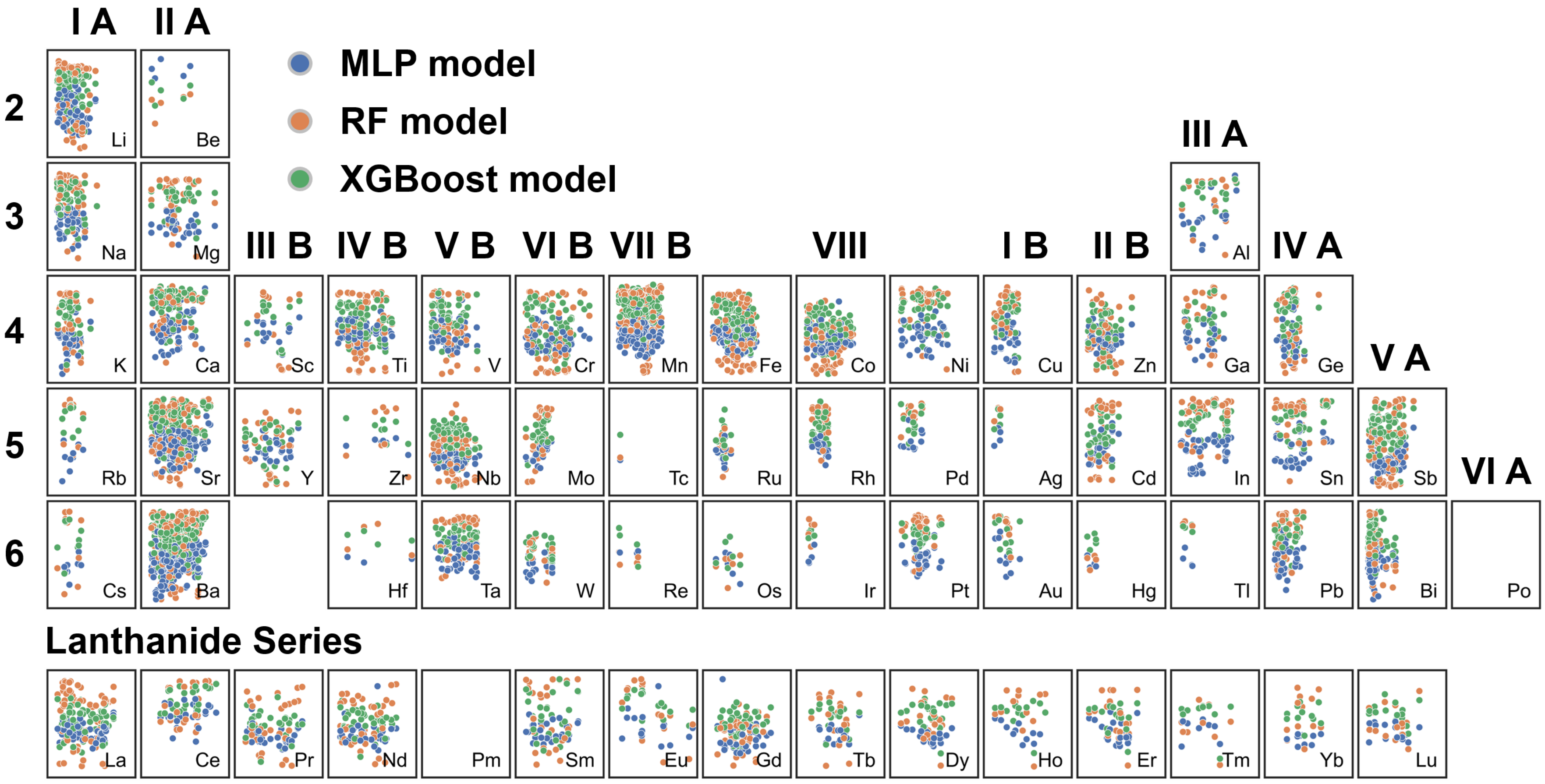

**Figure S2:** The distribution of predicted melting points and predicted normalized PFs for the set of 1227 materials in the periodic table.

**Supplementary Note 2. List of descriptor collection**

311 descriptors were collected, including the temperature descriptor of power factors (PFs), 20 crystal structure and property descriptors obtained from the Materials Project (MP) database API using pymatgen[1] package (refer to Table S2), and 290 element-compositional descriptors obtained through XenonPy[2] package. These 290 descriptors were calculated from 58 element-level property data (refer to Table S3) using five calculation methods, including weighted average (Ave), weighted sum (Sum), weighted variance (Var), max-pooling (Max), min-pooling (Min), and the details could be found at https://xenonpy.readthedocs.io/en/latest/features.html.

**Table S2:** The list of 20 crystal structure and property descriptors from MP database

| **Descriptor Name** | | |
|---|---|---|
| spacegroup | volume | volume_per_atom |
| a | b | c |
| alpha | beta | gamma |
| band_gap | e_above_hull | e_above_hull_per_atom |
| efermi | energy | energy_per_atom |
| density | final_energy_per_atom | formation_energy_per_atom |
| nelements | nsites | |

**Table S3:** The list of 58 element-level property data from XenonPy

| **Element-level Property Name** | | |
|---|---|---|
| atomic_number | first_ion_en | num_unfilled |
| atomic_radius | fusion_enthalpy | num_valance |
| atomic_radius_rahm | gs_bandgap | num_d_unfilled |
| atomic_volume | gs_energy | num_d_valence |
| atomic_weight | gs_est_bcc_latcnt | num_f_unfilled |
| boiling_point | gs_est_fcc_latcnt | num_f_valence |
| bulk_modulus | gs_mag_moment | num_p_unfilled |
| c6_gb | gs_volume_per | num_p_valence |
| covalent_radius_cordero | hhi_p | num_s_unfilled |
| covalent_radius_pyykko | hhi_r | num_s_valence |
| covalent_radius_pyykko_double | heat_capacity_mass | period |
| covalent_radius_pyykko_triple | heat_capacity_molar | specific_heat |
| covalent_radius_slater | icsd_volume | thermal_conductivity |
| density | evaporation_heat | vdw_radius |
| dipole_polarizability | heat_of_formation | vdw_radius_alvarez |
| electron_negativity | lattice_constant | vdw_radius_mm3 |
| electron_affinity | mendeleev_number | vdw_radius_uff |
| en_allen | melting_point | sound_velocity |
| en_ghosh | molar_volume | Polarizability |
| en_pauling | | |

**Supplementary Note 3. Process of filtering descriptors**

Descriptors were filtered in two stages: Removing descriptors with low variance, the 297 reserved descriptors are called VarCond. Set. Then filtering was based on different correlation coefficient conditions, with one condition are met referred to as CorCond.1 Set and two referred to as CorCond.2 Set, as shown in Figure S3. The 18 descriptors from the CorCond.2 Set will be used for feature creation through Symbolic Regression[3, 4] (SR). The detailed information about these descriptors is presented in Table S4.

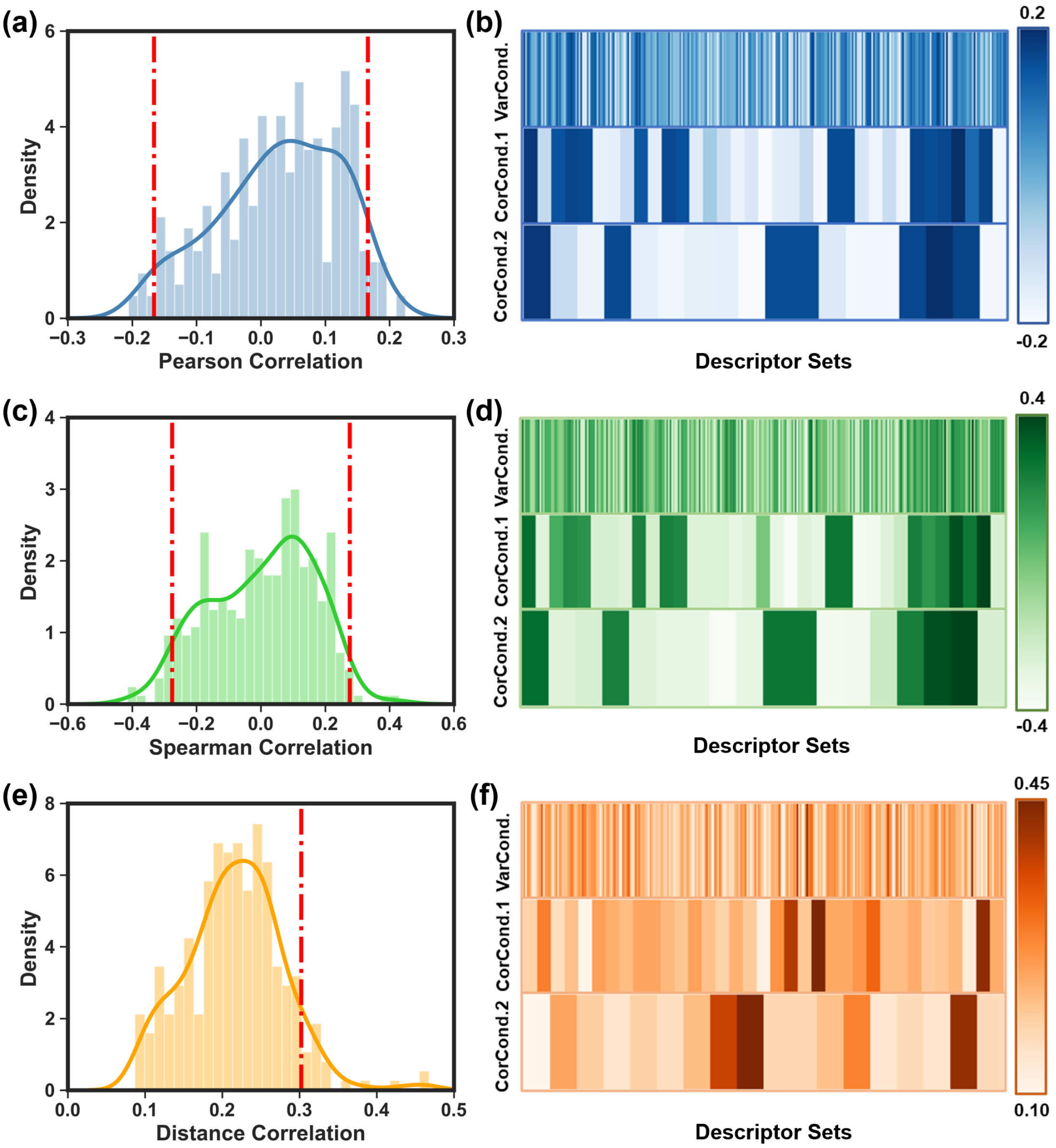

**Figure S3:** Process and analysis of filtering descriptors. (a) The distribution of Pearson correlation in the initial set of descriptors, and the red line represent the values of the filtering criterion. (b) Heatmaps of Pearson correlation for different sets. (c) and (d), (e) and (f) are the same as (a) and (b), but represent Spearman and Distance correlations, respectively.

**Table S4:** The list of 18 descriptors from CorCond.2 Set

| No. | Name | Description | SR Variable |
|---|---|---|---|
| 1 | T | Temperature of power factors | $x_0$ |
| 2 | Density | Density of materials from MP API | $x_1$ |
| 3 | Av.enallen | Ave value of Allen's scale of electronegativity (en_allen) | $x_2$ |
| 4 | Av.vdwrad | Ave value of Van der Waals radius (vdw_radius) | $x_3$ |
| 5 | Su.unf | Sum value of Total unfilled electron (num_unfilled) | $x_4$ |
| 6 | Va.boilp | Var value of Boiling temperature (boiling_point) | $x_5$ |
| 7 | Va.hhip | Var value of Herfindahl−Hirschman Index (HHI) production values (hhi_p) | $x_6$ |
| 8 | Va.unf | Var value of Total unfilled electron (num_unfilled) | $x_7$ |
| 9 | Va.punf | Var value of Unfilled electron in p shell (num_p_unfilled) | $x_8$ |
| 10 | Va.sunf | Var value of Unfilled electron in s shell (num_s_unfilled) | $x_9$ |
| 11 | Va.sval | Var value of Valance electron in s shell (num_s_valence) | $x_{10}$ |
| 12 | Ma.hhip | Max value of Herfindahl−Hirschman Index (HHI) production values (hhi_p) | $x_{11}$ |
| 13 | Ma.unf | Max value of Total unfilled electron (num_unfilled) | $x_{12}$ |
| 14 | Ma.punf | Max value of Unfilled electron in p shell (num_p_unfilled) | $x_{13}$ |
| 15 | Ma.sunf | Max value of Unfilled electron in s shell (num_s_unfilled) | $x_{14}$ |
| 16 | Mi.elecaff | Min value of Electron affinity (electron_affinity) | $x_{15}$ |
| 17 | Mi.lattc | Min value of Physical dimension of unit cells in a crystal lattice (lattice_constant) | $x_{16}$ |
| 18 | Mi.sval | Min value of Valance electron in s shell (num_s_valence) | $x_{17}$ |

**Supplementary Note 4. Symbolic regression for feature creation**

The mathematical formulae were acquired and selected using an efficient stepwise strategy with SR based on genetic programming (GPSR) as implemented in the gplearn[5] code. Pearson, Spearman, and Distance correlations were used as evaluation metrics of training fitness to generate new descriptors, and the grid search strategy with the hyperparameters listed in Supplementary Table S5 was applied in GPSR. For the feature creation with different correlations, we employed symbolic regression (SR) with 4293 hyperparameter combinations to identify new features with stronger correlations to the PF data. As illustrated in Figure S4a, the scatter plots represent the distribution of Pearson correlation fitness versus formula complexity (formula length) across different hyperparameter combinations. Generally, the greater the complexity of the formula, the higher the correlation fitness, as shown in Figure S4b. However, these formulas are usually lengthy, which is not conducive to computation and analysis. Only formulas that have high fitness and low complexity are considered suitable. We further selected formulas with a fitness greater than 0.7 and complexity less than 20, and we applied the same selection criteria to Spearman and Distance correlations (as shown in Figure S5 and Figure S6). Additionally, we counted the selection frequency of original descriptors in the formulas generated by SR (as shown in Figure S4c). Discrete descriptors, such as temperature and electron occupancy, were frequently extracted for combinations of formulas. This facilitated the integration of discrete features into more continuous new descriptors, thereby benefiting the training of machine learning (ML) models. From the Pareto frontier, we selected 12 formulas corresponding to the correlations (as shown in Figure S4d), with the details of these formulas listed in Table S6.

**Table S5:** Setup of hyperparameters in gplearn software for SR

| Parameter | Value | Combination |
|---|---|---|
| Generations | 300 | 1 |
| Population size in every generation | 5000 | 1 |
| Probability of crossover (pc) | [0.30, 0.85], (step = 0.05) | |
| Probability of subtree mutation (ps) | [(1-pc)/3, (1-pc)/2] (step = 0.01) | 477 |
| Probability of hoist mutation (ph) | [(1-pc)/3, (1-pc)/2] (step = 0.01) | |
| Probability of point mutation (pp) | 1-pc-ps-ph | |
| Function set | $\{+, -, \times, \div, \sqrt{x}, \ln x, \lvert x\rvert, -x, 1/x\}$ | 1 |
| Parsimony coefficient | auto | 1 |
| Metric | Pearson, Spearman, Distance Cor. | 3 |
| Stopping criteria | 0.90 | 1 |
| Random_state | 0, 1, 2 | 3 |
| Init_depth | [2, 6], [4, 8], [6, 10] | 3 |

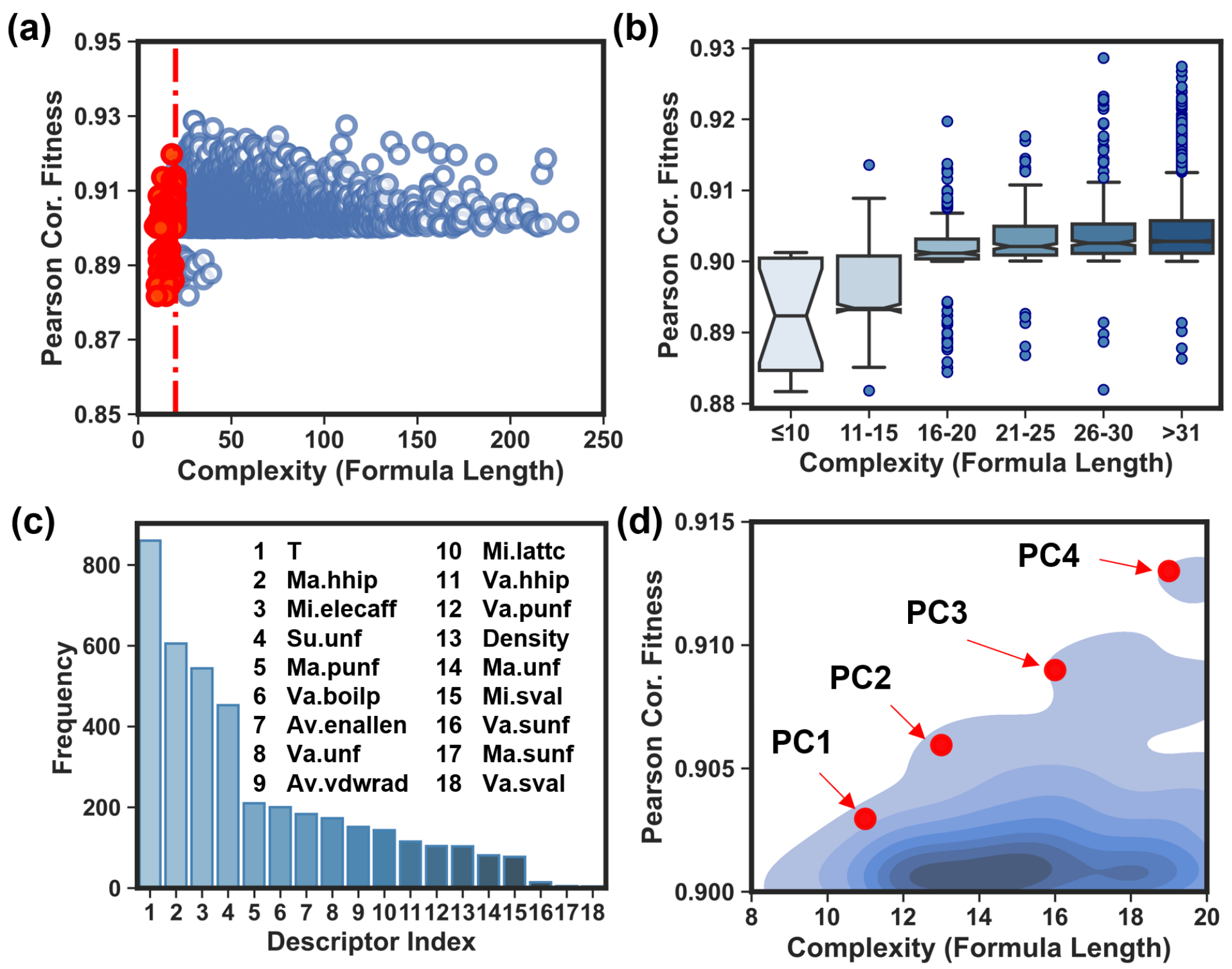

**Figure S4:** Symbolic regression process analysis and formula selection with Pearson correlation (PC). (a) Mathematical formula complexity versus PC fitness. (b) Statistics of formulas with PC fitness not less than 0.7 by complexity. (c) Frequency of original feature occurrence in mathematical formulas. (d) Pareto front of PC fitness versus complexity.

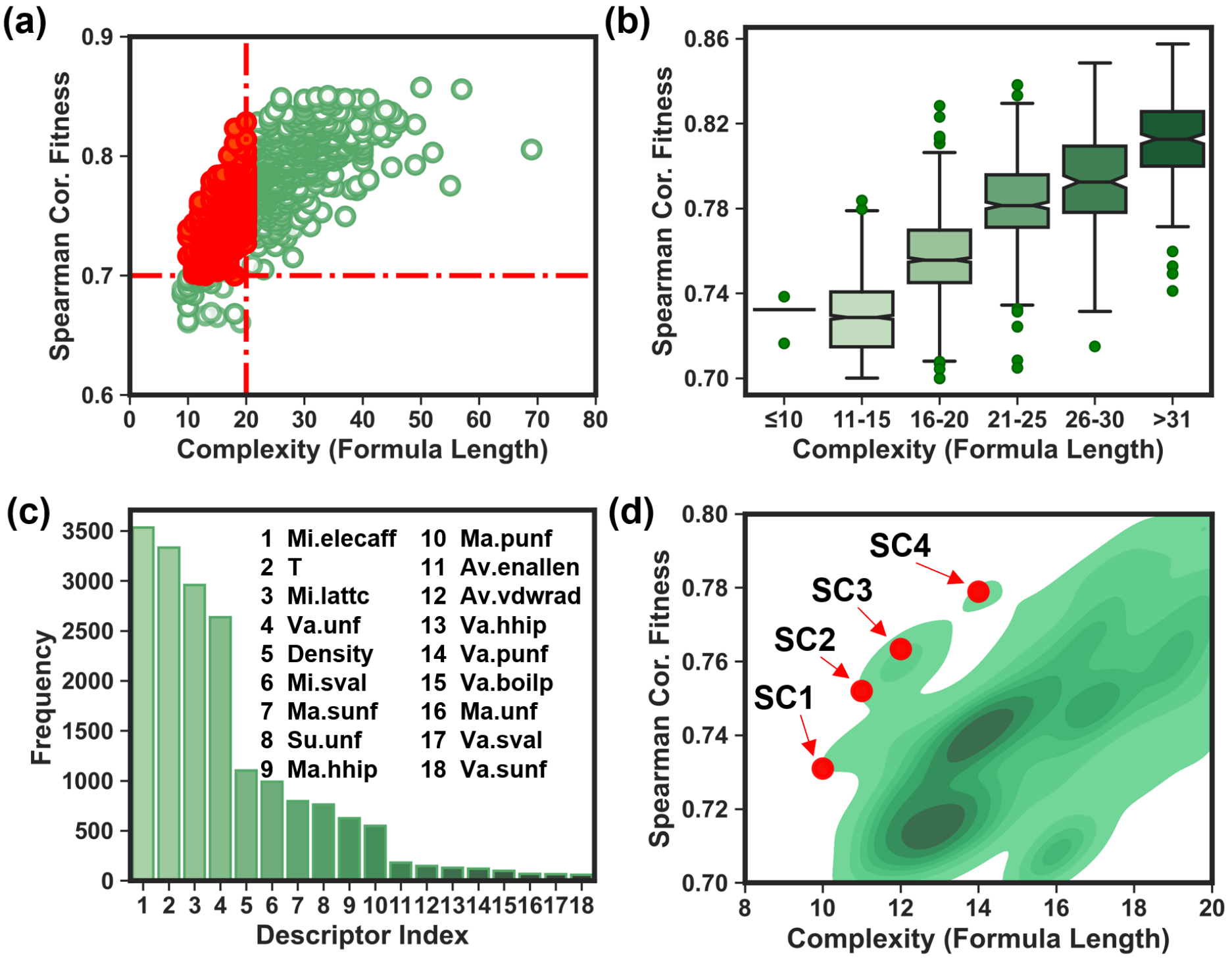

**Figure S5:** Symbolic regression process analysis and formula selection with Spearman correlation (SC). (a) Mathematical formula complexity versus SC fitness. (b) Statistics of formulas with SC fitness not less than 0.7 by complexity. (c) Frequency of original feature occurrence in mathematical formulas. (d) Pareto front of SC fitness versus complexity.

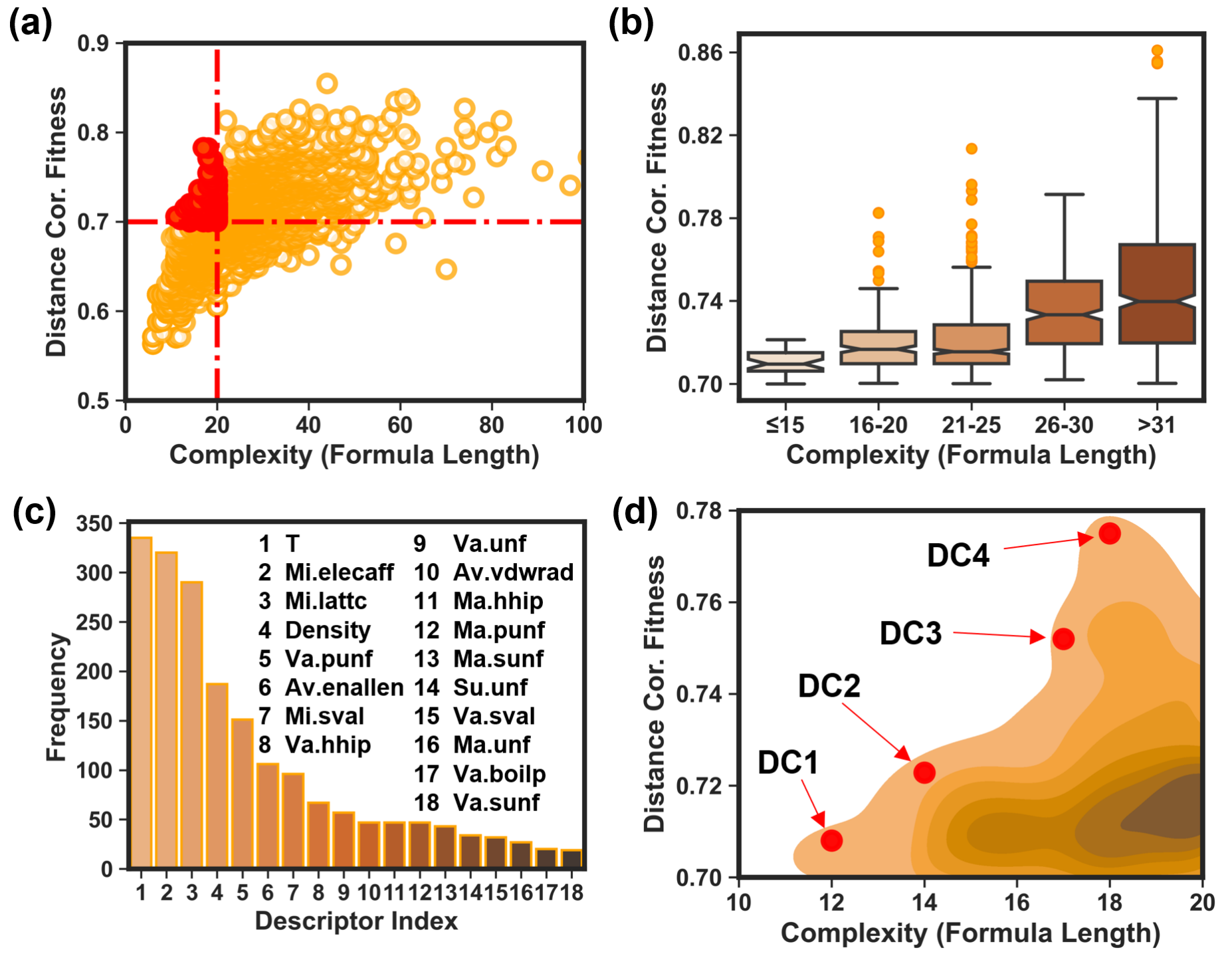

**Figure S6:** Symbolic regression process analysis and formula selection with Distance correlation (DC). (a) Mathematical formula complexity versus DC fitness. (b) Statistics of formulas with DC fitness not less than 0.7 by complexity. (c) Frequency of original feature occurrence in mathematical formulas. (d) Pareto front of DC fitness versus complexity.

**Table S6:** The 12 mathematical formulas at the Pareto front in Figure S4d, S5d and S6d

| Formula | Format | Complexity | PC value | SC value | DC value |
|---|---|---|---|---|---|
| PC1 | $\frac{1}{\ln\left[(x_0+x_{11})-x_0\times x_3/x_2\right]}$ | 11 | -0.863 | -0.531 | 0.593 |
| PC2 | $\frac{x_6}{x_{11}\times\left\|x_{11}/x_4+x_0\times x_{15}\right\|}$ | 13 | -0.883 | -0.347 | 0.546 |
| PC3 | $\frac{1}{\ln\left\|x_6/\left[x_0^3\times\ln\left(x_{16}-x_{12}\right)\right]\right\|}$ | 16 | 0.899 | 0.476 | 0.558 |
| PC4 | $\frac{-\sqrt{x_5}}{x_0\times\ln\left(\frac{1}{\sqrt{x_{13}-x_{15}}+\ln x_2-\ln x_0}\right)}$ | 19 | -0.771 | -0.498 | 0.543 |
| SC1 | $\frac{1}{x_0\times x_1\times x_8/x_2-x_0\times x_{16}}$ | 10 | 0.125 | 0.704 | 0.434 |
| SC2 | $\frac{x_7/\ln\left(\ln x_{15}\right)-x_{14}}{x_0\times x_{16}}$ | 11 | 0.246 | 0.706 | 0.542 |
| SC3 | $\ln\left[x_0\times x_{17}\times\left(x_{15}+2x_{16}-x_1\right)\right]$ | 12 | 0.327 | 0.729 | 0.664 |
| SC4 | $\frac{\left(x_2+x_{11}/x_0\right)}{x_{16}/x_3-\sqrt{x_{12}/x_5}}$ | 14 | -0.225 | -0.737 | 0.648 |
| DC1 | $\ln\left[\left(x_{15}-x_1\right)+1.792x_{16}\right]+2\ln x_0$ | 12 | 0.513 | 0.720 | 0.719 |
| DC2 | $\ln\left(x_1-x_{15}-x_{16}-2.96\right)+\ln x_0$ | 14 | 0.404 | 0.699 | 0.706 |
| DC3 | $\sqrt{x_0\times x_{17}\times\left(x_{16}\times\ln x_2+x_{15}-x_1-x_8\right)}$ | 17 | 0.375 | 0.768 | 0.731 |
| DC4 | $\ln\left(x_{13}-x_{16}\right)-\ln\left[\ln\left[\frac{x_{16}}{x_{17}\times\ln\left(x_1+0.909\right)}\right]\right]$ | 18 | -0.424 | -0.721 | 0.782 |

## Supplementary Note 5. Construction of different ML models

The training effects of various descriptor combinations, including INI, COR, SR, and PCA, on ML models were explored in this section. The INI descriptor set comprised 311 generated initial descriptors. The COR descriptor set consisted of 18 descriptors filtered by feature selection, referred to as the CorCond.2 Set in Supplementary Note 2. Considering the breadth of data coverage and to prevent feature collinearity, we selected three descriptors, PC2, SC4, and DC1 (see Table S6), for inclusion in the subsequent training of the ML models. The SR set comprised 18 COR descriptors and 3 new descriptors created by symbolic regression. Additionally, the extra principal component analysis (PCA) technique with 99.9% variance was used to compare.

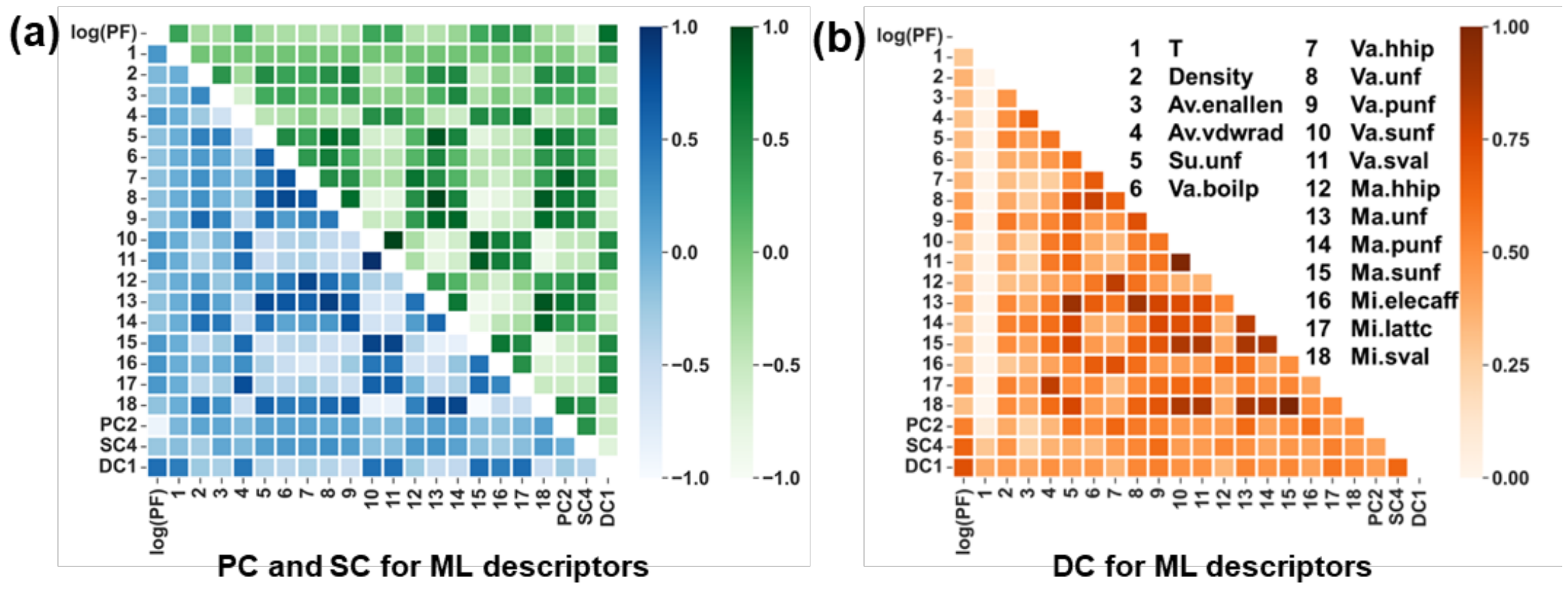

**Figure S7:** Correlation heatmap for the SR sets. (a) The PC (blue part) and SC (green part) heatmap. (c) The DC (orange part) heatmap

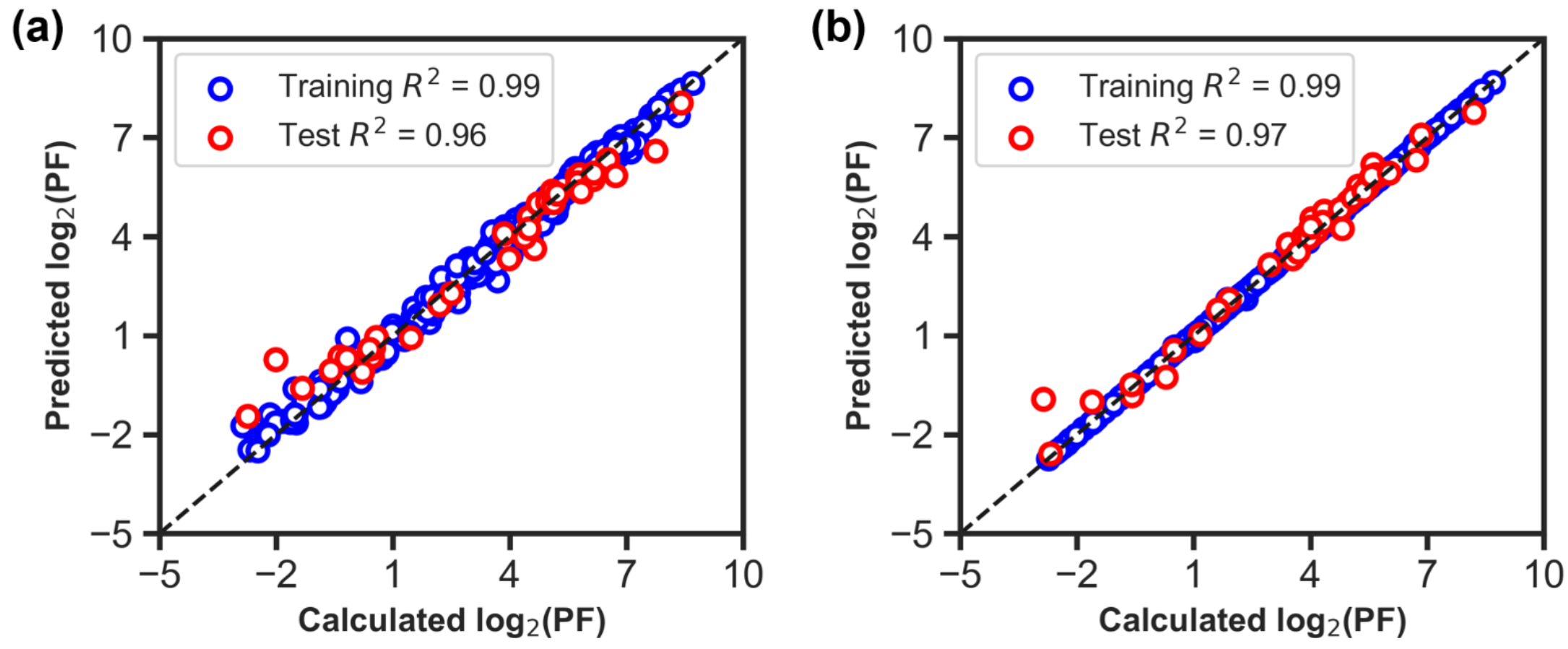

**Figure S8:** Construction of different ML models. (a) Performance of the Extreme Gradient Boosting (XGBoost) model based on SR descriptors, where training $R^2$ is 0.98 and test $R^2$ is 0.96. (b) Performance of the Multi-Layer Perceptron (MLP) model based on SR descriptors, where training $R^2$ is 0.99 and test $R^2$ is 0.97.

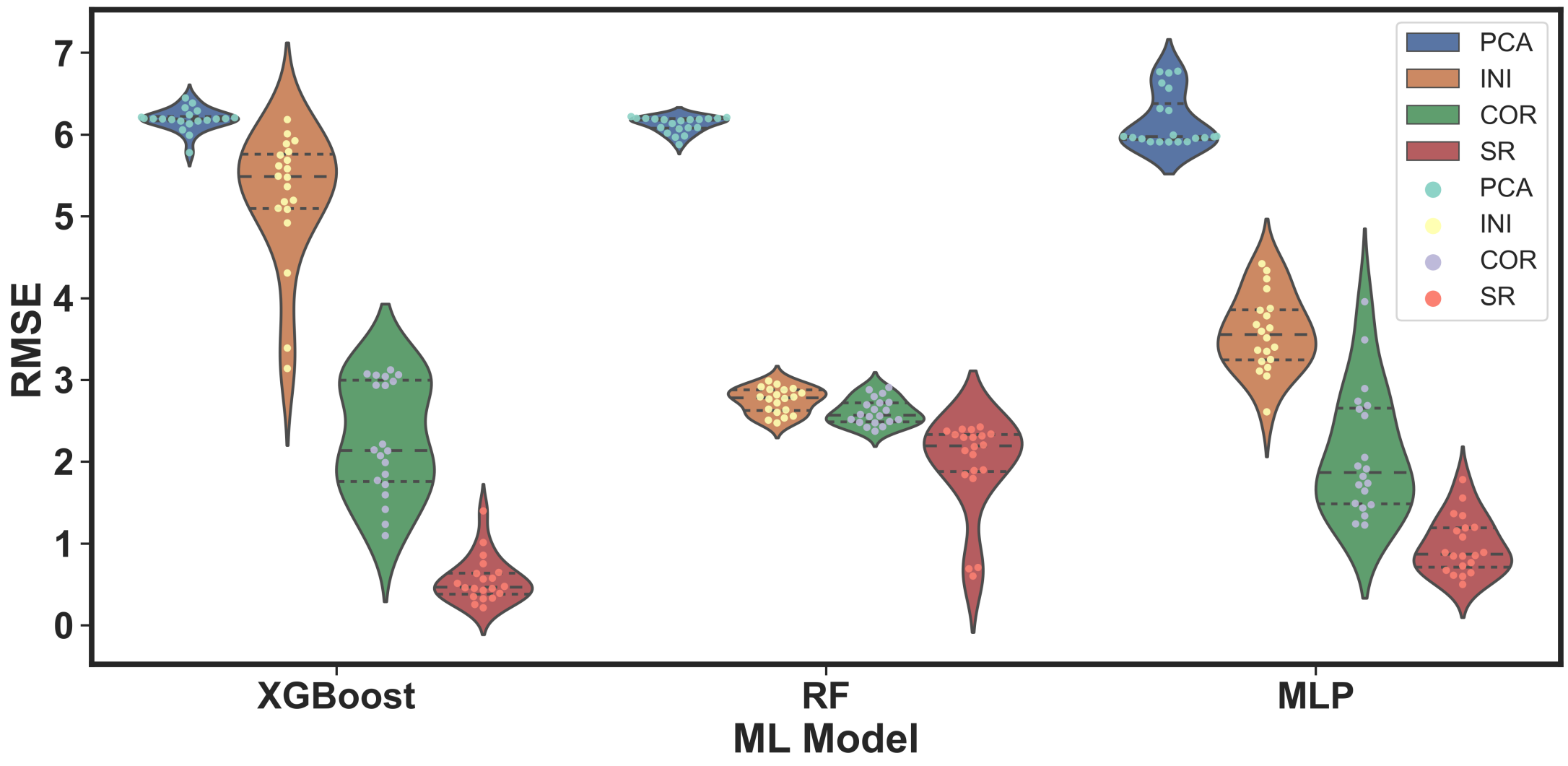

**Figure S9:** RMSE of ML models at different feature engineering processes, including initial (INI), mathematical correlation (COR) coefficients filtering, and feature creation via SR stages. And, an additional PCA approach was applied to compare.

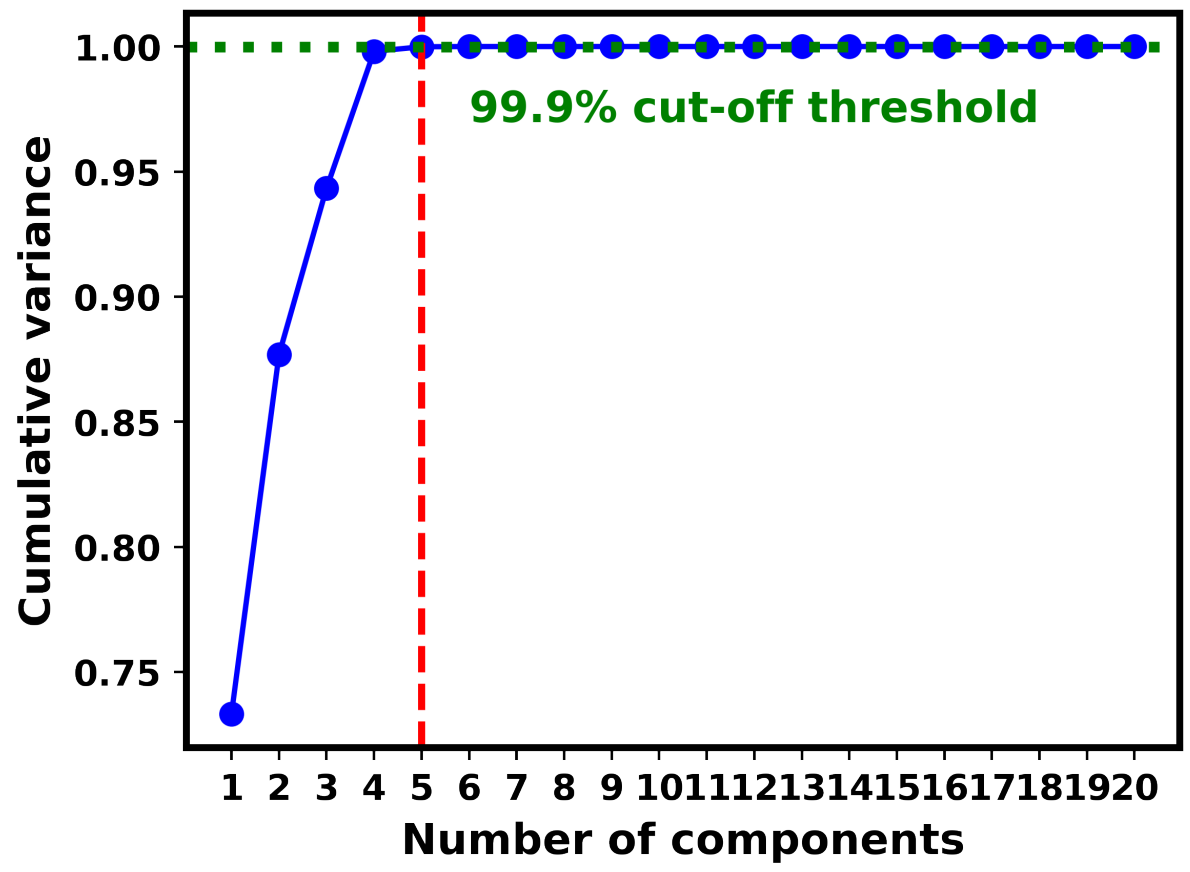

**Figure S10:** Relationship between the number of principal components and cumulative variance

**Table S7:** The correlation between the PCA descriptor set and PF

| Formula | PC value | SC value | DC value |
|---|---|---|---|
| PCA1 | 0.030 | 0.003 | 0.144 |
| PCA2 | -0.159 | -0.324 | 0.341 |
| PCA3 | 0.128 | 0.141 | 0.211 |
| PCA4 | 0.051 | 0.023 | 0.147 |
| PCA5 | -0.022 | -0.011 | 0.208 |

Here, the PCA with 5 principal components of 99.9% variance was performed. Although the data underwent dimensionality reduction, they still remained unsuitable for training ML models due to low data correlation issues.

**Supplementary Note 6. SHAP analysis of the feature importance**

We rank the feature importance based on the impurity of Random Forest (RF) model (as shown in Figure S11), and the results are similar to the SHAP analysis mentioned in the main text. Both analysis methods indicate that descriptors created through SR are more valuable for machine learning models. We present the dependence plot the original descriptors and the descriptors created by SR, shown in Figure S12. It can be seen that the discretization of the original descriptors is very common, while the SR features exhibit better continuity.

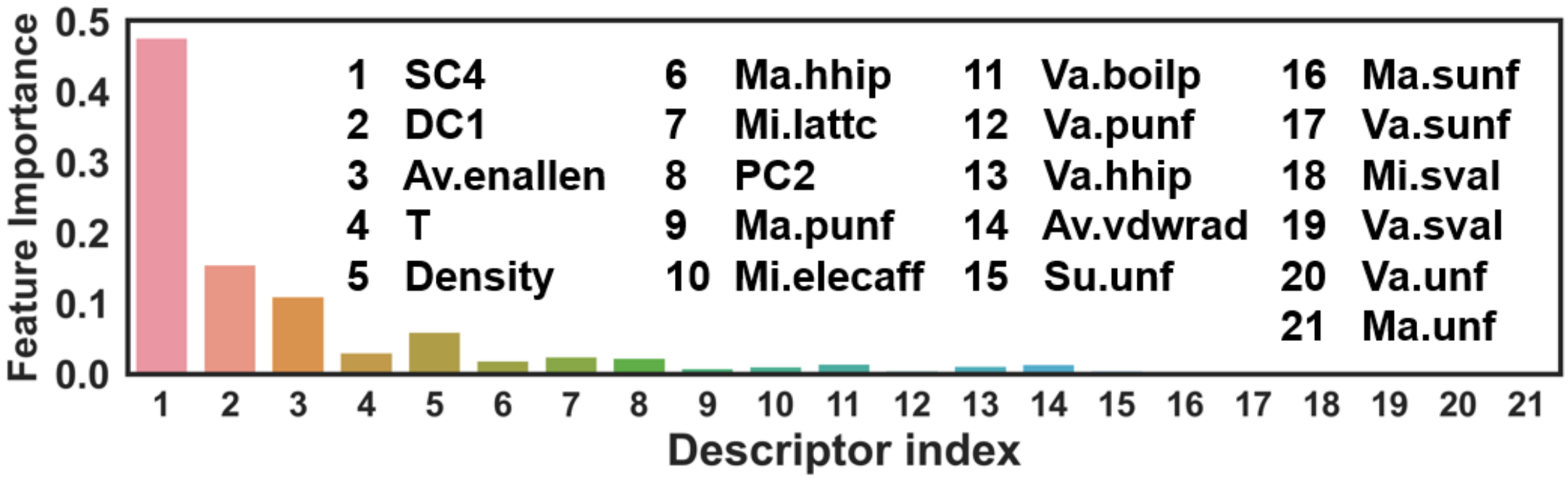

**Figure S11:** Feature importance based on the impurity of Random Forest (RF) model.

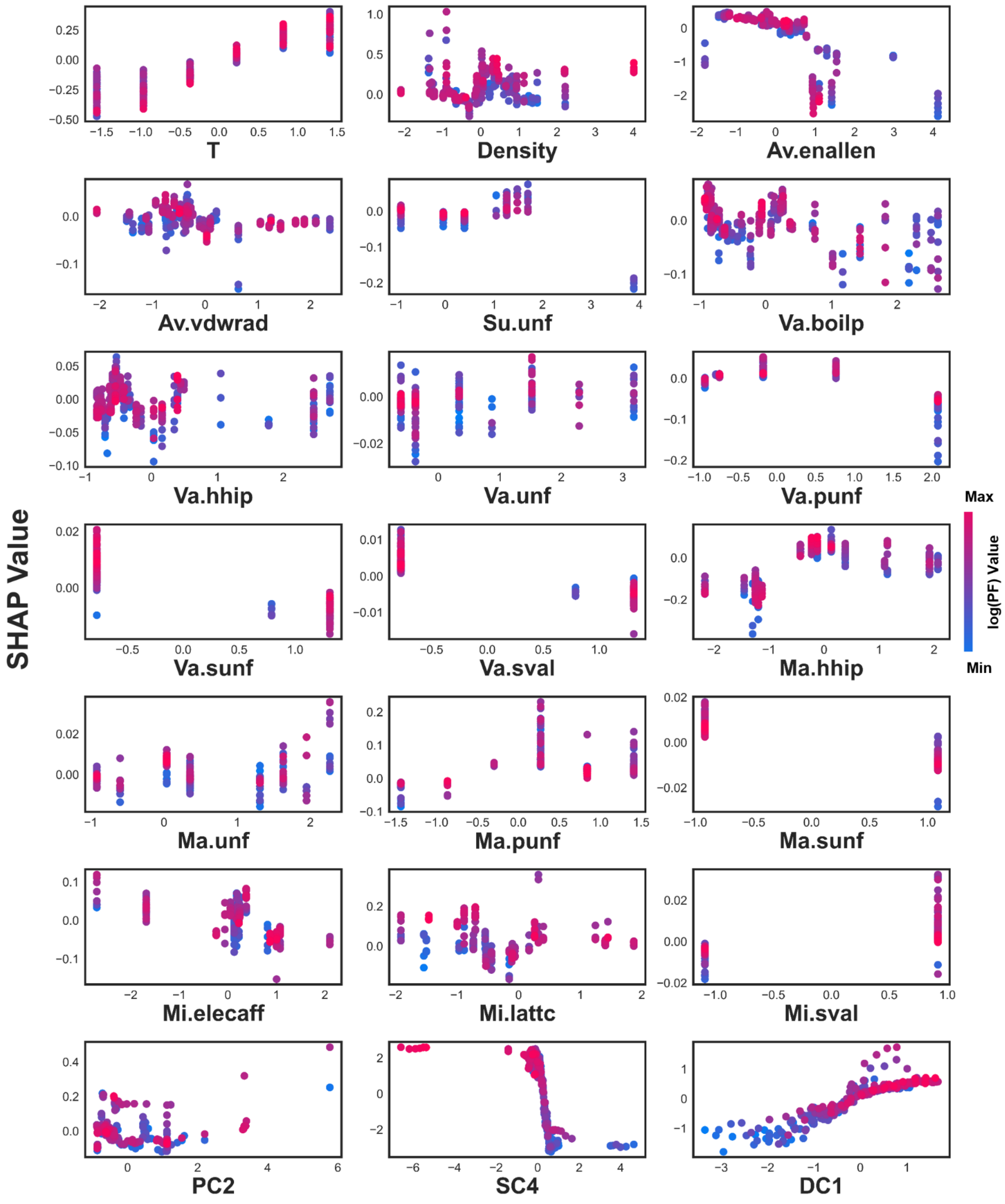

**Figure S12:** SHAP values of each descriptor in a dependence plot.

**Supplementary Note 7. Calculation and comparison of deformation potentials (DPs)**

As shown in Figure S13, we perform linear fitting of the DP under different strains, with fitting accuracy greater than 0.9. Additionally, we compare the deformation potentials obtained using the bare eigenvalues of the electronic structures for some compounds from the MatHub-3d database[6] (Figure S14).

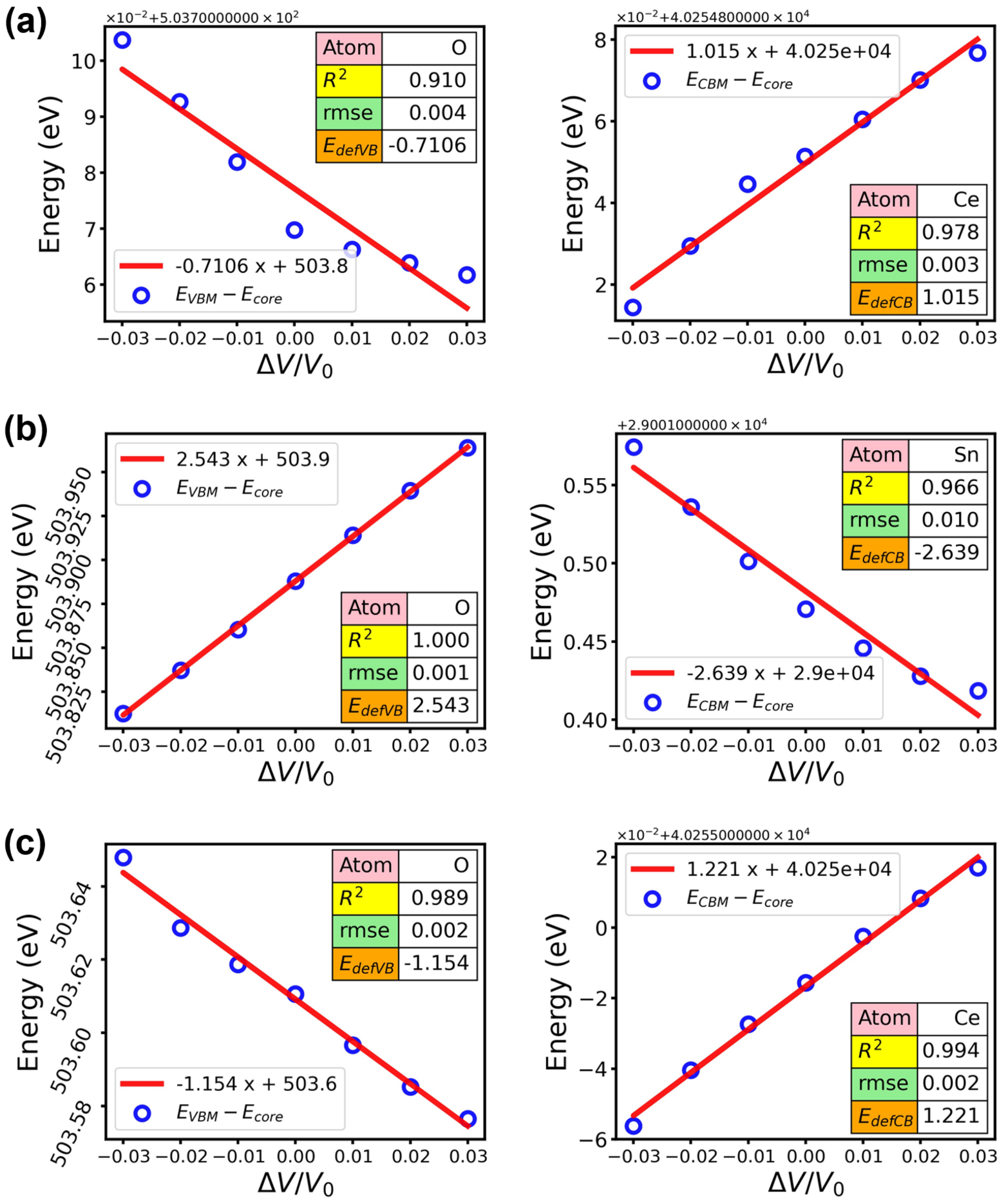

**Figure S13:** The linear fitting of DPs at VBM and CBM. (a) DP fitting of $Ba_2CePbO_6$ (mp-1228392). (b) DP fitting of $BaSnO_3$ (mp-1178513). (c) DP fitting of $BaCeO_3$ (mp-3187).

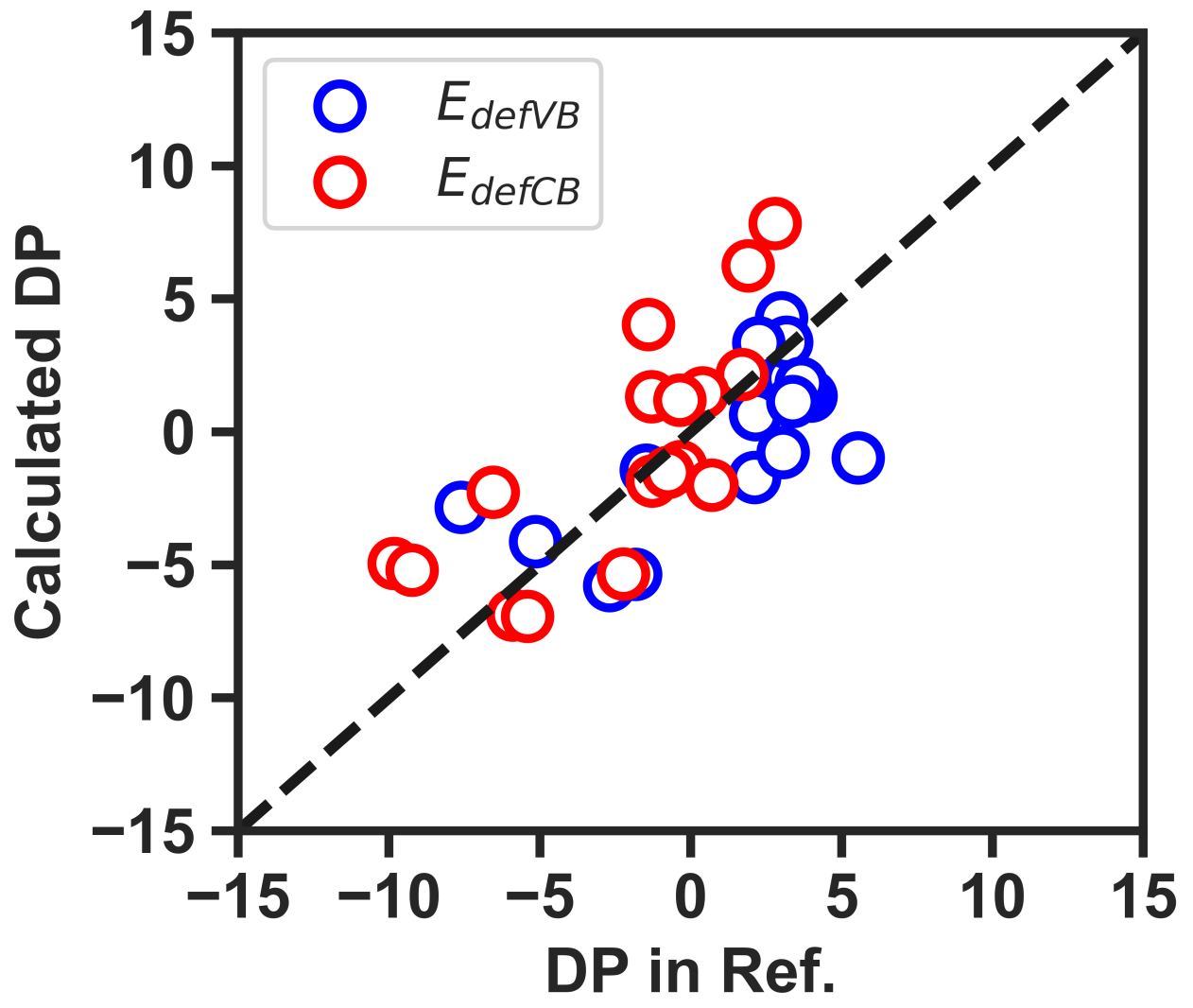

**Figure S14:** Comparison of deformation potentials with the bare eigenvalues method.

**Supplementary Note 8. Comparison of electrical transport calculations by CRTA and CEPCA**

we compared the electrical transport calculations under the constant relaxation time approximation (CRTA) and the constant electron-phonon coupling approximation (CEPCA) using $Cu_2O$ as an example. As shown in Figure S15, the work related to Ref1[7] employed hybrid density functional theory to calculate the electrical transport properties of $Cu_2O$. However, due to the use of the CRTA, it overestimated the electrical conductivity, which is closely related to relaxation time, resulting in an inflated PF. We also conducted calculations (using the PBE-GGA functional) on $Cu_2O$ within CRTA, and our results were similar to Ref1. The work related to Ref2[6] (fitting DP using eigenvalues, replacing Young's modulus with bulk modulus) applied the CEPCA based on the DP theory, which eliminates the influence of relaxation time compared to CRTA. We used CEPCA to perform calculations on $Cu_2O$, and our results (fitting DP with the core energy level as a reference level, calculating Young's modulus directly) were consistent with Ref2.

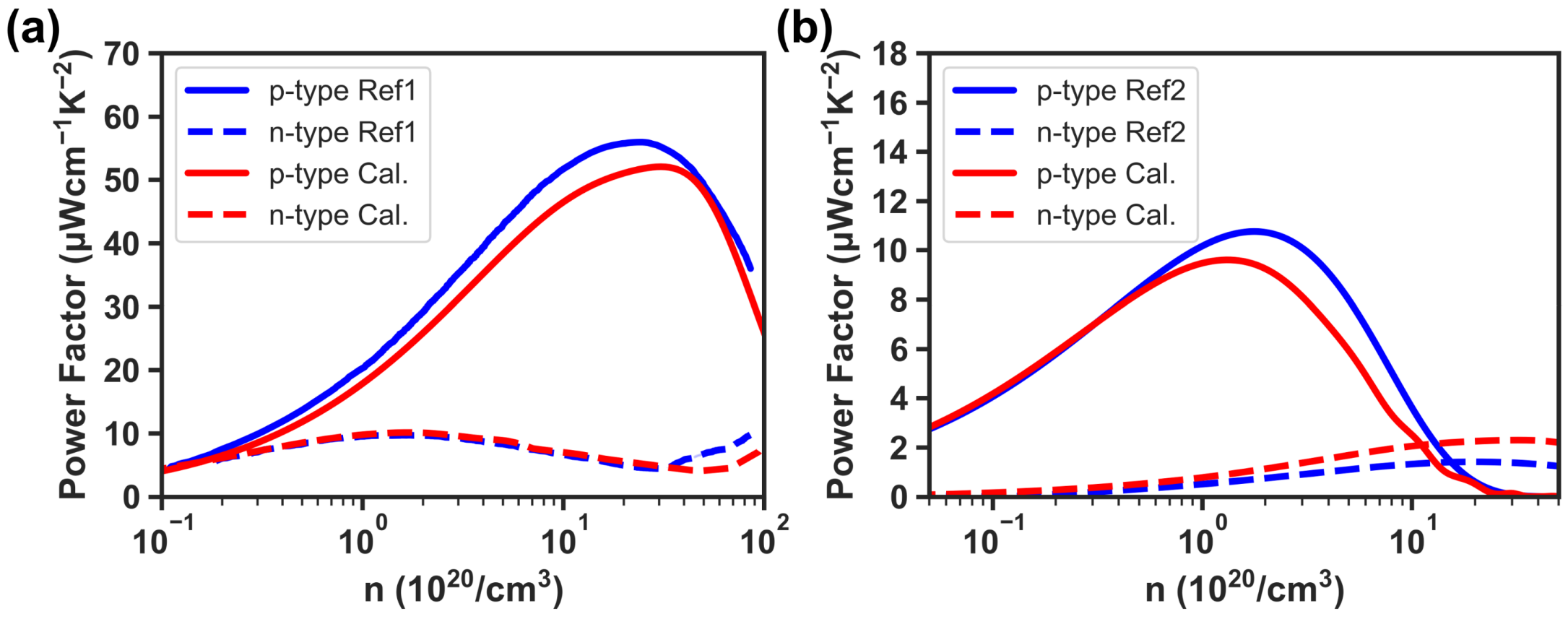

**Figure S15:** Comparison of electrical transport calculations. (a) Electrical transport calculations by CRTA. (b) Electrical transport calculations by CEPCA.

**Supplementary Note 9. Band structures and density of states of materials**

1. $CeO_2$ (mp-20194)

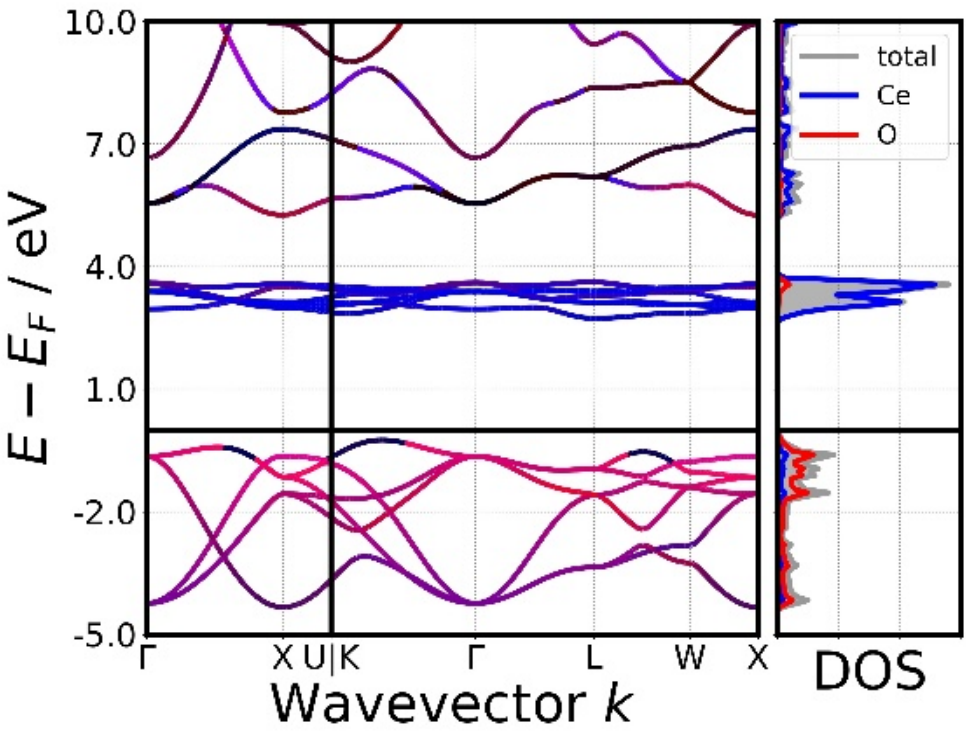

2. $Ba_2InSbO_6$ (mp-1206536)

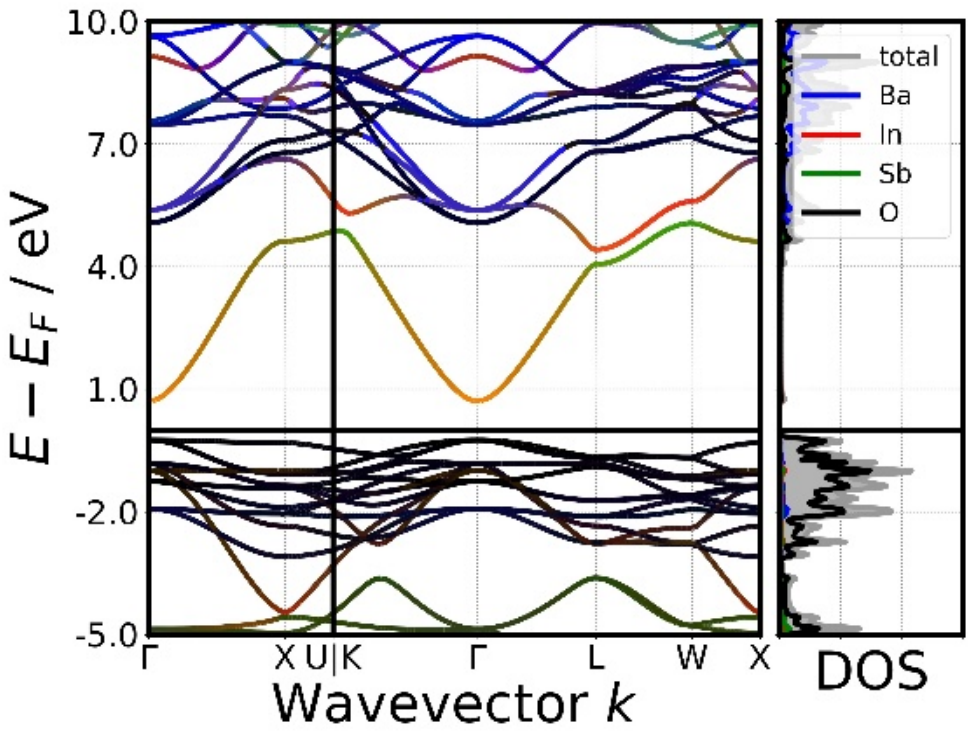

3. $Ba_2BiSbO_6$ (mp-545603)

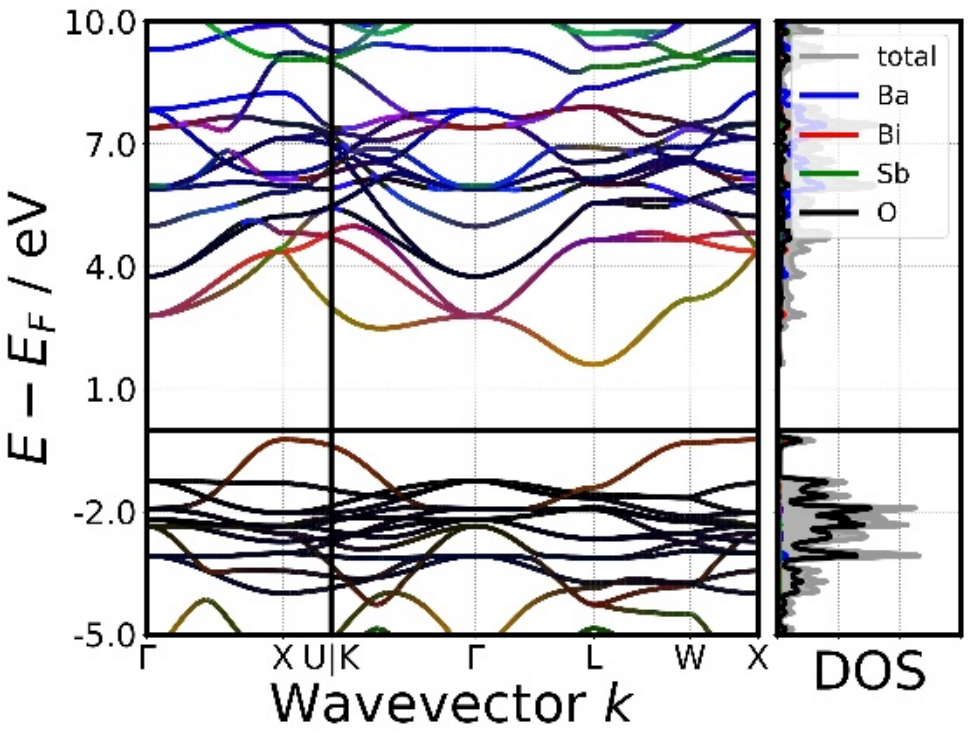

4. $Cu_2O$ (mp-361)

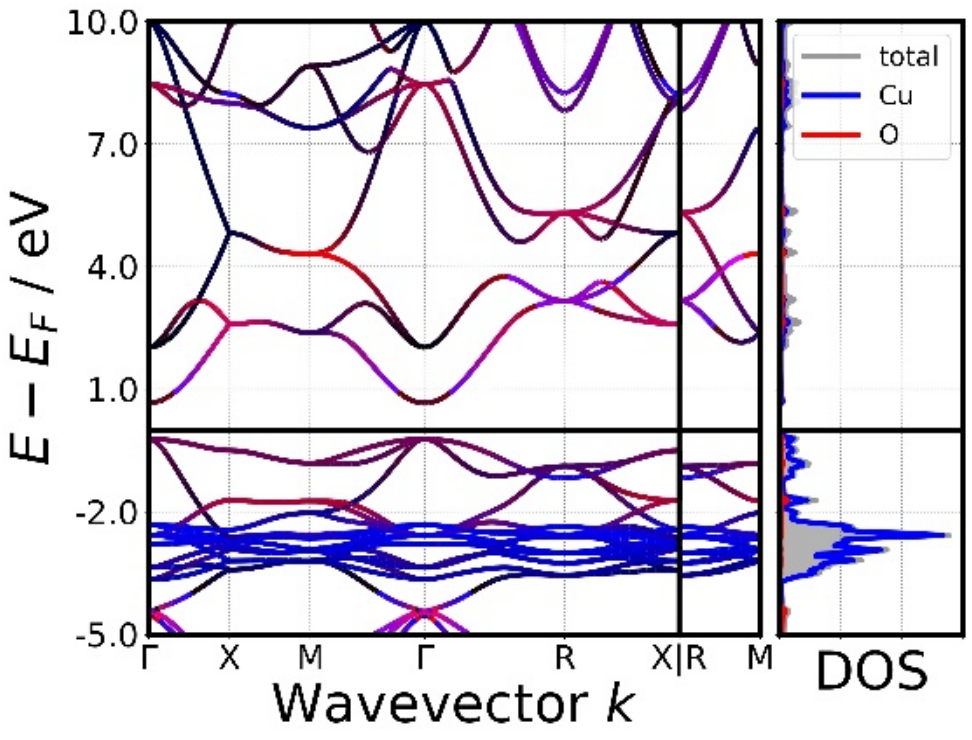

5. $BaSnO_3$ (mp-3163)

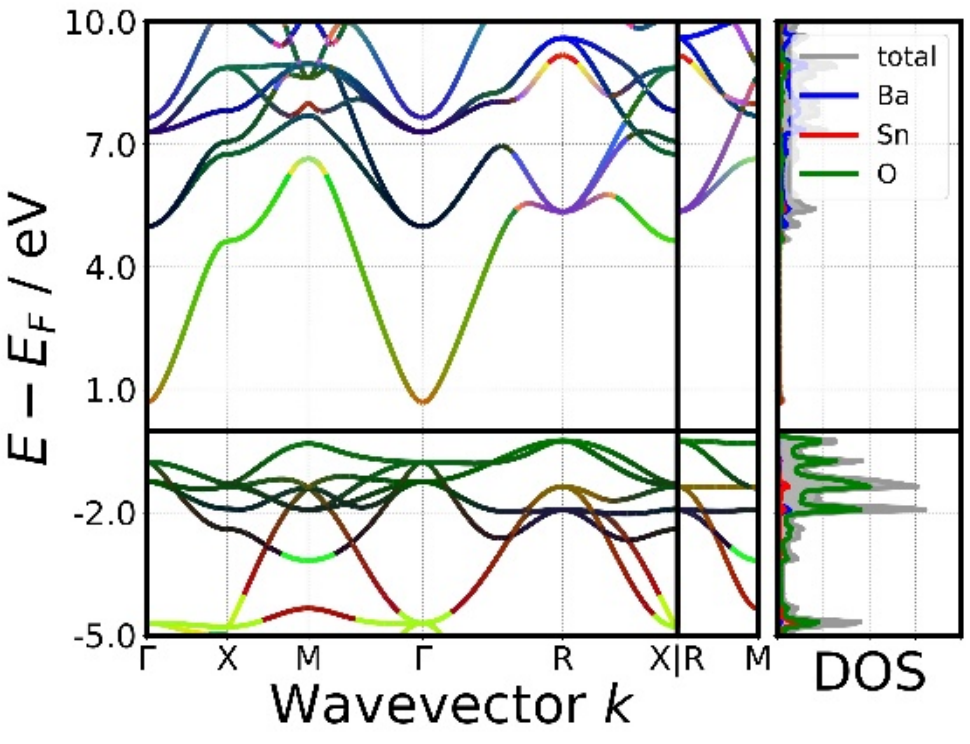

6. $BaPb_3O_4$ (mp-1214389)

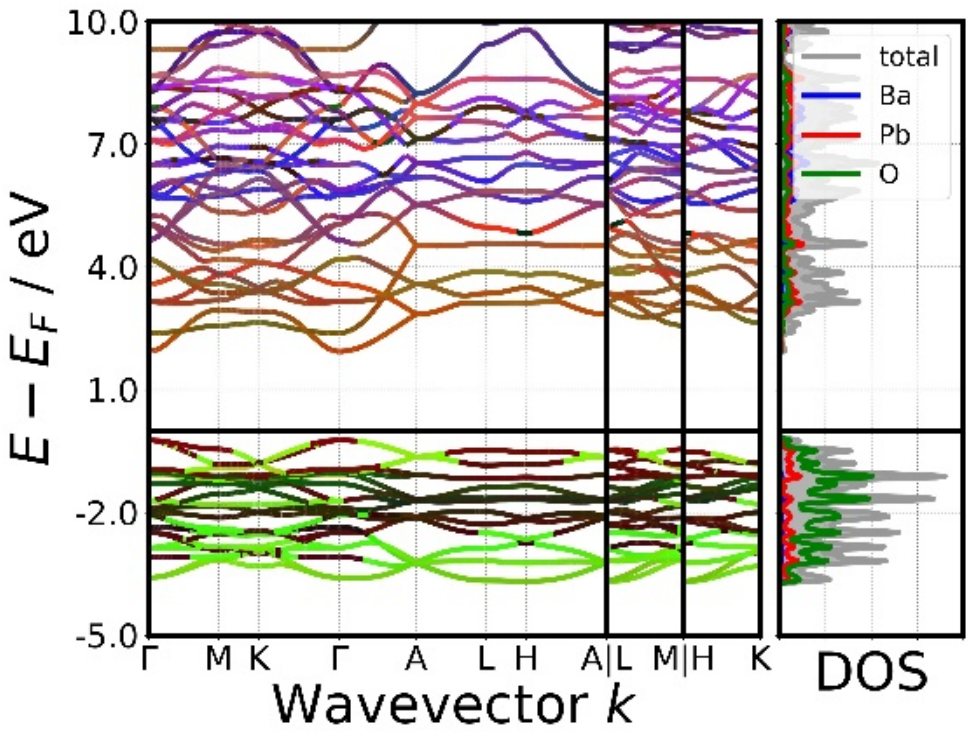

7. $BaCeO_3$ (mp-4900)

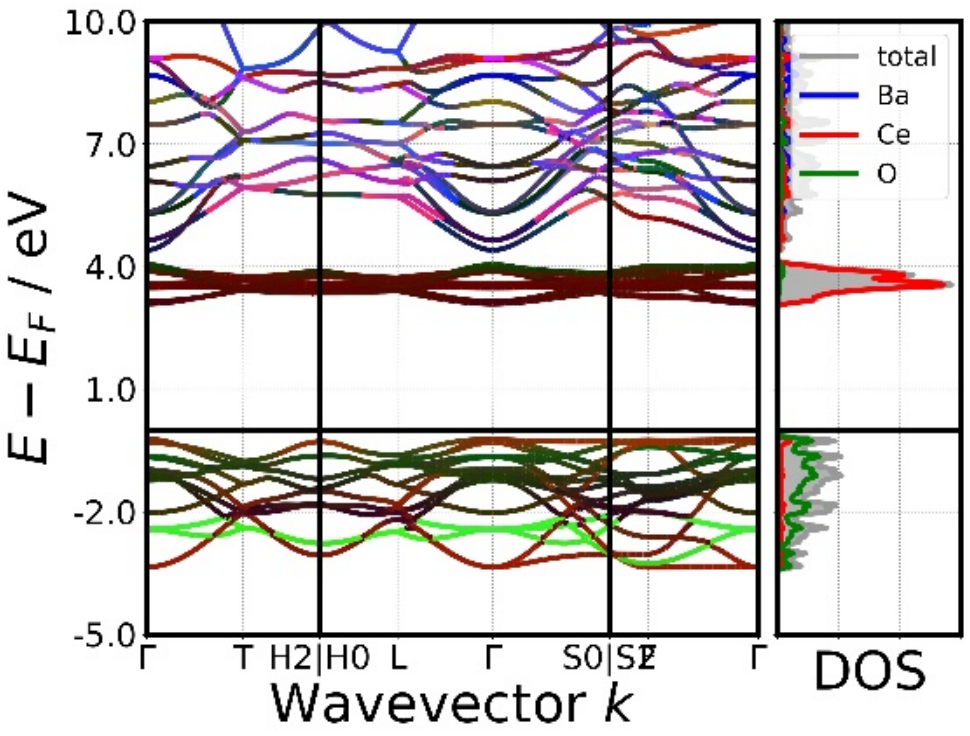

8. $Ba_2CePbO_6$ (mp-1228392)

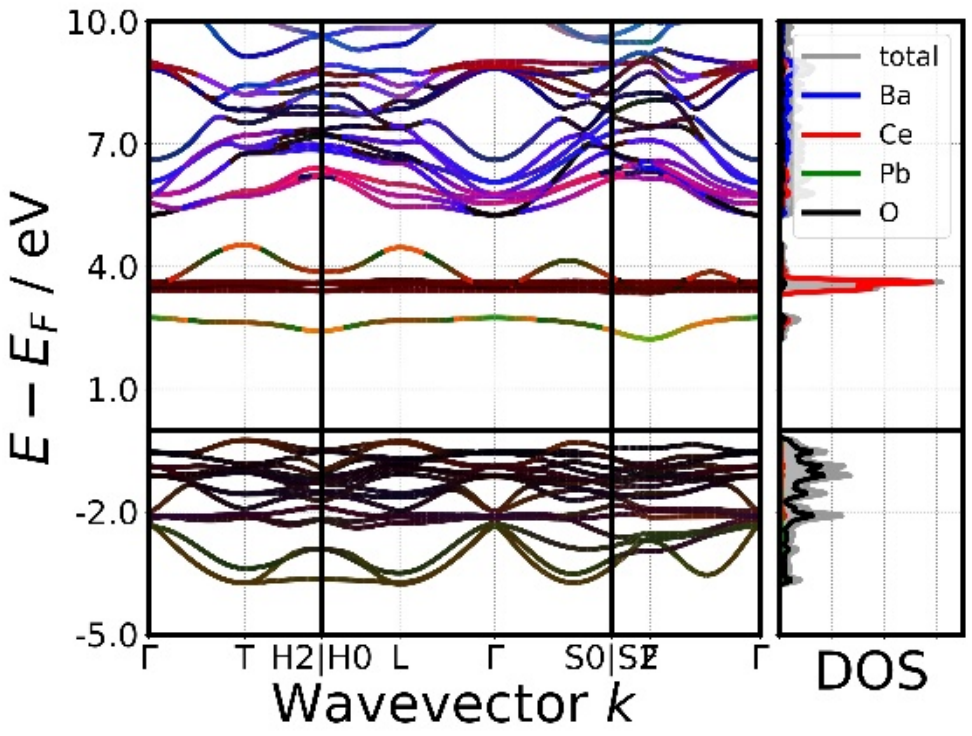

9. $Ba_2BiSbO_6$ (mp-23091)

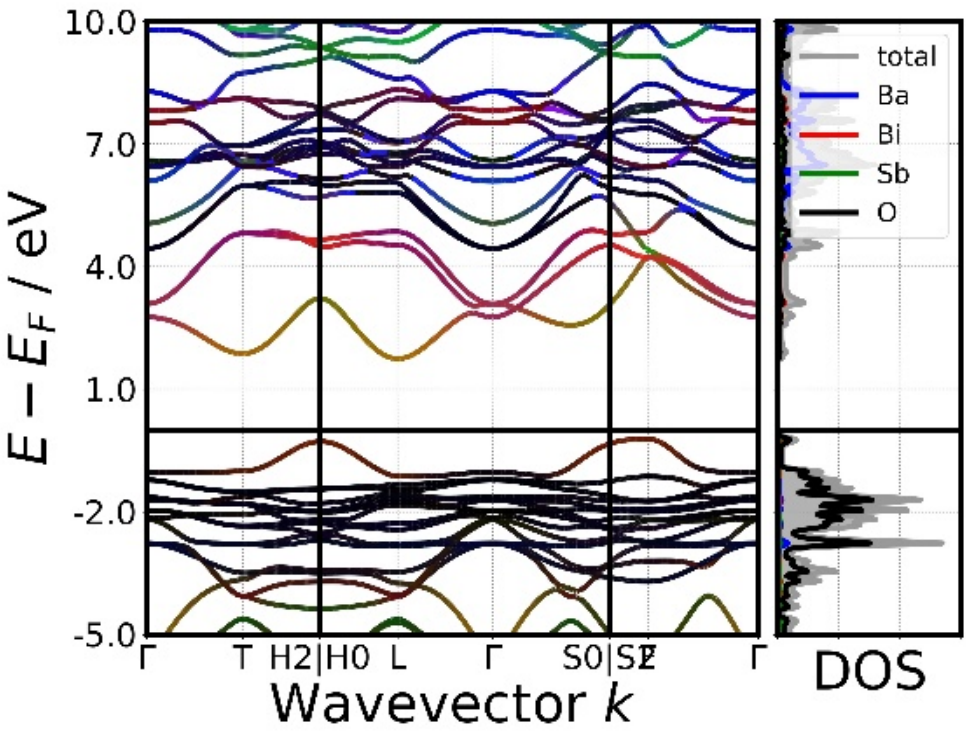

10. $Sm_2TeO_2$ (mp-16033)

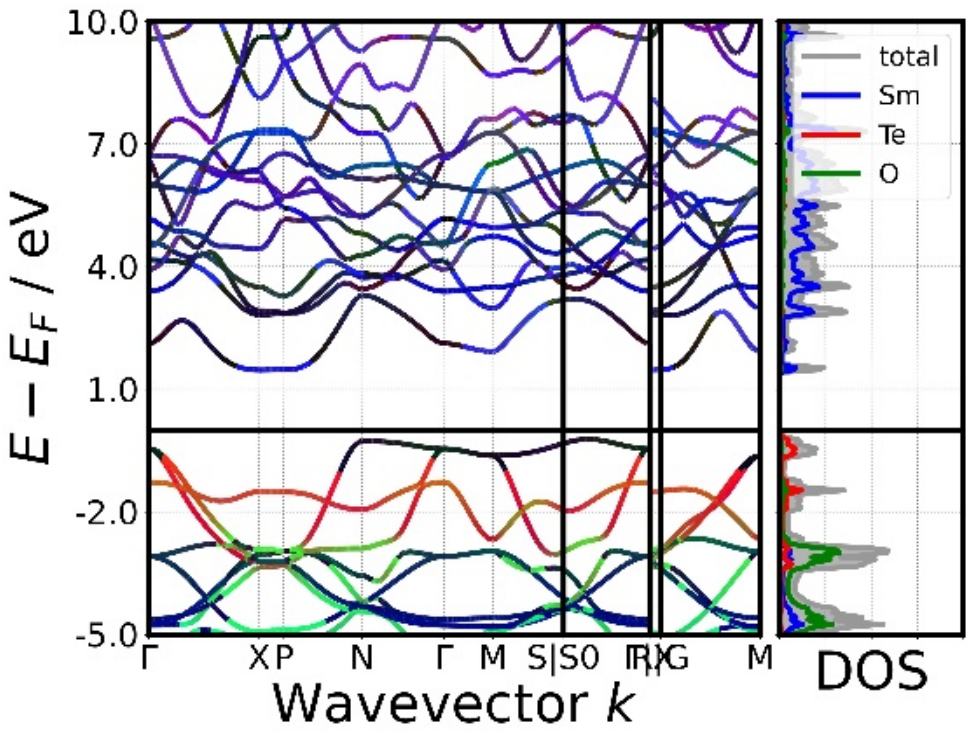

11. $Ba_2PbO_4$ (mp-20098)

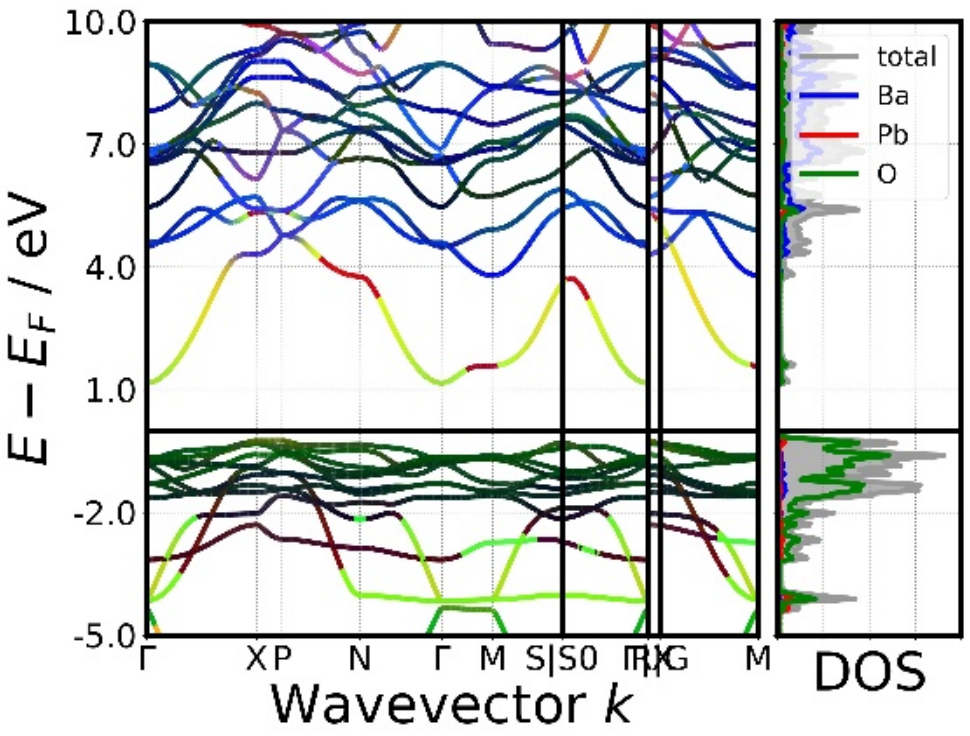

12. SnO (mp-2097)

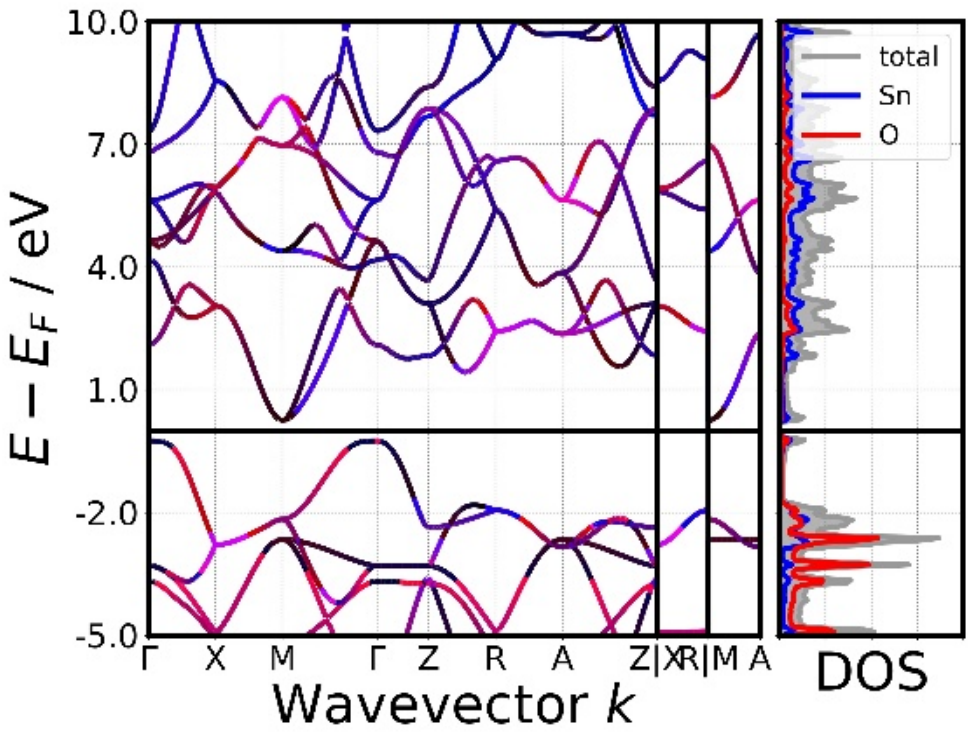

13. $SmBi_2ClO_4$ (mp-546152)

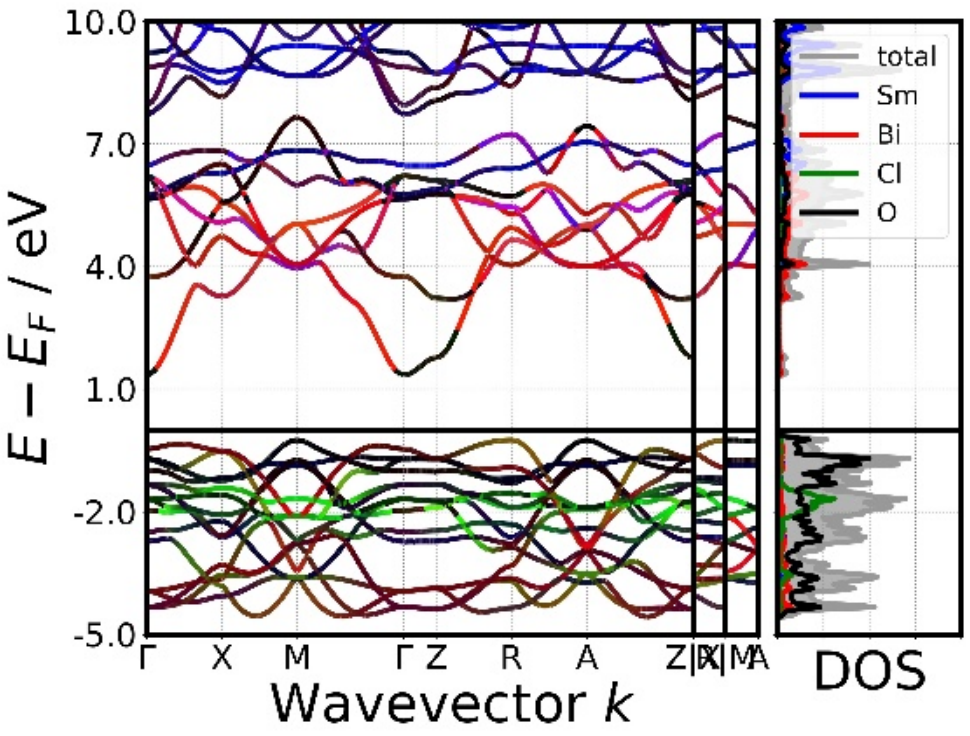

14. $Ca(AuO_2)_2$ (mp-2898)

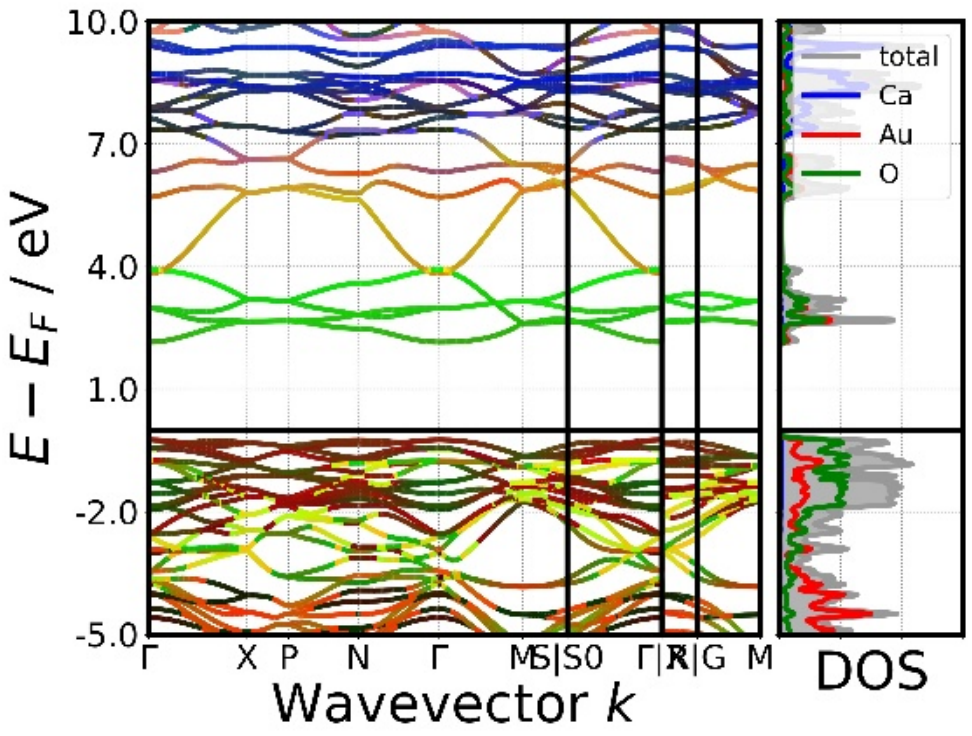

15. $Ba(AuO_2)_2$ (mp-9297)

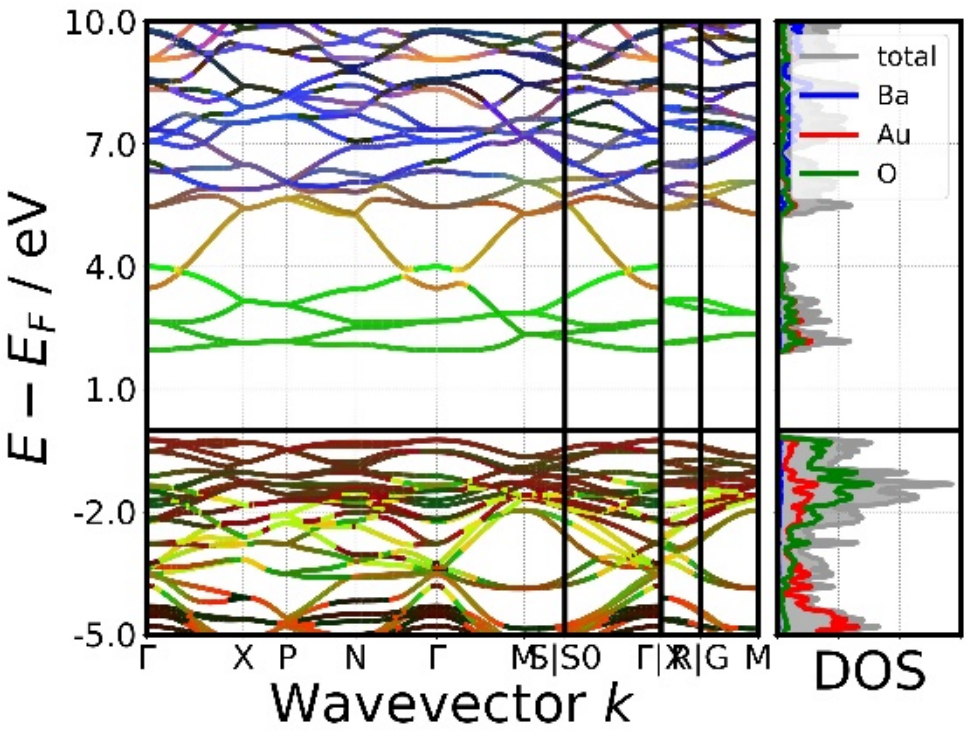

16. $Sr(AuO_2)_2$ (mp-9298)

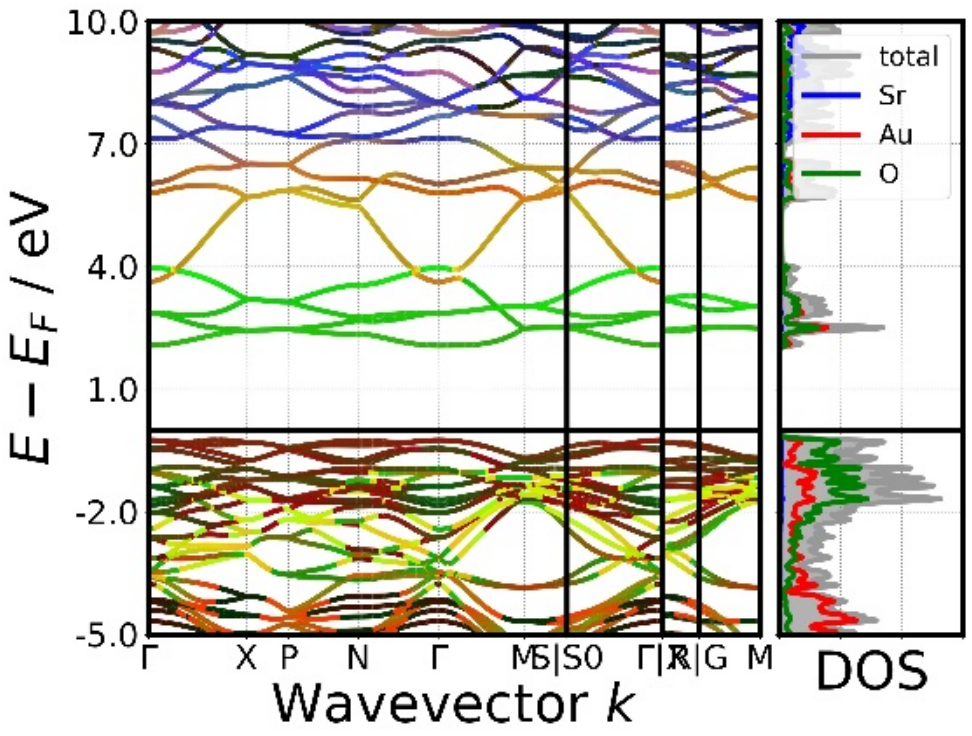

17. $BaSnO_3$ (mp-1178513)

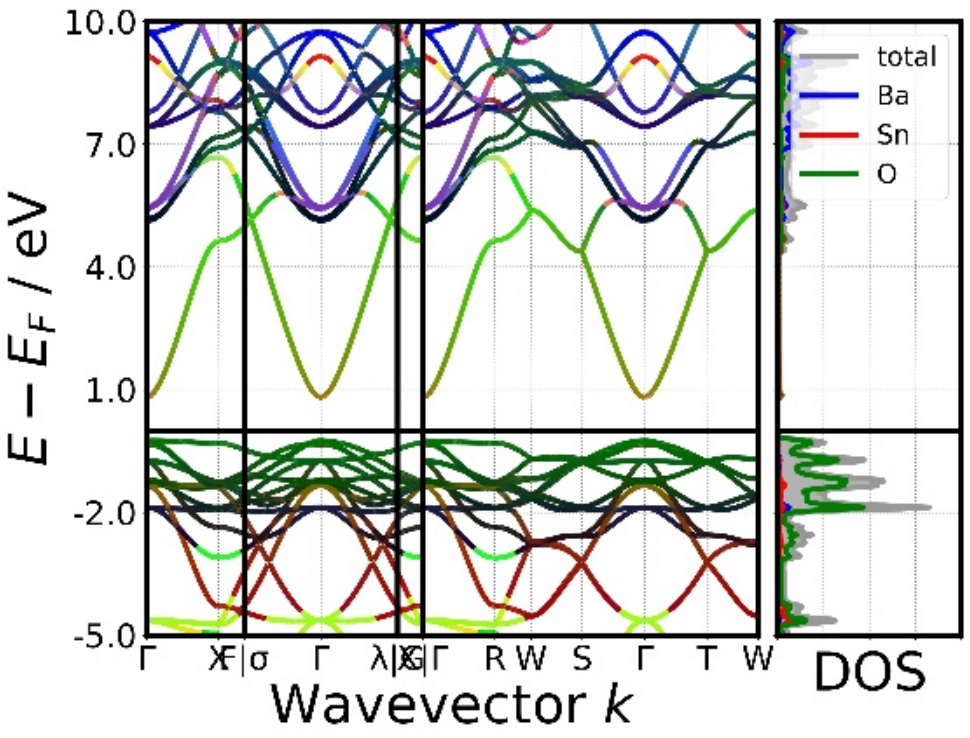

18. $BaPbO_3$ (mp-22230)

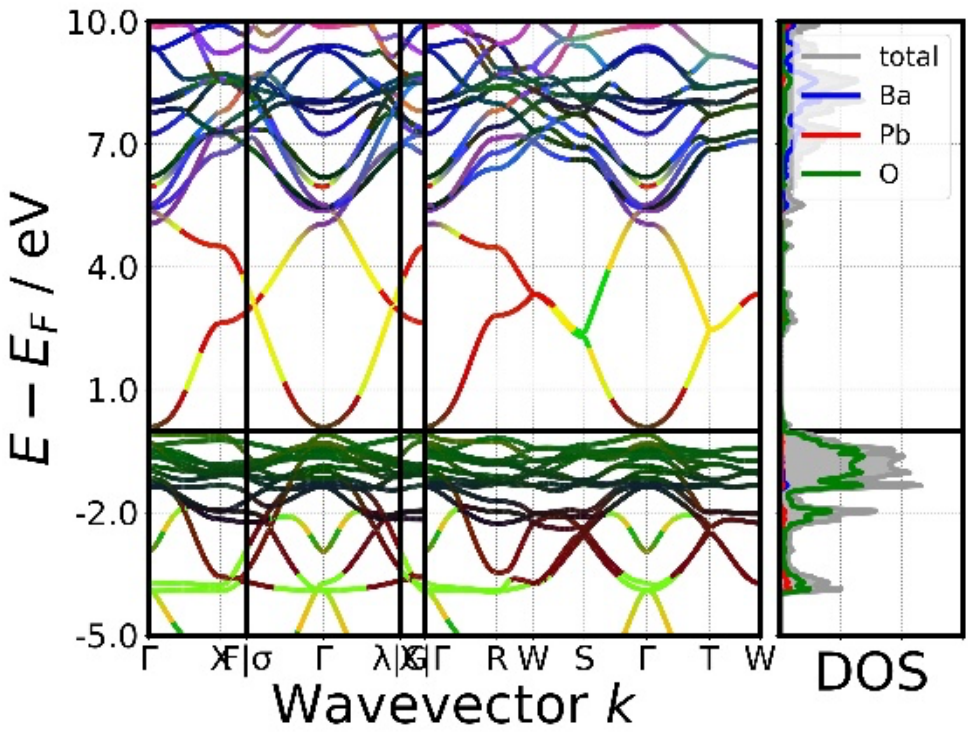

19. $BaCeO_3$ (mp-3316)

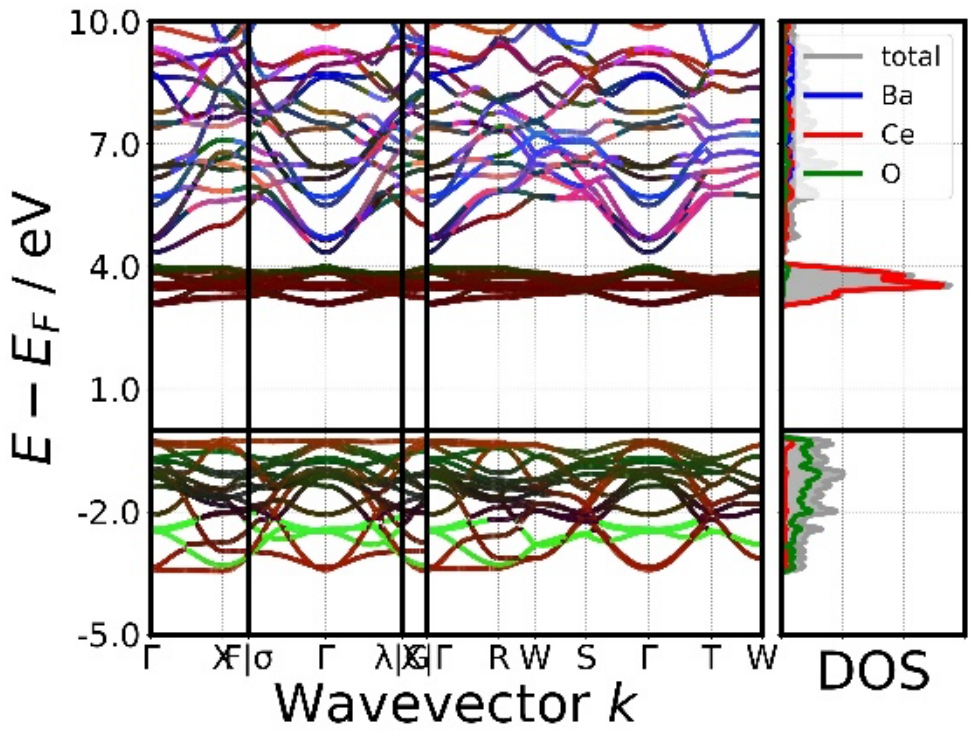

20. $Sr_2PdO_3$ (mp-4359)

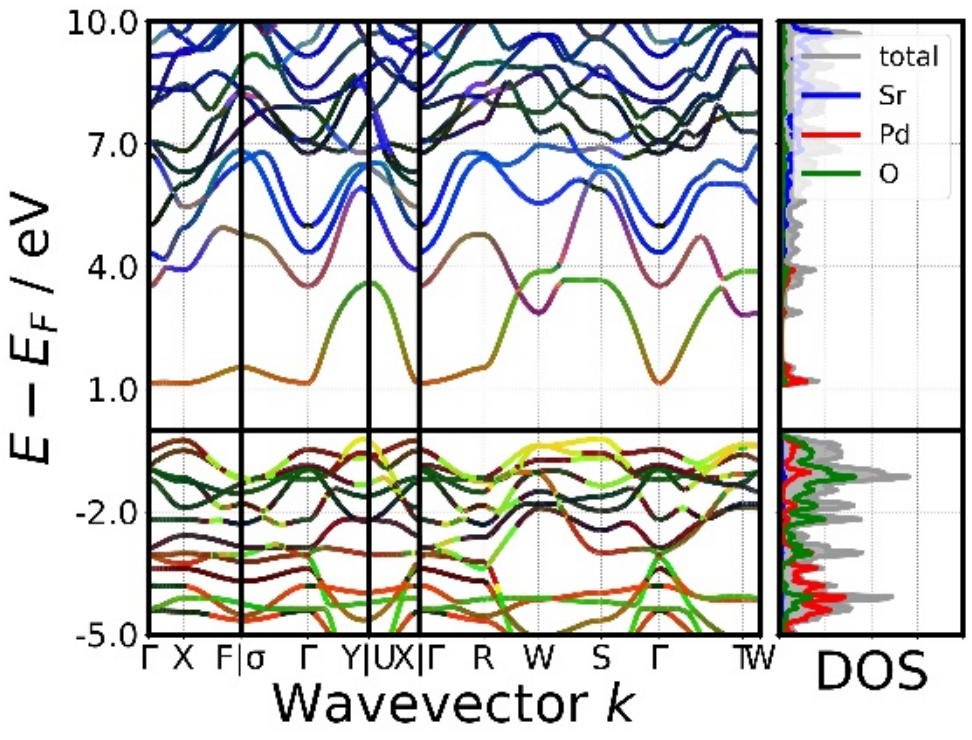

21. $SrPbO_3$ (mp-20489)

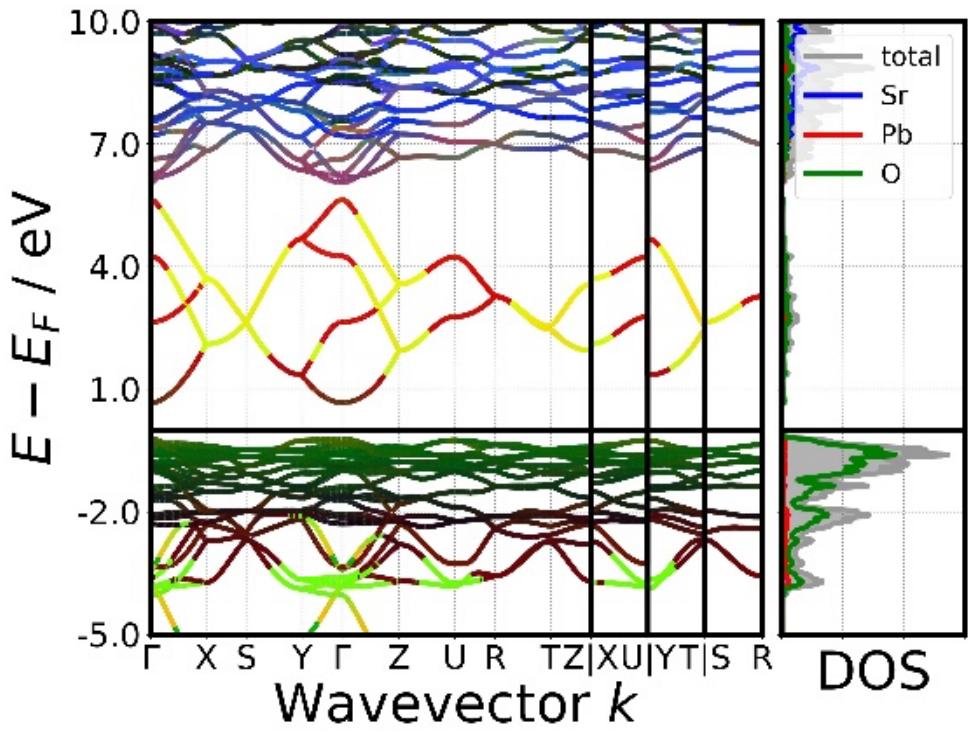

22. $SrCeO_3$ (mp-22428)

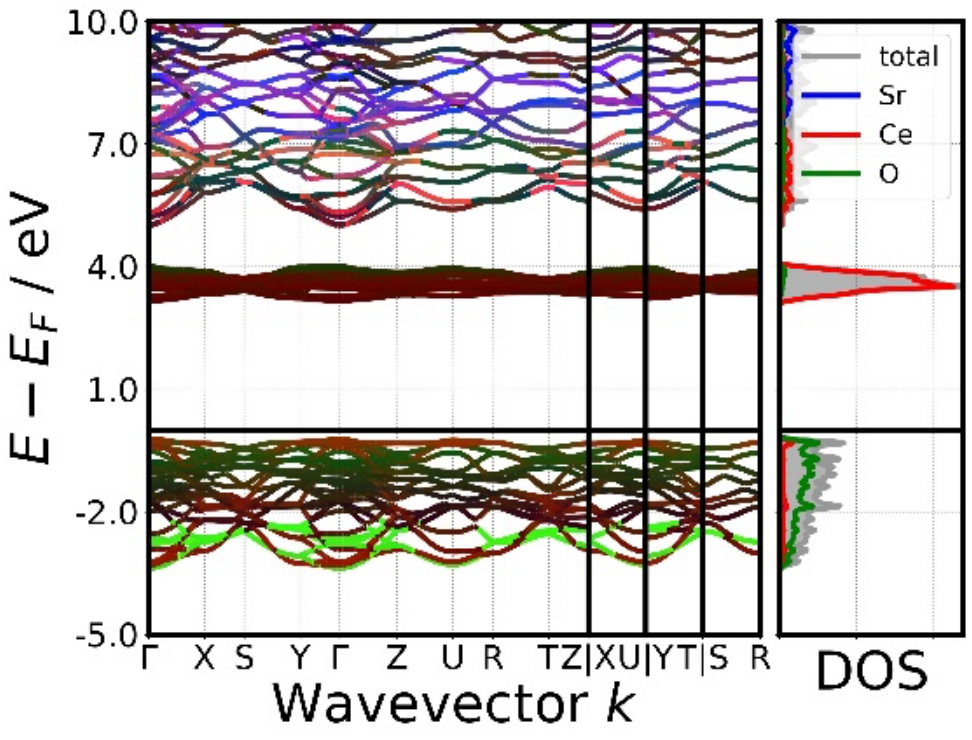

23. $BaCeO_3$ (mp-3187)

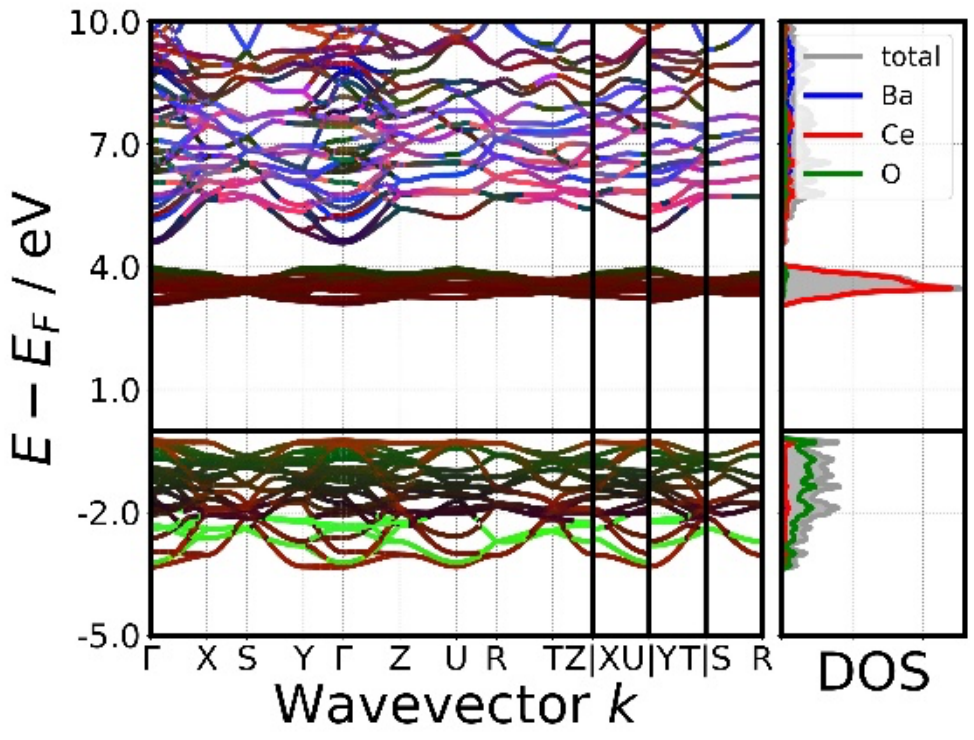

24. $Sr_2CeO_4$ (mp-15743)

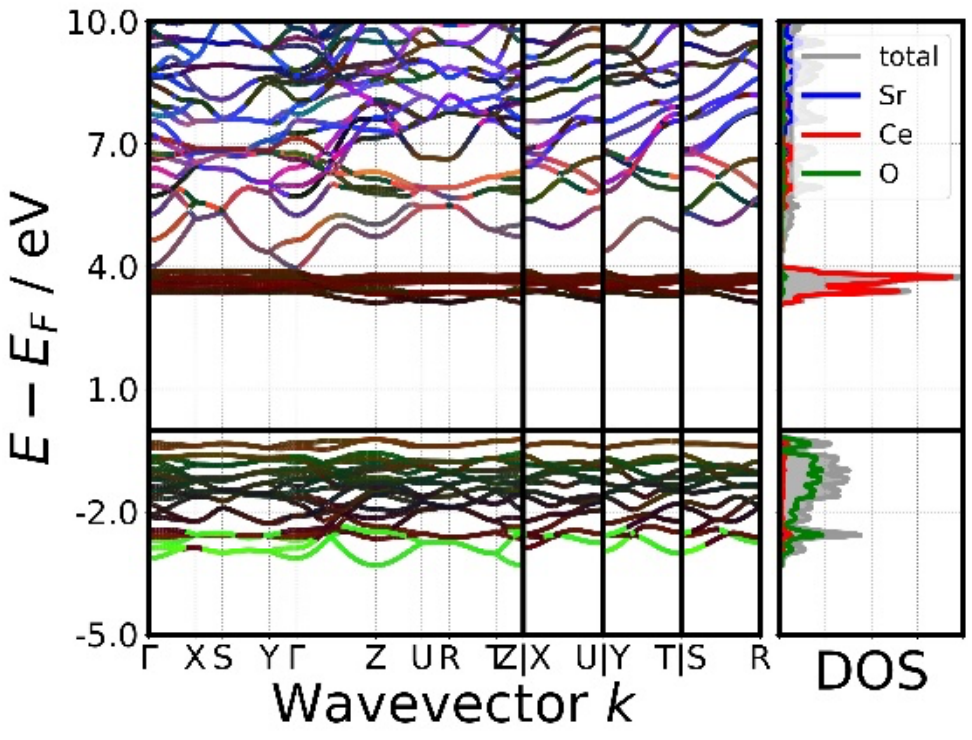

25. $Cd_2SnO_4$ (mp-5966)

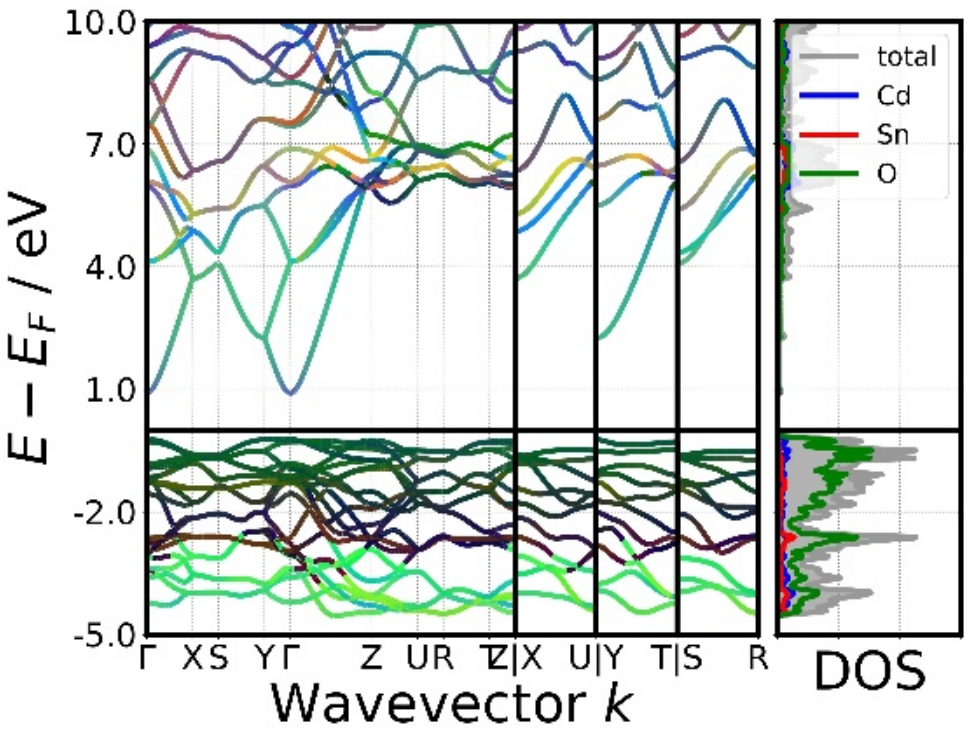

26. $Ba_2InGaO_5$ (mp-1106089)

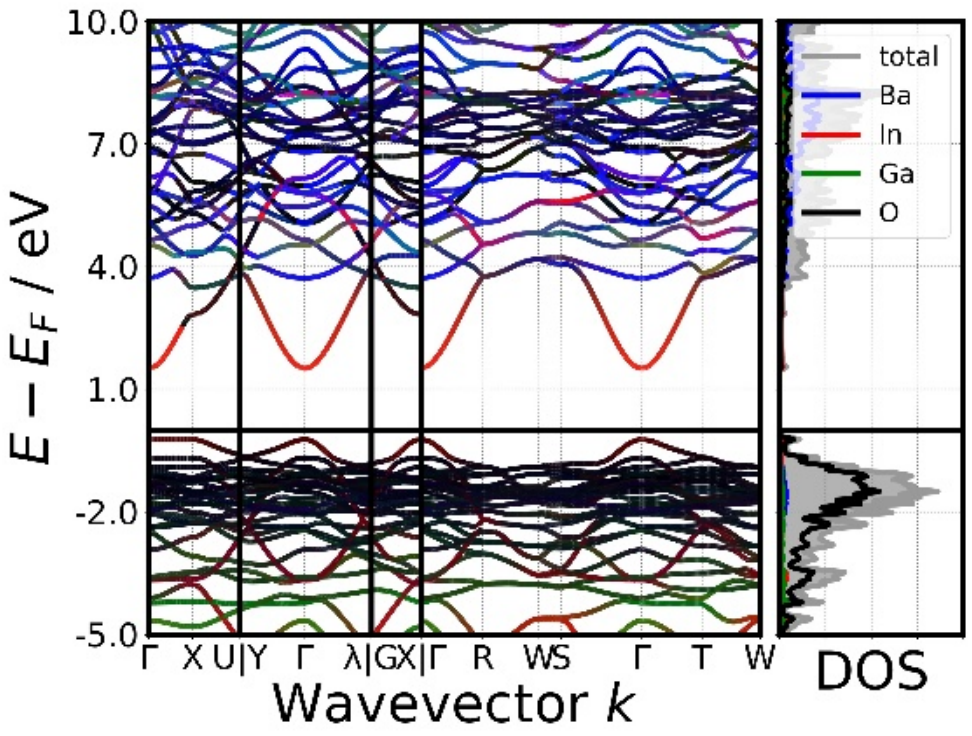

27. $Ba_2In_2O_5$ (mp-20546)

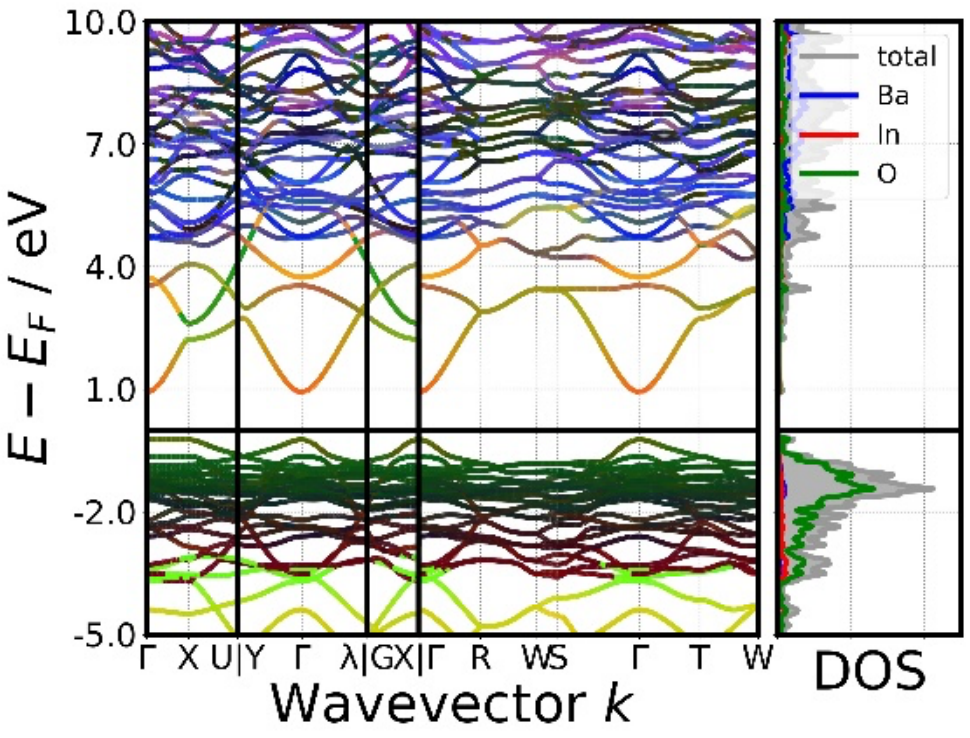

28. $BaSr(PbO_3)_2$ (mp-1227843)

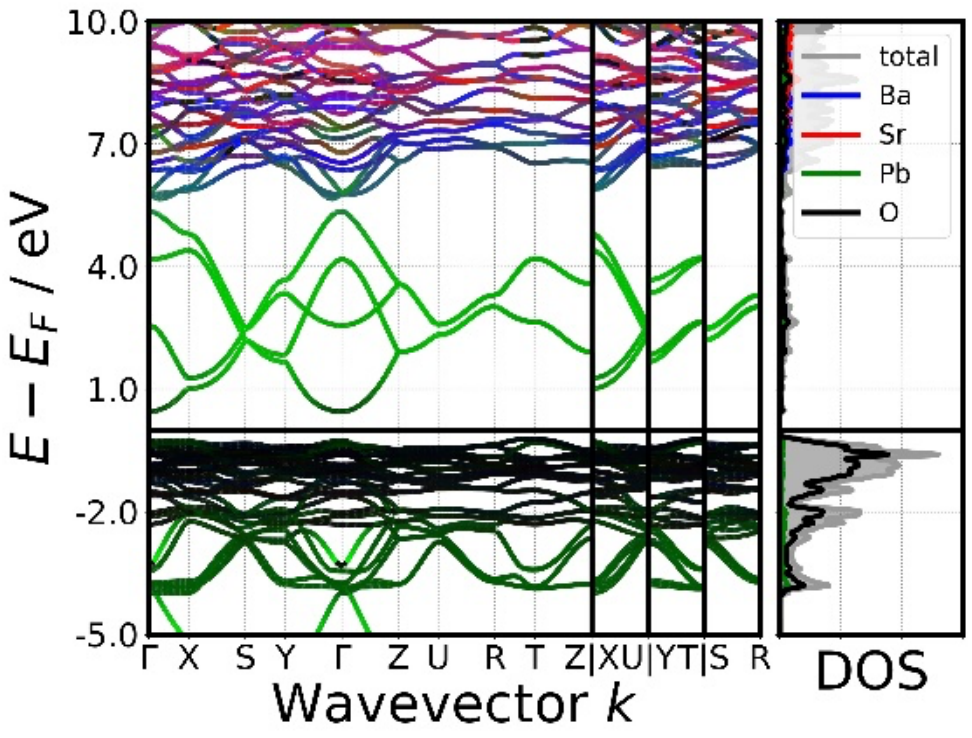

29. $Mn(RhO_2)_2$ (mp-554354)

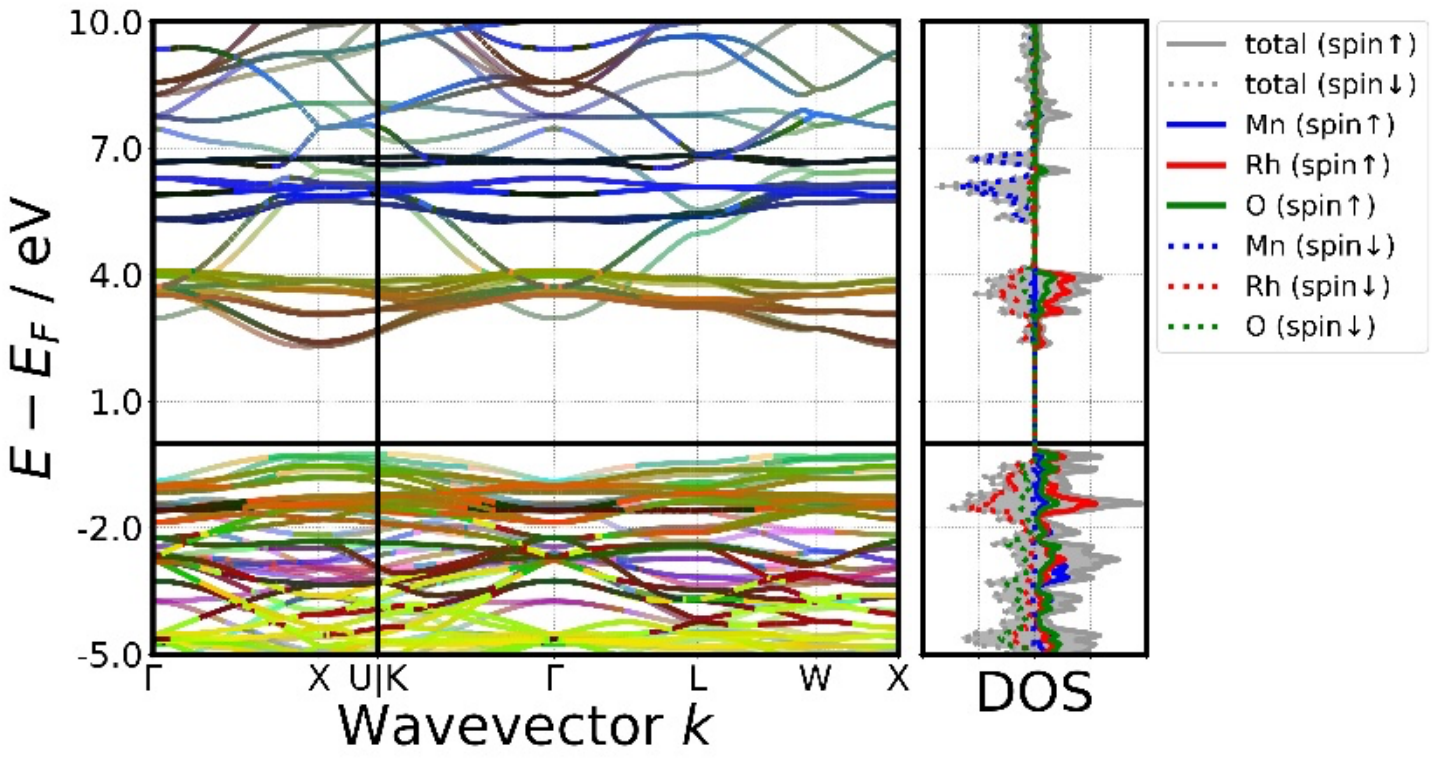

30. NiO (mp-19009)

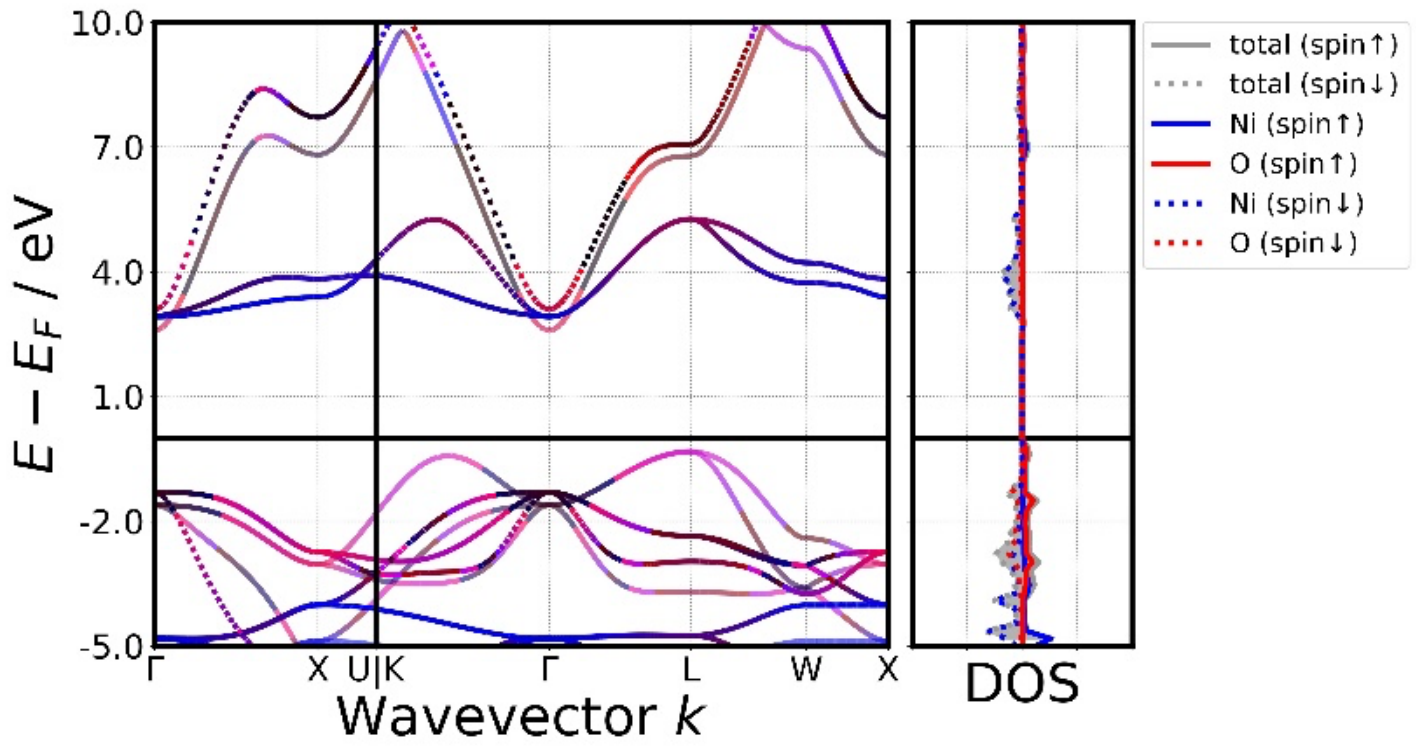

31. $Ba_4CeMn_3O_{12}$ (mp-561553)

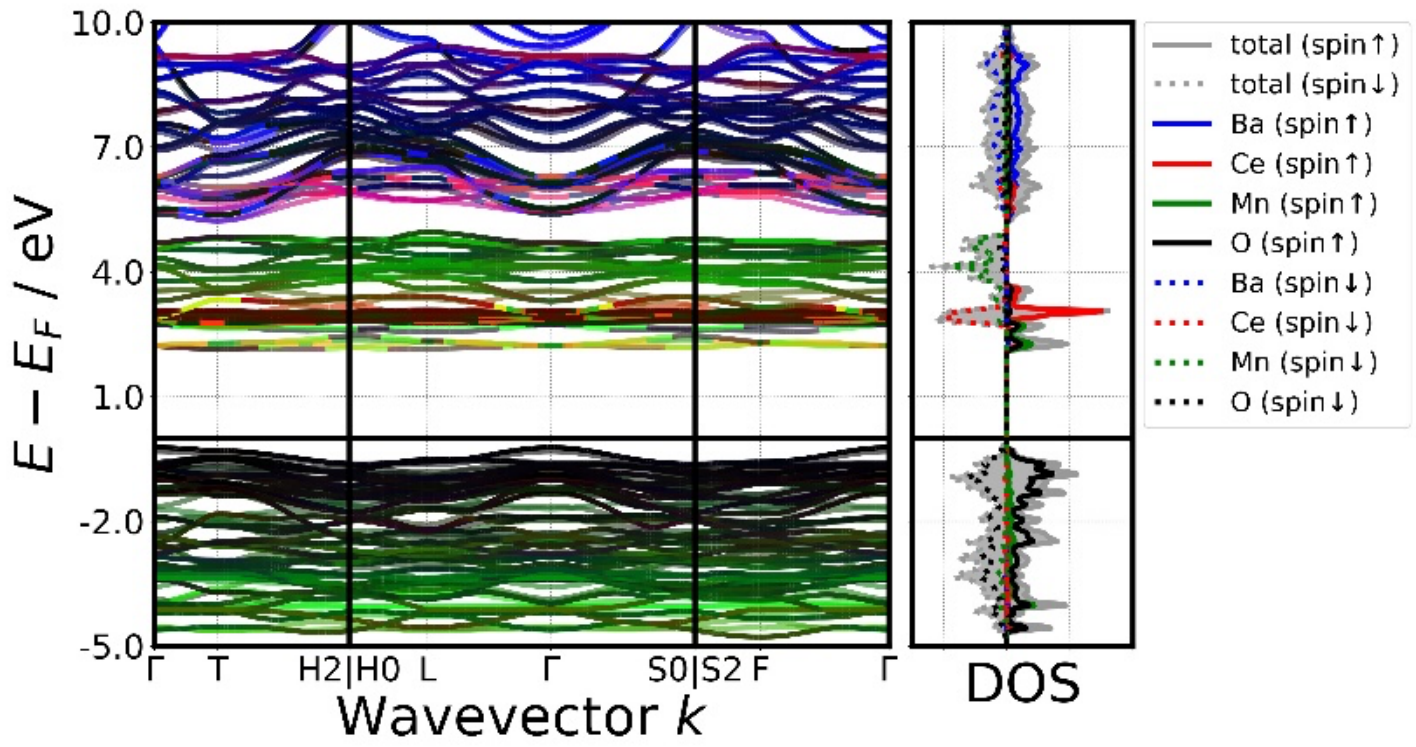

32. $Mn(PtO_2)_3$ (mp-18971)

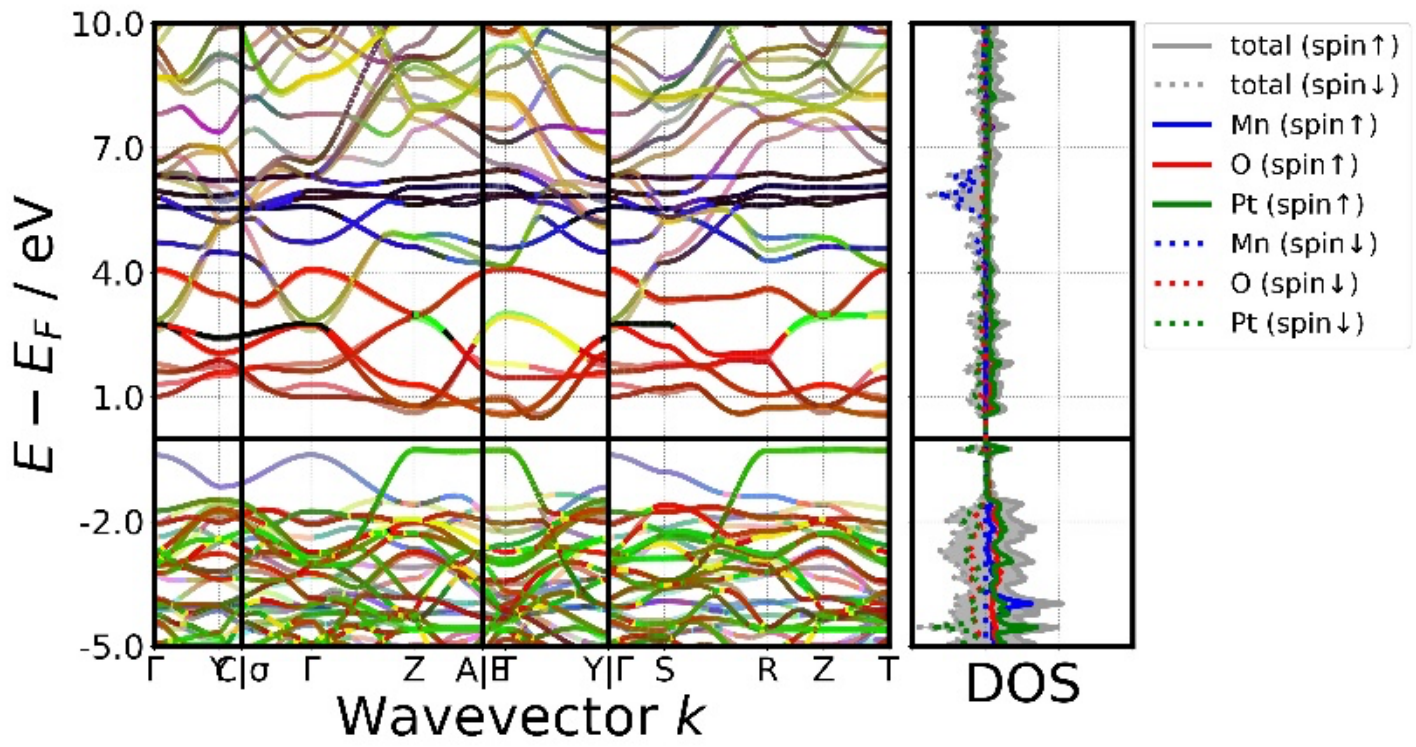

33. $SmMnO_3$ (mp-22203)

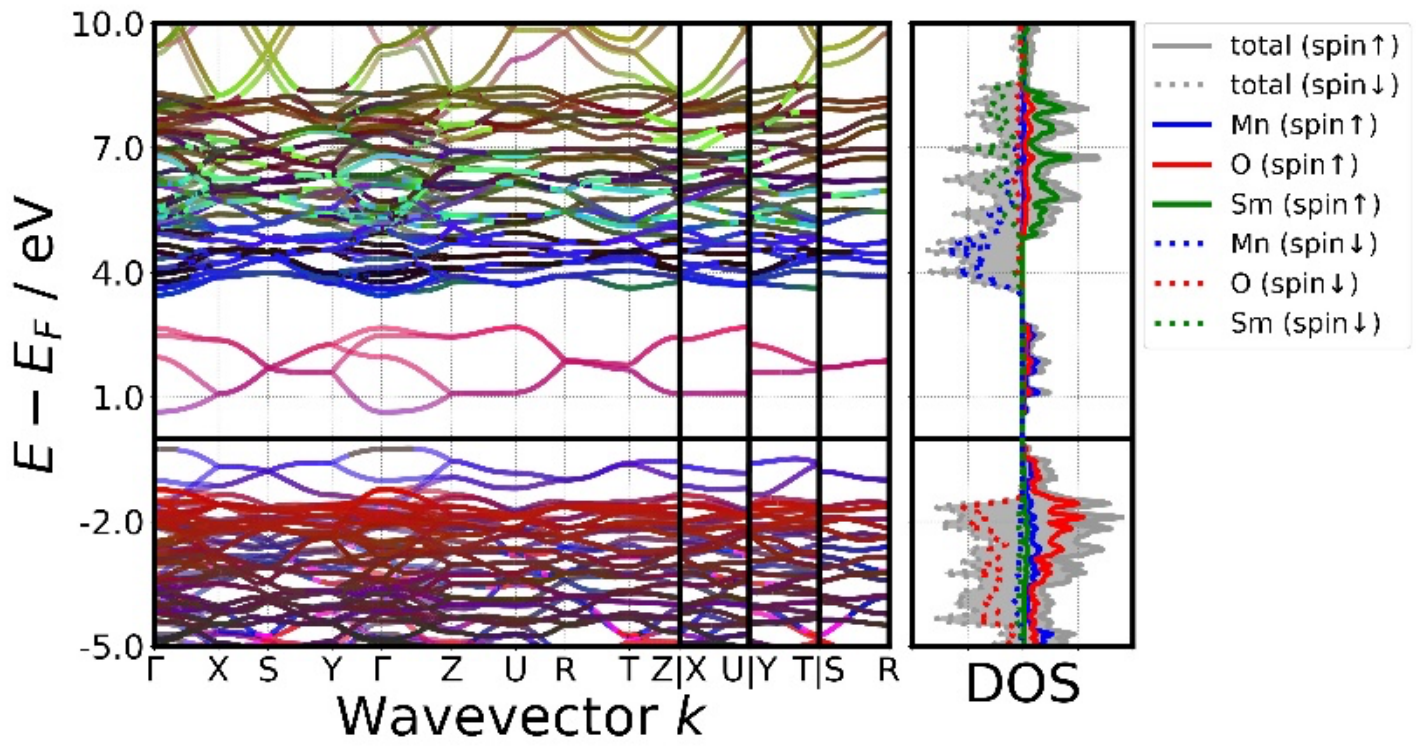

**Supplementary Note 10. Statistical performance of electrical transport properties for different metal elements in P-type and N-type**

we present the statistical mean performance of electrical transport properties for different metal elements in P-type and N-type, with the elements being ordered according to their position in the periodic table.

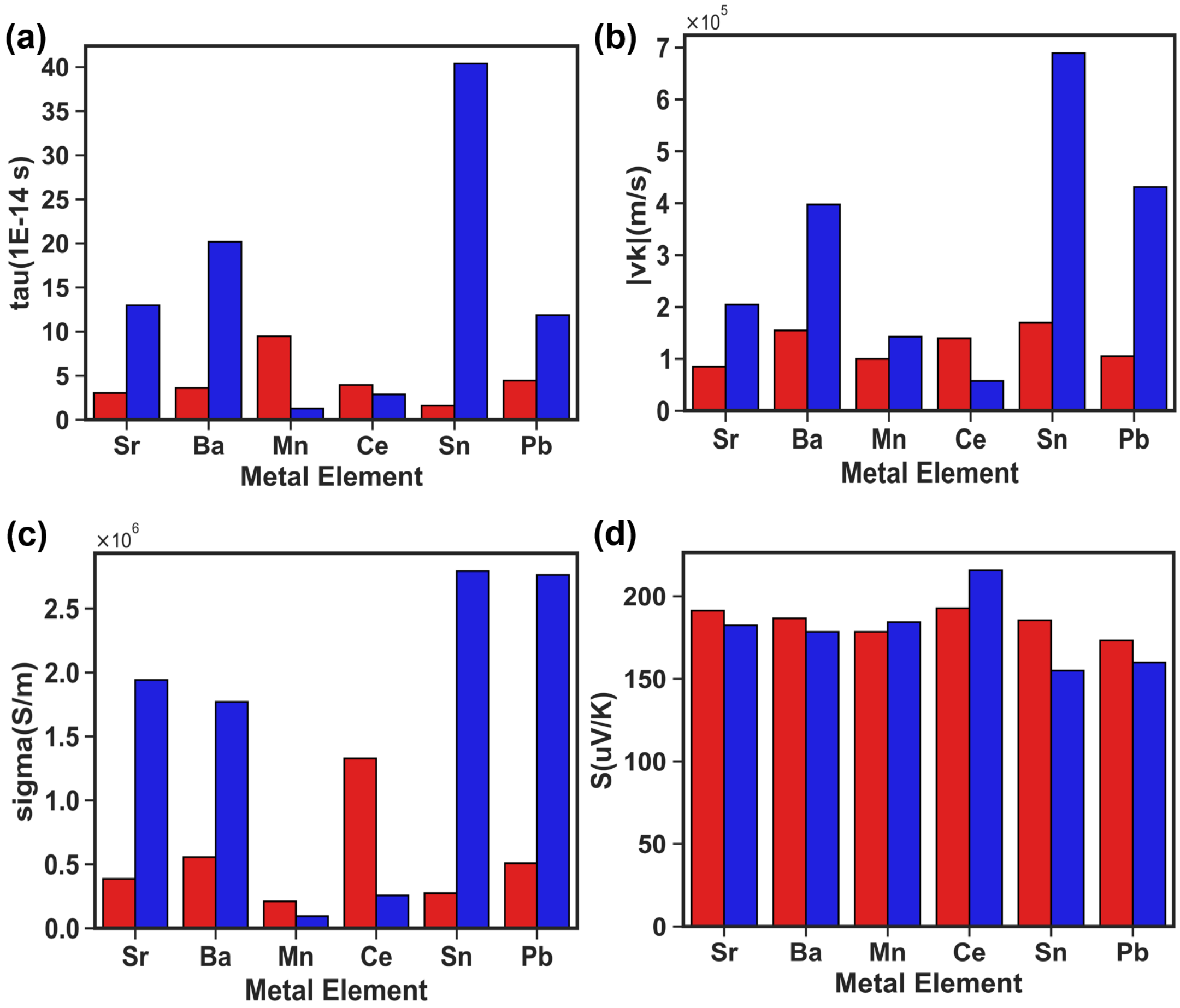

**Figure S16:** Statistical mean performance of electrical transport properties for different metal elements in P-type and N-type. (a) Statistics of electronic relaxation time. (b) Statistics of electron group velocity. (c) Statistics of sigma. (d) Statistics of Seebeck coefficient.

**Supplementary Note 11. Electrical transport performance for 33 candidate materials**

**Table S8:** Electrical transport information at $PF_{max}$ for 33 candidate materials

| MPID | compound | Band gap (eV) | DP (eV) VBM | Fitting atom VBM | DP (eV) CBM | Fitting atom CBM | Young's modulus (GPa) | σ (S/m) p-type | Seebeck (μV/K) p-type | PF ($\mu W cm^{-1} K^{-2}$) p-type | σ (S/m) n-type | Seebeck (μV/K) n-type | PF ($\mu W cm^{-1} K^{-2}$) n-type |
|---|---|---|---|---|---|---|---|---|---|---|---|---|---|
| mp-20194 | $CeO_2$ | 2.98 | 0.92 | O | 1.21 | Ce | 186.91 | 4047165.70 | 179.56 | 1304.93 | 95099.49 | -187.75 | 33.52 |
| mp-1206536 | $Ba_2InSbO_6$ | 0.95 | 2.06 | O | -2.86 | In | 196.88 | 331475.04 | 182.80 | 110.76 | 3546081.10 | -159.10 | 897.65 |
| mp-545603 | $Ba_2BiSbO_6$ | 1.83 | -5.78 | O | -6.88 | O | 152.76 | 158517.77 | 188.73 | 56.46 | 208510.28 | -171.81 | 61.55 |
| mp-361 | $Cu_2O$ | 0.88 | -2.84 | Cu | -4.95 | Cu | 22.71 | 8211.59 | 165.22 | 2.24 | 34793.93 | -165.24 | 9.50 |
| mp-3163 | $BaSnO_3$ | 0.93 | 2.01 | O | -5.20 | Sn | 220.28 | 469811.16 | 180.06 | 152.32 | 1926820.30 | -131.75 | 334.47 |
| mp-1214389 | $BaPb_3O_4$ | 2.17 | 0.41 | O | 2.25 | Pb | 32.18 | 955552.36 | 189.11 | 341.73 | 57063.33 | -176.02 | 17.68 |
| mp-4900 | $BaCeO_3$ | 3.31 | -1.71 | O | 1.33 | Ce | 130.93 | 718339.91 | 194.04 | 270.46 | 29097.32 | -199.91 | 11.63 |
| mp-1228392 | $Ba_2CePbO_6$ | 2.47 | -0.71 | O | 1.02 | Ce | 140.60 | 1346718.60 | 193.48 | 504.13 | 1854636.50 | -194.70 | 703.08 |
| mp-23091 | $Ba_2BiSbO_6$ | 1.96 | -5.36 | O | -6.93 | O | 147.15 | 110468.57 | 189.11 | 39.51 | 187487.85 | -164.12 | 50.50 |
| mp-16033 | $Sm_2TeO_2$ | 1.69 | 4.29 | Te | 7.84 | Sm | 95.93 | 244565.05 | 180.32 | 79.52 | 36459.77 | -189.20 | 13.05 |
| mp-20098 | $Ba_2PbO_4$ | 1.40 | 1.35 | O | -5.35 | O | 111.87 | 239352.69 | 187.76 | 84.38 | 164330.24 | -168.26 | 46.52 |
| mp-2097 | $SnO$ | 0.50 | -4.13 | Sn | 4.05 | Sn | 49.96 | 120377.65 | 176.27 | 37.40 | 136473.27 | -168.95 | 38.95 |
| mp-546152 | $SmBi_2ClO_4$ | 1.59 | -0.98 | O | 6.25 | Bi | 76.64 | 497123.40 | 188.25 | 176.18 | 130466.20 | -171.20 | 38.24 |
| mp-2898 | $Ca(AuO_2)_2$ | 2.39 | 1.42 | O | -1.30 | Au | 122.16 | 139955.68 | 198.19 | 54.98 | 292896.07 | -185.35 | 100.62 |
| mp-9297 | $Ba(AuO_2)_2$ | 2.20 | 0.65 | O | -1.84 | Au | 87.40 | 617169.85 | 192.68 | 229.12 | 80405.24 | -175.77 | 24.84 |
| mp-9298 | $Sr(AuO_2)_2$ | 2.31 | -0.78 | O | -1.52 | Au | 108.22 | 462180.51 | 196.26 | 178.02 | 139151.37 | -195.25 | 53.05 |
| mp-1178513 | $BaSnO_3$ | 1.06 | 2.54 | O | -2.64 | Sn | 215.64 | 258280.15 | 182.61 | 86.12 | 4718803.60 | -166.44 | 1307.17 |
| mp-22230 | $BaPbO_3$ | 0.18 | 3.37 | O | 2.16 | O | 104.21 | 88097.03 | 105.86 | 9.87 | 4250806.00 | -99.54 | 421.19 |
| mp-3316 | $BaCeO_3$ | 3.32 | -1.28 | O | 1.39 | Ce | 121.00 | 1174737.40 | 194.68 | 445.24 | 23539.89 | -204.75 | 9.87 |
| mp-4359 | $Sr_2PdO_3$ | 1.36 | 3.36 | Pd | 1.47 | Pd | 142.08 | 67877.18 | 181.00 | 22.24 | 1138181.60 | -183.52 | 383.33 |
| mp-20489 | $SrPbO_3$ | 0.88 | 1.84 | O | -2.00 | O | 127.60 | 66332.01 | 195.29 | 25.30 | 2207616.60 | -155.05 | 530.72 |
| mp-22428 | $SrCeO_3$ | 3.38 | -0.86 | O | 1.18 | Ce | 133.16 | 1323832.70 | 196.19 | 509.53 | 26149.92 | -182.23 | 8.68 |
| mp-3187 | $BaCeO_3$ | 3.33 | -1.15 | O | 1.22 | Ce | 127.15 | 1280273.60 | 197.10 | 497.34 | 32657.69 | -186.89 | 11.41 |
| mp-15743 | $Sr_2CeO_4$ | 3.36 | 1.15 | O | 1.19 | Ce | 117.61 | 69338.41 | 200.63 | 27.91 | 23532.43 | -185.34 | 8.08 |
| mp-5966 | $Cd_2SnO_4$ | 1.13 | -1.44 | O | -2.27 | Cd | 116.66 | 185226.33 | 191.36 | 67.83 | 3942626.40 | -143.04 | 806.66 |
| mp-1106089 | $Ba_2InGaO_5$ | 1.74 | -1.69 | O | -1.86 | In | 114.32 | 202969.01 | 220.17 | 98.39 | 3192146.00 | -173.38 | 959.56 |
| mp-20546 | $Ba_2In2O_5$ | 1.17 | -0.92 | O | -2.62 | In | 98.44 | 562043.67 | 210.23 | 248.41 | 1131729.70 | -174.23 | 343.57 |
| mp-1227843 | $BaSr(PbO_3)_2$ | 0.67 | 1.31 | O | -1.17 | O | 124.92 | 152030.97 | 171.48 | 44.70 | 7624574.00 | -142.91 | 1557.13 |
| mp-554354 | $Mn(RhO_2)_2$ | 2.54 | -2.87 | O | -2.89 | O | 137.48 | 58830.01 | 196.45 | 22.70 | 170457.09 | -171.52 | 50.15 |
| | | | -1.98 | O | -3.27 | O | | | | | | | |
| mp-19009 | NiO | 2.91 | -5.84 | O | 2.67 | Ni | 126.79 | 38661.91 | 176.03 | 11.98 | 1043913.70 | -172.14 | 309.33 |
| | | | -5.97 | O | 2.76 | Ni | | | | | | | |
| mp-561553 | $Ba_4CeMn_3O_{12}$ | 2.37 | 1.58 | O | -2.16 | Mn | 107.56 | 182301.90 | 162.93 | 48.39 | 43001.71 | -204.27 | 17.94 |
| | | | 1.23 | O | 1.39 | Mn | | | | | | | |
| mp-18971 | $Mn(PtO_2)_3$ | 0.73 | -2.94 | Pt | -3.76 | O | 197.07 | 500756.58 | 168.03 | 141.38 | 123738.22 | -175.24 | 38.00 |
| | | | 0.99 | Pt | -3.30 | O | | | | | | | |
| mp-22203 | $SmMnO_3$ | 0.88 | -2.85 | O | -7.31 | Mn | 175.36 | 31291.08 | 178.20 | 9.94 | 16385.30 | -177.32 | 5.15 |
| | | | -1.79 | O | 2.68 | Mn | | | | | | | |

*Note: The deformation potential of the magnetic material is calculated separately for spin up and spin down. The upper part of the table presents the DP results for spin up, while the lower part shows the DP results for spin down.